\documentclass[preprints,article,accept,pdftex,10pt,a4paper]{Definitions/mdpi}

\usepackage{amsmath}
\usepackage{amsthm}
\usepackage{amscd}
\usepackage{amsfonts}
\usepackage[psamsfonts]{amssymb}
\usepackage{graphicx,epsfig,subfigure}
\usepackage{hyperref}
\usepackage{bm}

\def\bb#1{{\pmb{#1}}}

\def\+{\!+\!}
\def\-{\!-\!}
\def\={\!=\!}

\def\<{\langle}
\def\>{\rangle}

\firstpage{1} 
\pubvolume{xx}
\issuenum{1}
\articlenumber{5}
\pubyear{2018}
\copyrightyear{2018}
\history{J. Phys.: Condens. Matter 31 (2019) 383001 \hspace{1.2in} \href{https://iopscience.iop.org/article/10.1088/1361-648X/ab273f}{https://doi.org/10.1088/1361-648X/ab273f}}

\Title{Topical Review on Skyrmions and Hall Transport}

\Author{Bom Soo Kim $^{1,\dagger}$\orcidA{}*}

\AuthorNames{Bom Soo Kim}

\address{%
$^{1}$ \quad bkim2@loyola.edu}

\firstnote{Department of Physics, Loyola University Maryland,  
	4501 N. Charles Street, Baltimore, MD 21210, USA} 

\abstract{We review recent progresses towards an understanding of the Skyrmion Hall transport in insulating as well as conducting materials. First, we consider a theoretical breakthrough based on the quantum field theory Ward identity, {\it a first principle analysis}, relying on symmetries and conservation laws. Broken parity (inversion) symmetry plays a crucial role in Skyrmion Hall transport. In addition to the well known thermal and electric Hall conductivities, our analysis has led us to the discovery of a new and unforeseen physical quantity, Hall viscosity - an anti-symmetric part of the viscosity tensor. We propose a simple way to confirm the existence of Hall viscosity in the measurements of Hall conductivity as a function of momentum. We provide various background materials to assist the readers to understand the quantum field theory Ward identity. \\
\indent In the second part, we review recent theoretical and experimental advancements of the Skyrmion Hall effects and the topological (Magnon) Hall effects for conducting (insulting) magnets. For this purpose, we consider two enveloping themes: spin torque and thermo-electromagnetic effect. First, we overview various spin torques, such as spin transfer torque, spin-orbit torque, and spin Hall torque, and generalized Landau-Lifshitz-Gilbert equations and Thiele equations using a {\it phenomenological approach}. Second, we consider irreversible thermodynamics to survey possible thermo-electromagnetic effects, such as Seebeck, Peltier and Thompson effects in the presence of the electric currents, along with the Hall effects in the presence of a background magnetic field. Recently developed spin Seebeck effects are also a significant part of the survey. We also accommodate extensive background materials to make this review self-contained. Finally, we revisit the Skyrmion Hall transport from the Ward identity view point.  }


\begin{document}
	
\tableofcontents

\bigskip

\section{A comprehensive guide to tackle the complexity of Skyrmion transports.}

Recent years have witnessed rich and novel physical phenomena in the systems with broken parity invariance. For example, chiral magnetic Skyrmions have been discovered experimentally in real condensed matter systems \cite{SkyrmionExp1}\cite{SkyrmionExp2}\cite{Jonietz2010}\cite{Heinze2011} based on earlier theoretical studies \cite{SkyrmionTheory1}\cite{SkyrmionTheory2}\cite{SkyrmionTheory3}\cite{SkyrmionTheory5}\cite{SkyrmionTheory6}\cite{SkyrmionTheory4}\cite{SkyrmionTheory7}. Theoretically, Hydrodynamics has been re-evaluated in the absence of parity and boost symmetries and various new transport coefficients have been added to the elementary Hydrodynamics \cite{Jensen:2011xb}\cite{Bhattacharya:2011tra}\cite{Hoyos:2013eza}. Parity breaking hydrodynamics confirms that there exists a new hydrodynamic transport coefficient, Hall viscosity, in the absence of parity symmetry \cite{Avron:1995}. This mysterious Hall viscosity has provoked extensive theoretical investigations especially in Quantum Hall systems. In particular, the Hall viscosity has been proposed to account for half of the angular momentum in systems with a mass gap, such as the integer and fractional quantum Hall systems \cite{Read:2008rn}\cite{Read:2010epa}. Once one considers a Galilean invariant system with a background magnetic field, there is a similar relation between the Hall viscosity and Hall conductivity \cite{Hoyos:2011ez}. The theoretical understanding for the Hall viscosity and its relation to various other physical quantities have been firmly established \cite{Bradlyn:2012ea}\cite{Hoyos:2014lla}\cite{Hoyos:2015yna}. Nevertheless, its experimental confirmation is still lacking. 
  
This review serves as an attempt to understand various thermo-electromagnetic transport phenomena of Skyrmions in conducting and insulating materials in view of searching for Hall viscosity in the Skyrmion physics. Numerous ongoing Skyrmion experiments could be used to confirm or measure the Hall viscosity. Relevant transport coefficients include electric (Hall) conductivity, thermal (Hall) conductivity and Seebeck effects along with spin Hall conductivity and spin Seebeck effect. They are reviewed below in \S \ref{sec:ThermoEMBackground} in a systematic fashion following the framework of irreversible thermodynamics \cite{Callen1948}\cite{CallenThermodynamics}. We also extend this to include the role of spins by using two different electron species, up-spin and down-spin electrons. Spin Seebeck effect is a new research area that has been established around 2008. We also review them in a self-contained manner. 

\begin{figure}[h]
	\begin{center}
		\includegraphics[width=0.38\textwidth]{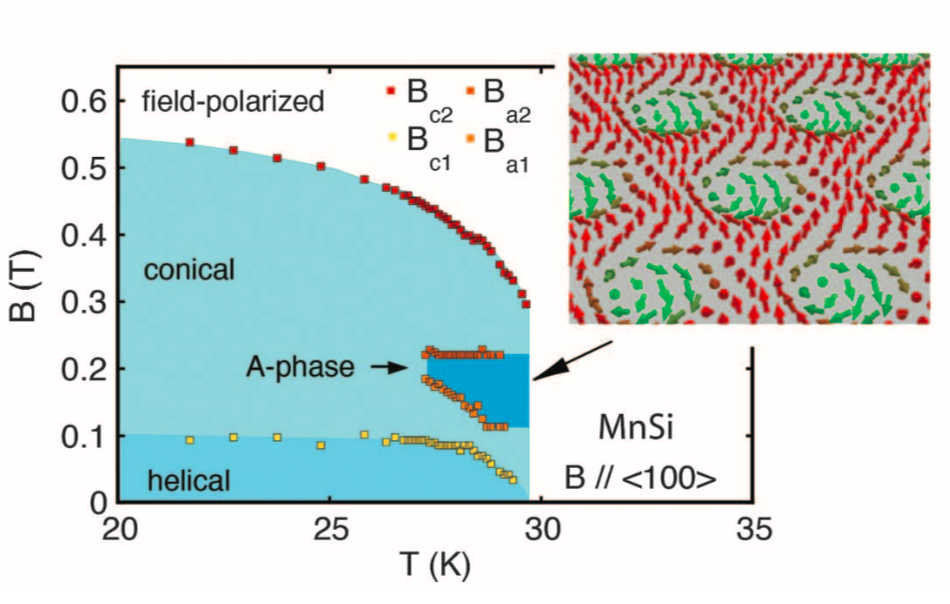} \qquad \includegraphics[width=0.56\textwidth]{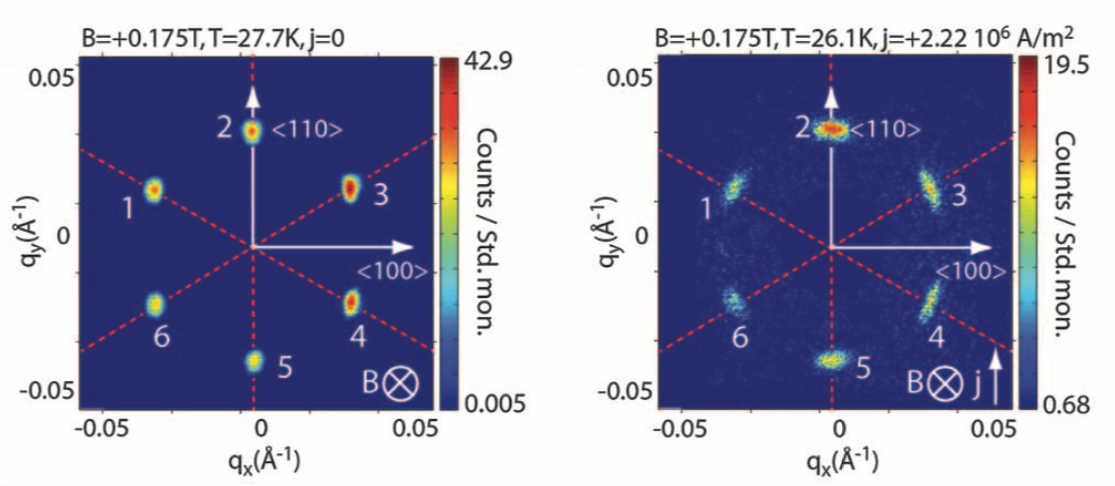}  \\
		\quad \includegraphics[width=0.34\textwidth]{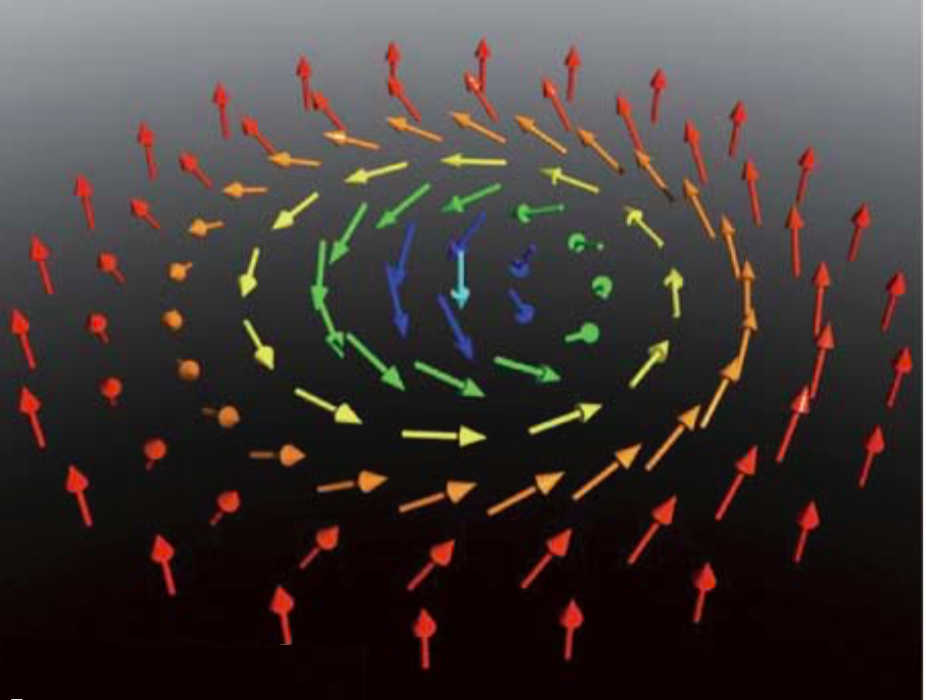} \qquad\quad \includegraphics[width=0.56\textwidth]{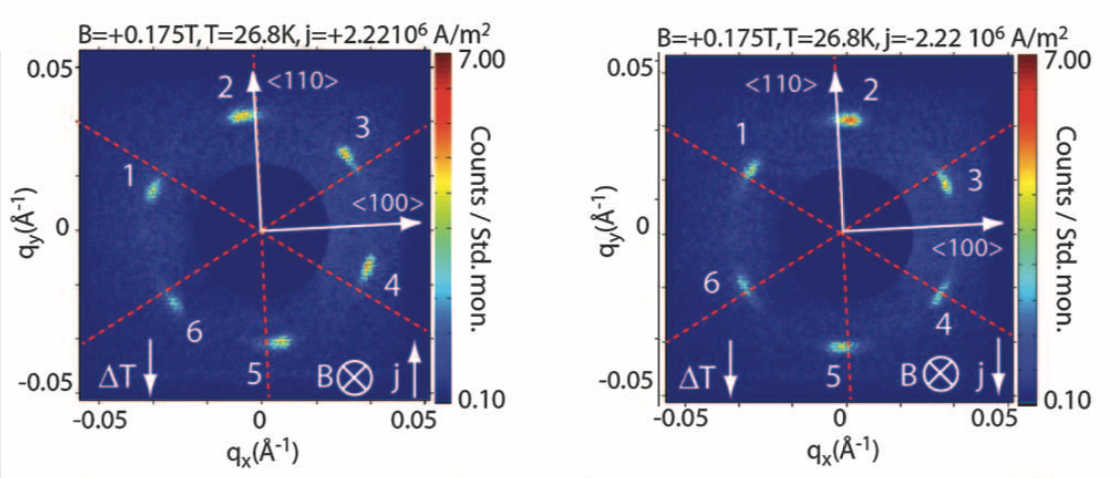} 
		\caption{\footnotesize\small (\textbf{a}): Top left inset. Skyrmions crystal in the A-phase of the $MnSi$ discovered in neutron scattering experiment \cite{SkyrmionExp1}. (\textbf{b}): Bottom left inset. Illustration of the Skyrmion crystals revealed in a thin film of $ Fe_{0.5} Co_{0.5}Si$ using Lorentz transmission electron microscopy \cite{SkyrmionExp2}. (\textbf{c}): Right. Rotation motion of Skyrmion crystals under the temperature gradient along with the current \cite{Jonietz2010}. Reproduced with permission from \cite{SkyrmionExp1} \cite{SkyrmionExp2} \cite{Jonietz2010}. }
		\label{fig:RotationSK}
	\end{center}
\end{figure}

Along the way, we study the spin dynamics known as spin torques, for Skyrmions are made with a bunch of spins tightly arranged in a beautiful way as illustrated in the figure \ref{fig:RotationSK} (\textbf{a}) and (\textbf{b}). Spin torques are important tools to control the motion of Skyrmions. There are various different kinds of spin torques such as spin transfer torque (STT) and spin-orbit torque (SOT), that are related to the interaction with spins of electrons in the same layer, and spin Hall torque (SHT) that stems from the spins of another layer that interacts with the spins of the Skyrmions. In insulating magnets, Skyrmions interact with Magnons, the low energy excitations of magnetization, that provides another possible ways to manipulate the Skyrmions motion. These spin torques are packaged into the Landau-Lifshtiz-Gilbert equation or its generalizations that include various deformations due to the internal structure of the Skyrmions. To understand the center of motion of Skyrmions as a rigid object, Thiele equation is useful. The construction is insightful and we review them below as well. These are done in \S \ref{sec:SpinTorqueBackground}.   

Due to the extended nature of Skyrmions, understanding the physical properties of Skyrmions is challenging. Typically, the experimental setup is complex as illustrated in the figure \ref{fig:RotationSK} (\textbf{c}). It involves ambient magnetic field and temperature to establish the Skyrmion phase in addition to the electric current and/or temperature gradient. Due to the difficulties, direct observation of the Skyrmion Hall effect has been done very recently \cite{Jiang2017}\cite{Litzius2017}. Accordingly, most studies have been done with phenomenological approaches. 

Now, do we know {\it a priori} what kind of physical quantities play role in Skyrmion physics? It will be a great help if there are guides from first principle methods. Quantum field theory Ward identity is an example that solely relies on symmetries and conservation laws. The Ward identities systematically reveal all the relevant physical quantities available in the system with allowed symmetries. These identities can be used to relate different physical quantities with each other.  \\

Here we review three enveloping themes to shed lights on the understanding these complex transport phenomena of the magnetic Skyrmions: a first principle quantum field theoretic method, a phenomenological approach, and the experimental facts.

In the first part, Part \ref{sec:PartI}, we review this useful Ward identities and their constructions step by step in \S \ref{sec:WI} along with background materials in \S \ref{sec:WIBackground}, such as Central extension of Skyrmions, Parity breaking Hydrodynamics, Hall viscosity, Angular momentum and Kubo formula, so that the review would be self-contained. We also provide an underlying geometrical picture using area preserving shear transformations in 2 spatial dimensions in \S \ref{sec:GeometricPictureWI}. We construct more general and powerful Ward identities \cite{Kim:2015qsa} by incorporating the topological Skyrmion charge through the central extension of the momentum-momentum commutation relation \cite{Watanabe:2014pea}\cite{Toma1991}. Previous Ward identities would not able to handle the topological objects like Skyrmions in a systematic fashion. 

We directly apply these Ward identities for the systems with Skyrmions. For example, we consider the thermal Hall conductivity $\bb{\kappa}_\epsilon$ as a function of momentum $q$. 
\begin{align}
	\bb{\kappa}_\epsilon = \bb{\kappa}^{(0)}_\epsilon + q^2 \bb{\kappa}_\epsilon^{(2)} + q^4 \bb{\kappa}_\epsilon^{(4)} + \cdots \;, 
\end{align}	
where we present the momentum dependence of the conductivity explicitly. In {\it insulating magnets} without other mobile carriers, we find that the momentum independent Hall conductivity $ \bb{\kappa}^{(0)}_\epsilon$ is directly related to the topological charge density of Skyrmions, which is explained in \eqref{HallWI} of \S \ref{sec:WIInsulating}. 
Moreover, we note a simple way to measure the Hall viscosity in terms of thermal Hall conductivity. Hall viscosity $\eta_H$ is directly proportional to the Skyrmion density and also the ratio between the slope $ \bb{\kappa}_\epsilon^{(2)} $ and intercept $\bb{\kappa}^{(0)}_\epsilon $ of thermal Hall conductivity. 
\begin{equation} 
	\begin{split}
		\eta_H = c \frac{ \bb{\kappa}_\epsilon^{(2)} }{\bb{\kappa}^{(0)}_\epsilon}    \;,
	\end{split}
\end{equation} 
where we consider the zero momentum limit $q^2 \to 0$. This is presented in \eqref{MeasuringHV}. This amazingly simple result provides insights on the dynamics of Skyrmions and does not rely on time dependent measurements that are usually more challenging. We also outline a clean way for the Hall viscosity measurements even in the presence of the low energy spin wave excitations, Magnons, in insulating magnets in \S \ref{sec:Outlook}.     

In the second part, Part \ref{sec:PartII}, we collect the background materials that were mentioned above, such as the background materials for spin torque, LLG equation and Thiele equation in \S \ref{sec:SpinTorqueBackground} and for thermo-electromagnetic effects including Seebeck, Peltier and Thompson effects along with the spin Seebeck effects and Magnon contributions in \S \ref{sec:ThermoEMBackground}. Finally, we discuss the topological Hall and Skyrmion Hall effects in conducting magnets and Skyrmion Seebeck effects in insulating magnets in \S \ref{sec:SkyrmionHallMeasurements}. We also provide an appendix that contains the full details of the thermo-electromagnetic transport coefficients in \S \ref{appsec:TEME}. 

To make this manuscript accessible to a broader audience, we have tried to include much of the relevant background materials and focus on a small number of references that provides intuitive insights on the subjects. This review is based on the results developed in \cite{Kim:2015qsa}. Part \ref{sec:PartI} is a fully revised and updated version of a recent review \cite{Kim:2017mlj} based on the invited talks at the APS March Meeting 2018 and the MMM conference 2017. We have also added some results that have not been published before.

\part{Ward Identities \& Skyrmion Hall Transport} \label{sec:PartI}

\section{Ward identities: Background Materials} \label{sec:WIBackground}

We would like to introduce several different topics in this section: Parity breaking Hydrodynamics (in \S \ref{sec:Hydrodynamics}), the mysterious Hall viscosity (\S \ref{sec:HallViscosity}) and the associated theoretical understanding (\S \ref{sec:HallViscosityApplication}) followed by symmetry considerations along with the spontaneously generated angular momentum (\S \ref{sec:AngularMomentum}). Underlying theme of all these subjects is the broken parity symmetry in 2+1 dimensions. They have witnessed exciting developments in recent years. Understanding of topological Skyrmion charge as a central charge is also well qualified for this section, yet we postpone presenting this until \S \ref{sec:SkyrmionCentralExtension} because the subject fits nicely there.  

These concepts are introduced in a simplified context to gain familiarity and understanding. We consider more realistic scenarios with broken parity symmetry in the following section \S \ref{sec:WI}. In general, the larger the number of broken symmetries, the larger the number of transport coefficients that are expected to be present. These universal transport coefficients mentioned here are expected to play important roles in \S \ref{sec:WI} and Part \ref{sec:PartII}. 

We also review the Kubo formula that connects transport coefficients to the retarded Green's functions in quantum field theories in \S \ref{sec:Kubo}. The Ward identities reveal their full power when the retarded Green's functions are rewritten in terms of transport coefficients. Along the way, we find a natural way to incorporate thermodynamic quantities through the so-called contact terms. We will be able to collect all the relevant physical quantities systematically in Ward identity approach.  

We start this background materials by introducing the energy momentum tensor, which is the basic mathematical ingredient and facilitates our discussion greatly.  

\subsection{Energy Momentum tensor} \label{sec:EMTensor}

As the energy momentum tensor plays the central role in Part I, we introduce some of its basic facts on the energy momentum tensor. In 2+1 dimensions, the energy momentum tensor $T^{\mu\nu}$, with $\mu, \nu = 0, i$ and $i, j= 1, 2$, has the following matrix form.  
\begin{equation}
T^{\mu\nu} (x^0, x^i)= \left( \begin{array}{ccc}
T^{00} & T^{01} & T^{02} \\
T^{10} & T^{11} & T^{12} \\
T^{20} & T^{21} & T^{22} \end{array} \right) \;. 
\end{equation}
Here $T^{00}$ is the energy density, $T^{0i}$ the energy flux across a surface perpendicular to $i$ direction, 
$T^{i0}$ the momentum density along the direction $i$, $T^{ij}$ the flux of $i$ directional momentum across a surface perpendicular to $j$ direction. It is straightforward to generalize the tensor to the general $d+1$ dimensions. 

There are different ways to construct energy momentum tensor. If a theory has an established action $\mathcal S= \int d^{d+1}x \sqrt{-g} \mathcal L$, a simple way is to take a variation of the action with respect to the background metric $g_{\mu\nu}$ as 
\begin{align} 
	T_{\mu\nu} = \frac{2}{\sqrt{-g}} \frac{\delta (\sqrt{-g} \mathcal L)}{\delta g^{\mu\nu}} \;,
\end{align}
where $g$ is the trace of metric tensor $ g_{\mu\nu} $, which has the form $\eta_{\mu\nu} = \text{diag} \{-1, 1, \cdots, 1\}$ in the flat space time. This is particularly useful for the systems with Lorentz invariance, whose action is manifestly symmetric and thus the energy momentum tensor is symmetric. For the boost invariant systems, the energy flux in a particular coordinate is the momentum density in the same direction, $T^{0i}=T^{i0}$. For the system with the rotation invariance, $T^{ij}=T^{ji}$. Thus when a system has a large symmetry, the energy momentum tensor is tightly constrained. This is especially the case for the rotational invariance that is useful to constrain the transport coefficients in a fairly general fashion. 

In passing, we comment on the renormalizability of a local quantum field theory, meaning that resulting physical quantities are independent of cut-off that is introduced to compute them. To achieve this to be manifest in the renormalizable theories, one can introduce more terms to the already symmetric energy momentum tensor \cite{Callen} \cite{Coleman} \cite{Jackiw}. It turns out that the new tensors have the same momentum and Lorentz generators. 

A more conventional way to construct the energy momentum tensor uses the Noether's theorem, which establishes the relation between the symmetries and conservation laws in a precise way. One of the simplest action is $ \mathcal S = \int d^4 x \mathcal L = \int d^4 x \frac{1}{2} ( \partial_\mu \psi \partial^\mu \psi - m^2 \psi^2)$ given for a scalar field $\psi$ in the flat $3+1$ dimensions with $\mu = 0,1,2,3$. To construct the energy momentum tensor, we consider a infinitesimal translation, $x^\mu \to x'^\mu = x^\mu - a^\mu $ with a constant $a^\mu$. The scalar field transforms under this as $\psi (x) \to \psi (x') = \psi(x+a) = \psi(x) +a^\mu \partial_\mu \psi(x) $. The system has the space time translational symmetry, meaning that the action is invariant up to a surface term that does not change the equation of motion. Thus the Lagrangian $\mathcal L$ is invariant up to a surface term as  $ \mathcal L \to \mathcal L + a^\mu \partial_\mu \mathcal L = \mathcal L +a^\nu \partial_\mu ( \eta^{\mu}_{~ \nu} \mathcal L ) $. 

This variation can be computed with the variations of the scalar field and its derivative of the Lagrangian. 
$ a^\mu \partial_\mu \mathcal L = \frac{\partial \mathcal L}{\partial \psi} (a^\mu \partial_\mu \psi) + \frac{\partial \mathcal L}{\partial (\partial_\nu \psi)} \partial_\nu (a^\mu \partial_\mu \psi) = a^\mu  \partial_\nu (\frac{\partial \mathcal L}{\partial (\partial_\nu \psi)}  \partial_\mu \psi) + a^\mu [ \frac{\partial \mathcal L}{\partial \psi} - \partial_\nu ( \frac{\partial \mathcal L}{\partial (\partial_\nu \psi)}) ]  \partial_\mu \psi  $, where differential by parts is used. The terms in the square bracket vanishes due to the equation of motion of the scalar field. Identifying the remaining term with the surface term with appropriate indices, we arrive at the conservation equation $ \partial_\mu T^\mu_{~~ \nu} = 0$. After raising a index with a flat metric $\eta^{\mu\nu}$,
\begin{align}
	T^{\mu\nu}  = \frac{\partial \mathcal L}{\partial (\partial_\mu \psi)}  \partial^\nu \psi - \eta^{\mu\nu} \mathcal L \;.
\end{align}  
This tensor $T^{\mu \nu}$ is symmetric under the exchange of the indices $\mu$ and $\nu$. 

In general, the energy momentum tensor obtained from the Noether's method is not symmetric. This is the case even for the theories with Lorentz symmetry. When the energy momentum tensor is expected to be symmetric, it is desirable to construct the symmetric one. There exists a well known procedure to improve the tensor to be symmetric \cite{Belinfante}. For example, the action for the Maxwell fields has the form $ \mathcal S = \int d^4 x (-\frac{1}{4} F^{\mu\nu} F_{\mu\nu} )$ with $ F_{\mu\nu} = \partial_\mu A_\nu - \partial_\nu A_\mu$. Straightforward computation gives $ T^{\mu\nu} = F^{\rho \mu} \partial^\nu A_\rho + \frac{1}{4} \eta^{\mu\nu} F^{\rho\sigma} F_{\rho \sigma} $, which is not manifestly symmetric under the exchange the indices $\mu$ and $\nu$. Now one can modify the energy momentum tensor $\tilde T^{\mu\nu} = T^{\mu\nu} + \partial_\lambda K^{\lambda \mu\nu} $, by adding a total derivative with $K^{\lambda \mu\nu}$ that is anti-symmetric in the first two indices $\lambda$ and $\mu$. Then the conservation equation does not change, $\partial_\mu \tilde T^{\mu\nu} = \partial_\mu T^{\mu\nu} +\partial_\mu \partial_\lambda K^{\lambda \mu\nu} = \partial_\mu T^{\mu\nu}  $. Choosing $ K^{\lambda \mu\nu} = F^{\lambda \mu} A^\nu$ and using the equation of motion $ \partial_\mu F^{\mu\nu}=0$, one can see that 
\begin{align}
	\tilde T^{\mu\nu} = F^{\rho \mu} F^\nu_{~~ \rho} + \frac{1}{4} \eta^{\mu\nu} F^{\rho\sigma} F_{\rho \sigma} \;,
\end{align}
which is manifestly symmetric. These symmetric properties of the energy momentum tensors are the consequences of the symmetries, for example the Lorentz symmetry in the present case. 

When a system is invariant under a rotation, one can use a suitable set of orthogonal transformations to put the spatial parts of the energy momentum tensor to have only the diagonal components, $ T^{ij} \propto \delta^{ij}$. Thus the energy momentum tensor is symmetric. Once the rotation symmetry is broken, the energy momentum tensor also has the anti-symmetric components. Thus the tensor is no longer symmetric, $ T^{ij} \neq T^{ji}$. The anti-symmetric parts provide additional degrees of freedom because the system without rotation symmetry is less constrained. Similarly, when the boost symmetry is broken, $T^{0i}$ and $T^{i0}$ are no longer the same. These facts are reflected by additional transport coefficients in the context of 2+1 dimensional hydrodynamics without these symmetries \cite{Hoyos:2013eza}\cite{Hoyos:2013qna}. This tells us that we should not force the energy momentum tensor to be symmetric if the physical system does not require them to be symmetric. 

In the presence of the Skyrmions that we are interested in, the so-called party symmetry is broken. Parity is a discrete symmetry and one can still keep the Lorentz symmetry regardless of the parity symmetry. Nonetheless, due to the broken symmetry, more physical quantities can play role. We discuss this in the following subsection. 

\subsection{Parity breaking hydrodynamics in 2+1 dimensions}\label{sec:Hydrodynamics}

Why do we discuss hydrodynamics? Skyrmions are particle-like excitations and can move around under suitable conditions. Skyrmions can be controlled by the electric current or the temperature gradient. Thus their universal physical properties can be understood by Hydrodynamics. Interested readers can refer to \cite{Jensen:2011xb}\cite{Bhattacharya:2011tra} and \cite{LandauFluidMechanics}. 

Hydrodynamics is an effective and useful theory because it captures the universal features depending on the underlying symmetries of the system we are interested in. It describes dynamics at long distances and large time scales and incorporates dissipative effects, which the Lagrangian description is not very useful to describe. Thus hydrodynamics can be described at best by conservation equations. Its central object is the energy momentum tensor $T^{\mu\nu}$ that is introduced in \S \ref{sec:EMTensor}. In the relativistic hydrodynamics with Lorentz invariance, the hydrodynamic equations are given by the conservation equation $\partial_\mu T^{\mu\nu} = 0$. The variables are temperature $T$ and velocity $u_\mu$ with $u^\mu = (u^0, u^i)$. If we normalize the velocity as $ \eta_{\mu\nu} u^\mu u^\nu = -1$, there are equal numbers of equations and variables. Thus one can solve the problem. 

We solve the hydrodynamic equations order by order in a derivative expansion for the systems with a local thermal equilibrium, which enforces that the details of the thermal fluctuations can be smoothed out. Once one starts to solve the equations in this derivative expansion, one encounters the redundancies. One fix the redundancies with, for example, the Landau frame condition, 
\begin{align} 
	T^{\mu\nu} u_\nu = - \varepsilon u^\mu \;,
\end{align}
which signifies that the flow of the fluid is the energy flow. We further impose the local second law of thermodynamics that is described by the equation 
\begin{align}
	\partial_\mu T^{\mu\nu}u_\nu = 0 \;.
\end{align}
This equation leads to the so-called entropy current. The entropy current is useful because it is required to be positive definite and constrains the values of various transport coefficients. 

If we impose Lorentz invariance at leading (ideal) order, only two physical quantities, energy and pressure, are allowed in the energy momentum tensor $T^{\mu\nu}(T,u) = \varepsilon  u^\mu u^\nu + p P^{\mu\nu}$, where $ P^{\mu\nu} = u^\mu u^\nu + \eta^{\mu\nu}$ is the projection operator and is orthogonal to $u_\mu$, $ P^{\mu\nu} u_\mu = P^{\mu\nu} u_\nu = 0$. 

At the first derivative order, we can construct two symmetric Lorentz covariant second rank tensors using the velocity and a derivative, the shear tensor $\sigma^{\mu\nu} = P^{\mu\alpha} P^{\nu\beta} [\partial_\alpha u_\beta + \partial_\beta u_\alpha - P_{\alpha\beta} (\partial_\delta u^\delta) ]$ and the divergence of the velocity $P^{\mu\nu}  (\partial_\alpha u^\alpha)$. They are multiplied by the projection operator $ P^{\mu\nu}$ so that the definition of the energy $\varepsilon$ and pressure $p$ would not change after adding the first order derivative terms. This is the consequence of the Landau frame condition. By including these two tensors, we have two more terms in the energy momentum tensor. The corresponding coefficients are the shear and bulk viscosities, respectively. Combining them together,
\begin{align}
	T^{\mu\nu}(T,u)  & = \varepsilon  u^\mu u^\nu + p P^{\mu\nu}\-\eta \sigma^{\mu\nu} 
	- \zeta P^{\mu\nu}  (\partial_\alpha u^\alpha) \;. 
\end{align}
This is the relativistic hydrodynamics of Landau and Lifshitz \cite{LandauFluidMechanics}. 

In the presence of Skyrmions in $2+1$ dimensions, the parity symmetry is broken and one can use the totally anti-symmetric epsilon tensor $ \epsilon^{\mu\nu\rho}$ with $ \epsilon^{012} = -\epsilon^{210} = 1$ for their cyclic permutations and $0$ otherwise. One can create two new tensors using this epsilon tensor. Starting from the shear tensor, one can create $\tilde{\sigma}^{\mu\nu}  = \epsilon^{\alpha\beta(\mu}u_{\alpha} \sigma_{\beta}^{\ \ \nu)}$: the corresponding transport coefficient is called Hall viscosity $\eta_H$. This Hall viscosity is an interesting quantity, and we review this below in \S \ref{sec:HallViscosity}. One can generate another term called the vorticity, $\tilde \Omega = - \epsilon^{\beta\gamma\alpha} u_{\beta}\nabla_{\gamma}u_{\alpha}$. The corresponding coefficient is the Hall bulk viscosity, $\zeta_{H} $, the Hall analogue of the bulk viscosity if one subtracts the vorticity from the energy momentum tensor, in the so-called vortical-frame. Thus, one can add two more terms if the parity symmetry is broken. These exciting recent developments can be found in \cite{Jensen:2011xb}\cite{Bhattacharya:2011tra}. Combining all the terms, the parity breaking hydrodynamics upto the first derivative order can be described by the energy momentum tensor. 
\begin{align}\label{FirstOrderHydrodynamics}
	T^{\mu\nu}(T,u)  &= \varepsilon  u^\mu u^\nu + p P^{\mu\nu}-\eta \sigma^{\mu\nu} 
	-\zeta P^{\mu\nu}  (\partial_\alpha u^\alpha) 
	-\eta_{H} \tilde{\sigma}^{\mu\nu} -\zeta_{H} P^{\mu\nu} \tilde \Omega  \;.
\end{align} 
This program has been extended to the second order in the derivative expansion and also to the charged case (with an additional conserved current) that is slightly more complicated. 

An aside: if the boost symmetry is broken as mentioned in \S \ref{sec:EMTensor}, the energy momentum tensor is no longer symmetric, $T^{0i}(T,u)  \neq T^{i0}(T,u) $ along a space direction $i$. In addition to the transport coefficients present in \eqref{FirstOrderHydrodynamics}, there is one new transport coefficient in the first order of derivative expansion in the neutral Hydrodynamics and two additional transport coefficients in the charged Hydrodynamics \cite{Hoyos:2013eza}\cite{Hoyos:2013qna}. There can be many more transport coefficients upon breaking the spatial rotation symmetry as well because of the non-zero anti-symmetric parts of the spatial energy momentum tensor. 

Hydrodynamics solely relies on symmetries of a given system and is universal. The transport coefficients given in \eqref{FirstOrderHydrodynamics} can be present in the physical systems with the broken parity symmetry. In particular, the Hall viscosity $\eta_{H}$ can play an important role in the Skyrmion physics. 

\subsection{Hall viscosity} \label{sec:HallViscosity}

General considerations of 2+1 dimensional hydrodynamics in \S \ref{sec:Hydrodynamics} suggest the presence of the Hall viscosity $ \eta_H$ in physical systems with broken parity, especially in Skyrmion physics. We review this new physical quantity in this section. The references \cite{LandauElastiity}\cite{Avron:1995}\cite{Read:2010epa} are useful for understanding the Hall viscosity and some of the contents of this section. 

Let us review the Hall viscosity in a slightly different setting so that its physical meaning is more clearly displayed. We focus on our discussion for fluids. A small deformation, denoted by $\xi_i$, produces a stress $T_{ij}$ through the strain that is denoted by a symmetric derivative of the deformation, $ \xi_{ij} = \partial_i \xi_j + \partial_j \xi_i$, and the strain rate, a time derivative of strain, $\dot \xi_{ij} = \partial_t \xi_{ij} $. These can be connected to $u^\mu$ in  \S \ref{sec:Hydrodynamics} as $ u^0= \gamma \equiv 1/\sqrt{1-v^2}$, $ u^i = \gamma v^i$ and $ v^i =\dot \xi_i$. In this language, the energy momentum tensor has the form \cite{LandauElastiity}\cite{Read:2010epa}.
\begin{align} \label{EMTHall}
	T_{ij} & = p \delta_{ij} - \lambda_{ijkl}\xi_{kl} 
	- \eta_{ijkl} \dot \xi_{kl}  \;, 
\end{align} 
where $p$ is pressure, $ \lambda_{ijkl} $ the elastic modulus tensor, and $\eta_{ijkl}$ the viscosity tensor. For rotational invariant systems, one can constrain these four indices tensors efficiently. $\lambda_{ijkl} = \lambda \delta_{ij}\delta_{kl}$, where $\lambda = -V (\partial p/\partial V)$ is the elastic modulus that describes the change of pressure as the volume $V$ varies. The viscosity tensor has the general form.  
\begin{align} \label{ShearTensor11}
	\eta_{ijkl} = \eta (\delta_{ik}\delta_{jl}+\delta_{il}\delta_{jk}) 
	+(\zeta-\eta)\delta_{ij}\delta_{kl} \;,
\end{align} 
where $\eta$ and $\zeta$ are the shear and bulk viscosities.    

For systems with broken parity, for example in the presence of a background magnetic field, the odd part of the shear tensor is allowed \cite{Avron:1995}. Explicitly, in two spatial dimensions $d=2$ with indices $i,j=1,2$,
\begin{align}\label{HallViscosityTensor2}
	\eta_{ijkl}^A = -\eta_{klij}^A =-\frac{\eta_H}{2} 
	(\epsilon_{ik}\delta_{jl}+\epsilon_{jl}\delta_{ik} +\epsilon_{il}\delta_{jk} + 
	\epsilon_{jk}\delta_{il}) \;.  
\end{align} 
One can check that this quantity is symmetric under the exchange of the indices $i$ and $j$ as well as $k$ and $l$. On the other hand, it is anti-symmetric under the exchange of $ij$ and $kl$. Thus the Hall viscosity $\eta_H$ reveals itself in the anti-symmetric part of the shear tensor in the context of fluid dynamics \eqref{HallViscosityTensor2}. 

\begin{figure}[h!]
	\begin{center}
		\includegraphics[width=0.8\textwidth]{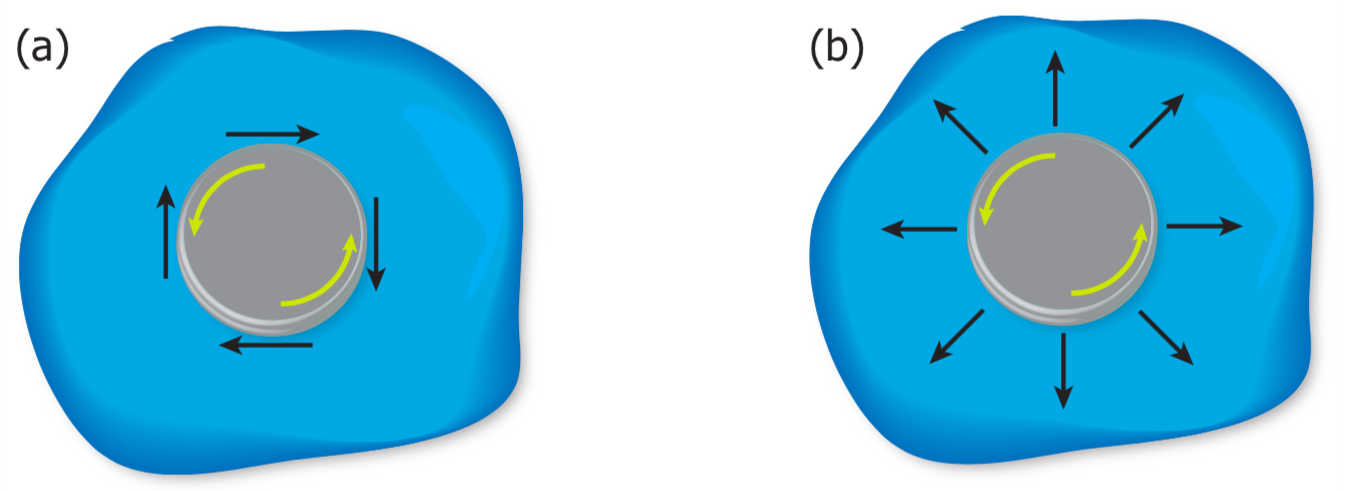} 
		\caption{\footnotesize\small (\textbf{a}): The direction of the shear viscosity, the black arrows, acts against that of the cylinder. (\textbf{b}): The Hall Viscosity acts perpendicular to the motion of the middle cylinder. Despite its name, it does not produce dissipation. Reproduced with permission from \cite{Hughes2013}.
		}
		\label{fig:HallViscosityDroplet}
	\end{center}
\end{figure}

We offer a geometric description of the Hall viscosity in a more realistic setting. Let us imagine a finite size cylinder surrounded by a fluid. The cylinder is rotating counterclockwise with a constant frequency in the middle of the fluid as the figure \ref{fig:HallViscosityDroplet}. The well known shear viscosity acts as a clockwise force along the surface of the cylinder, anti-parallel to the direction of the motion, trying to slow down the cylinder as described in figure \ref{fig:HallViscosityDroplet} (\textbf{a}). On the other hand, Hall viscosity produces a force perpendicular to the rotating direction of the cylinder, outward or inward depending on the situation. See the figure \ref{fig:HallViscosityDroplet} (\textbf{b}). In particular, the Hall viscosity does not provide a dissipation to the motion of the cylinder.   

One can compute the change of energy under a small deformation of the fluid. This is described by $\delta \varepsilon = - T_{ij} \delta \xi_{ij}$. After using the local first law of thermodynamics, one obtains  
\begin{align} \label{EntropyHallViscosity}
T\dot s =  \eta_{ijkl} \dot \xi_{ij} \dot \xi_{kl}  + \eta^A_{ijkl} \dot \xi_{ij} \dot \xi_{kl} \;, 
\end{align}
where $s$ is the entropy density. From the fact that the entropy is positive semi-definite, one can constrain the value of the shear and bulk viscosities in $ \eta_{ijkl}$, especially for the systems with the rotational invariance. Note that the second term $\eta^A_{ijkl} \dot \xi_{ij} \dot \xi_{kl}$ in \eqref{EntropyHallViscosity} actually vanishes because the strain rates are symmetric under the exchange of the indices $ij$ and $kl$, while $\eta^A_{ijkl}$ is anti-symmetric as in \eqref{HallViscosityTensor2}. This demonstrates that $\eta_H $ is dissipationless as mentioned above. Thus, the Hall viscosity can exist even at zero temperature. It has a better chance to be observed in low temperature experiments because other dissipative effects are suppressed. 

\subsection{Angular momentum \& compatible symmetries} \label{sec:AngularMomentum}

When the parity symmetry is broken, some physical systems can generate an angular momentum spontaneously \cite{Liu:2012zm}. This spontaneously generated angular momentum contains the expectation value of the linear momentum operator. Due to this, one can not keep the angular momentum and translation invariance at the same time, which puts specific constraints on maximum compatible quantities in Ward identities considered in \S \ref{sec:WI} \cite{Hoyos:2015yna}. Here we provide the basic idea.  

\begin{figure}[h!]
	\begin{center}
		\includegraphics[width=0.56\textwidth]{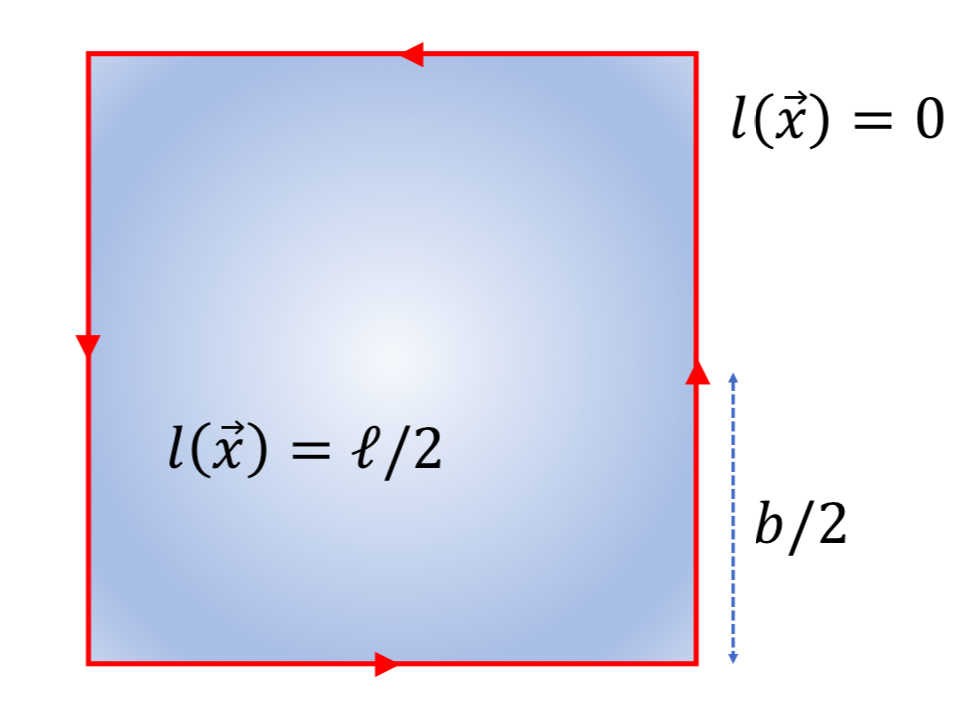} 
		\caption{\footnotesize\small Spontaneously generated angular momentum $l (\vec x) = \ell/2$ inside a square with length $b$. $l (\vec x) = 0$ outside. There is a momentum current going around along the boundary. Can we detect the angular momentum as the boundary is pushed to the infinity? 
		}
		\label{fig:Rotation}
	\end{center}
\end{figure}

For simplicity, we consider a 2 dimensional system without a time dependence so that a conservation equation is reduced to $ \partial_i T^{0i}(\vec x) = 0$. Reminded with the anti-symmetric property of the epsilon tensor $\epsilon^{ij}$, one can check the conservation equation has an obvious solution.  
\begin{align}\label{AngularMom}
	\langle T^{0i}(\vec x) \rangle = \epsilon^{ij} \partial_j l (\vec x) \;.
\end{align}
The solution is independent of the choice of $l(\vec x)$. Because the physical properties are independent of the details, we can choose $l (\vec x)$ such that the computation becomes simple. 
\begin{align}
	l (\vec x) =\begin{cases} 
		\ell/2 & \quad(|x|, |y| \leqslant b)\\
		0 & \quad(\text{otherwise}) \end{cases}\;,
\end{align}
where $\ell$ is a constant. This is illustrated in the figure \ref{fig:Rotation}. Thus $T^{0i} (\vec{x})$ vanishes both inside and outside of the square region with side $b$, and is only non-vanishing along the boundary. 
\begin{align}
	\< T^{0i}(\vec{x}) \> = (\ell/2) \epsilon^{ij} \left[ -\delta(x^{j} - b/2) + \delta(x^{i} + b/2)\right] \theta(b/2-|x|) \theta(b/2-|y|) \;.
\end{align}
This corresponds to an edge current, a momentum flow around the boundary of the square. The magnitude and direction of the edge current are $\ell$ and along the boundary, either clockwise or counter-clockwise, depending on the sign of $\ell$. This boundary contribution is related to the topological nature of the underlying field theory.

Now we consider the infinite volume limit, $b\to \infty$. Normally we discard the effects of $ T^{0i}$ because it is a boundary contribution. However, a careful treatment shows that it indeed makes a contribution to the total angular momentum \cite{Liu:2012zm}:
\begin{align}
	L = \int d^2 \vec x \epsilon_{ij} x^i \langle T^{0j} \rangle 
	= \int d^2\vec x\ell(\vec x)\partial_i x^i=\ell\int d^2\vec x=\ell V_2 \;,
\end{align} 
where $V_2$ is the volume enclosed by the boundary. The total angular momentum is independent of the shape of the boundary. This example illustrates that identifying spontaneously generated angular momentum is subtle in the system with a broken parity symmetry and, in particular, can exists even in a system without boundary.  

Note that there is even more subtle point when we consider maximum compatible symmetries in the presence of the rotational symmetry, the translation symmetry, and the angular momentum. The translation symmetry is not compatible with the presence of the spontaneously generated angular momentum, which is manifested with the expectation value of the momentum operator $\langle T^{0i}(\vec x) \rangle \neq 0$. (This is also true in the presence of a boundary. We do not explicitly consider the boundary effect in the Part \ref{sec:PartI} of this review.) Thus we have two independent options that are incompatible each other. 
\begin{itemize} 
\item Option A: keeping the rotation symmetry and the angular momentum without the translation symmetry. 
\item Option B: keeping both the translation and rotation symmetries without angular momentum. 
\end{itemize}
These two different options lead to two highly non-trivial and mutually exclusive sets of Ward identities as reviewed in quantum Hall systems in \S \ref{sec:HallViscosityApplication} and also in the Skyrmion physics considered in \S \ref{sec:WI} \cite{Hoyos:2015yna}.
 
\subsection{Hall viscosity: Applications} \label{sec:HallViscosityApplication}

After studying the basic properties of the Hall viscosity in \S \ref{sec:HallViscosity} and symmetry considerations in \S \ref{sec:AngularMomentum}, we review some interesting computations of the Hall viscosity in the quantum Hall system. 

Usual computation of the Hall viscosity \cite{Avron:1995} relies on the quantum mechanical adiabatic theory for the uniform geometric deformation (strain) $ \xi_{ij} = \partial_i \xi_j + \partial_j x_i$, where $ \xi_i = x_i' - x_i $ is the displacement vector for a linear deformation $x_i'$ of $x_i$. In two dimensions, there are two independent area preserving deformations, which are related to the shear transformations in the energy momentum tensor discussed in \S \ref{sec:EMTensor}. 

Consider the quantum systems with Hamiltonian $H$ and metric $g_{ij}= \delta_{ij} + \xi_{ij}$ with small deformations $\xi_{ij}$. The adiabatic deformations of the Hamiltonian with a constant strain give 
\begin{align} \label{HallViscosityComp}
	\Big\langle \frac{\partial H}{\partial \xi_{ij} } \Big\rangle =  \frac{\partial E}{\partial \xi_{ij} } + \Omega_{ijkl} ~\dot \xi_{kl} \;, 
\end{align}
where $\langle ~\rangle$ is the expectation value for a normalized quantum state $|\psi\rangle$. Now we identify the variation $\big\langle \frac{\partial H}{\partial \xi_{ij} } \big\rangle = - V T_{ij}$ and the first term on the right hand side $\lambda_{ijkl} = \frac{1}{2} \frac{\partial^2 E}{\partial \xi_{ij} \partial \xi_{kl} } $ as the energy momentum tensor with the volume $V$ and the elastic modulus tensor, respectively, in the context of fluid. Sometimes, the computation \eqref{HallViscosityComp} can be done more conveniently in terms of the background metric $g_{ij}$ instead of the small deformation $\xi_{ij}$. 

The last term $\Omega_{ijkl} $ in \eqref{HallViscosityComp} is the adiabatic or Barry curvature term and given by 
\begin{align}
	\Omega_{ijkl} = \text{Im} \Big[ \frac{\partial}{\partial \xi_{ij}} \Big\langle \psi \Big| \frac{\partial \psi}{\partial \xi_{kl}} \Big\rangle - \frac{\partial}{\partial \xi_{kl}} \Big\langle \psi \Big| \frac{\partial \psi}{\partial \xi_{ij}} \Big\rangle \Big] \;.
\end{align}
This adiabatic curvature is non-trivial if the phase of the state $\psi$ changes along a closed path through the geometric deformation. Note that it is related to the Hall viscosity as $ \Omega= \eta_H V$ with $V$ volume. As already discussed in \S \ref{sec:HallViscosity}, this adiabatic term is anti-symmetric under the exchange of the first and second pairs of indices $(ij)$ and $(kl)$, and thus non-dissipative. See also some earlier computations of the Hall viscosity in \cite{Levay}\cite{Tolkatly1}\cite{Tolkatly2}.

Here we mention two different sets of the Hall viscosity computations based on the two options discussed in the previous section \S \ref{sec:AngularMomentum}.   
In the presence of the angular momentum, we can not keep the translation symmetry. We still keep the rotation symmetry (Option A). With explicit computations with the paired and gapped super fluids and fractional quantum Hall wave functions in 2 dimensions, the authors of \cite{Read:2010epa}\cite{Read:2008rn} argue that the Hall viscosity has the general form 
\begin{align}
	\eta_H = \frac{1}{2} \hbar \bar s \bar n \;, 
\end{align}	
where $\bar n$ the average density of particles and $\bar s = -\ell $ is the average angular momentum, which is, for example, spin due to the cyclotron motion for non-interacting particles in a magnetic field. This non-trivial relation between the Hall viscosity and angular momentum is interesting because the average angular momentum is related to a topological quantity called the shift, $\bar {\mathcal S} = 2 \bar s$, the change in magnetic flux when the ground state is formulated on a sphere. In general, the shift $\bar {\mathcal S}$ is the offset $N_\phi = \nu^{-1} N - \bar {\mathcal S} $ that is required when the system is put in a curved space with a different genus compared to that of the plane geometry $N_\phi = \nu^{-1} N $, where $N_\phi$ is the number of magnetic flux quanta piercing the surface, $N$ the number of particle, and $\nu$ the filling factor \cite{Wen}.  

If one keeps both translation and rotation invariance without angular momentum (Option B), one can obtain another non-trivial relation stating that Hall viscosity $\eta_H$ is connected to the Hall conductivity $\sigma_H$. For a quantum Hall states with a background magnetic field ($B\neq 0$) with Galilean invariance and made up of particles of the same charge/mass ratio, the relation is given \cite{Hoyos:2011ez}
\begin{align}
\begin{split}
&\frac{\sigma_H(q)}{\sigma_H (0)} = 1 + C_2  (q \tilde l)^2 + \mathcal O (q \tilde l)^4\;, \qquad C_2 = \frac{\eta_H}{\bar n \hbar } - \frac{2\pi}{\nu} \frac{l^2}{\hbar \omega_c} B^2 \epsilon''(B) \;, 
\end{split}
\end{align}
where $\sigma_H (0)=\frac{\nu e^2}{2\pi \hbar}$ is the Hall conductivity at zero momentum $q=0$, $\tilde l^2=\hbar c/|e|B $ is the magnetic length squared, and $\omega_c= |e|B/mc$ is the cyclotron frequency. The quantity $ \epsilon''(B) $ is the energy density, energy per unit area, as a function of external magnetic field $B$ at a fixed filling factor. While the term with $ \epsilon''(B) $ is not universal, its magnitude can be extracted independently. For example, $  \epsilon(B) = (N^2/4\pi) \hbar \omega_c/\tilde l^2$ for the integer quantum Hall system with $\nu=N$ in the high magnetic field limit. Thus $C_2= -3N/4$ after using $\eta_H = \hbar \bar {\mathcal S} \bar n/4 $ and $  \bar {\mathcal S} = N$. Similarly, the $C_2$ factors haven been worked out for the fractional quantum Hall systems and 
Laughlin’s states in \cite{Hoyos:2011ez}.

These two symmetry choices, option A and option B, and the corresponding relations among physical quantities, so-called the Ward identities, are collectively considered in \cite{Bradlyn:2012ea}\cite{Hoyos:2014lla}\cite{Hoyos:2015yna}. The latter two references also extended the Ward identities to the relativistic case and also checked the consistency with the non-relativistic cases.  

Let us mention that the authors of \cite{Haldane:2009ke}\cite{GuidingCenter} argue that there exists another contribution to the Hall viscosity, called guiding-center Hall viscosity, in addition to the contribution associated with the geometry of the Landau orbits, that we discuss above. This seems to happen when the systems do not have the ``explicit'' rotation symmetry. This quantity characterizes correlated fractional quantum Hall fluids, and vanishes in uncorrelated integer quantum Hall fluids. Recent computations using matrix models show only the guiding-center contributions due to the projection of the quantum Hall states into the lowest Landau level \cite{Lapa:2018ubk}\cite{Lapa:2018els}. 

\subsection{Kubo formula} \label{sec:Kubo}

A transport coefficient could be viewed as parameterizing the response of a fluid to hydrodynamic perturbations. Kubo formula relates the transport coefficients to the retarded Green's function that is well developed in Quantum field theory. The Kubo formula for the Hall viscosity has been worked out in \cite{Saremi:2011ab}\cite{Bradlyn:2012ea}. 

There is a useful and systematic way to capture all the known transport coefficients and thermodynamic quantities. For an interesting system, one can construct the partition function that is coupled to possible external sources, such as background gauge fields $A_i$ and background metrics $g_{ij} $. This partition function ${\mathcal Z}[A_i , g_{ij} ]$ can be used to evaluate the variations with respect to the sources. In the linear response theory, these are currents and energy momentum tensors associated with the sources upon setting them to be trivial as $ A_i=0, g_{ij}=\eta_{ij}$. 
\begin{align}
	\begin{split}
		J_i &= \frac{\delta \log \mathcal Z}{\delta A^i} 
		= - \bar n A_i + \sigma_{ij} E_j + \cdots   \;, \\
		T_{ij} & = \frac{2}{\sqrt{-g}} \frac{\delta \log \mathcal Z}{\delta g^{ij}}  
		= p \delta_{ij} -\frac{\lambda}{2} \delta_{ij}\delta_{kl} h_{kl} 
		-\frac{1}{2} \eta_{ijkl} \dot h_{kl}  + \cdots \;,
	\end{split}
\end{align} 
where $\bar n$ is a charge density, $E_i = \partial_t A_i - \partial_i A_t$ is the electric field, $\sigma$ is the conductivity, $p$ is the pressure, $\lambda$ is the elastic modulus that is inverse compressibility, and $\eta$ is the viscosity we have defined previously. The first few term on the right side $\bar n$, $p$, and $ \lambda$ are the contact terms that can be obtained by an additional variation with the sources. For example, the diamagnetic term $\bar n$ can be obtained by second derivative of the partition function that has the $A^2$ term. These contact terms also play roles in the Ward identities. 

Now the conductivity $\sigma_{ij}$, one of the transport coefficients, can be obtained by taking a variation of the current $J_i$ with respect to $A_j$, followed by setting $A_i=0$. Then, 
\begin{align}
	\begin{split}
		\sigma_{ij} (\omega, {\bf q}) &= -i \frac{\bar n}{\omega} \delta_{ij} 
		+ \frac{i}{\omega} G_R^{ij} (\omega, {\bf q}) \;, \\
		G_R^{ij} (\omega, {\bf q}) & = \int dt\int d^d x e^{i\omega t - i {\bf q} \cdot {\bf x}}
		~i \theta (t) ~ \langle [J_i (t, {\bf x}), J_j (0, {\bf 0})]\rangle \;,
	\end{split}
\end{align} 
where we express the quantities in the momentum space $(\omega, \vec q)$. The factors $1/\omega$ come from the time derivative in $E_i$ and the Fourier transform to the momentum space. This demonstrates how the Kubo formulas relate between the transport coefficients and the retarded Green's functions. Note that there are also differences that are related to various contact terms. Similarly, The shear tensor $\eta_{ijkl}$ is related to the retarded Green's function of the commutator with the energy momentum tensor, which can be obtained by the metric perturbations $g_{ij}$. Some of these contact terms had been reported to be missing In the previous literature \cite{Bradlyn:2012ea}. It turns out that our Ward identity is powerful enough to capture all the possible terms including the contact terms. As we mention below, we double-checked our results coming from the Ward identities with available results to check the reliability \cite{Hoyos:2015yna}.  

To compute the Kubo formula for the Hall viscosity, it is sufficient to consider the local rest frame of the fluid at zero spatial momentum. Out of all the general metric perturbations $ g_{\mu\nu} = \eta_{\mu\nu} + \delta h_{\mu\nu}(t) + \mathcal O (\delta h^2)$ in 2+1 dimensions, one can work with the following minimal set to ensure a consistency condition: $ \delta h=\{\delta h_{xy}(t), \delta h_{xx}(t), \delta h_{yy}(t) \} $. Where $\mu, \nu = t, x, y$ and $\eta_{\mu\nu} = \text{diag} (-1,1,1)$ is the flat metric with only the diagonal components. For the tensor structures given in \eqref{EMTHall}, \eqref{ShearTensor11}, and \eqref{HallViscosityTensor2}, one obtains the corresponding response, the energy momentum tensor $T^{xy}$ \cite{Saremi:2011ab}
\begin{align}
	T^{xy} = - p \delta h_{xy} - \eta \frac{\partial \delta h_{xy}}{\partial t} + \frac{1}{2} \eta_H \frac{\partial (\delta h_{xx} - \delta h_{yy}) }{\partial t} + \mathcal O (\delta h^2) \;.
\end{align}

In terms of the retarded Green's function, we can write 
\begin{align}
\begin{split}
	\langle T^{\mu\nu}(t, \vec x)\rangle_{\delta h} &=\langle T^{\mu\nu}(t, \vec x)\rangle_{\delta h=0} \\
	&~~-\frac{1}{2}\int dt' \int d^{2}x'G_{R}^{\mu\nu,\alpha\beta}(t, \vec x; t', \vec x')\delta h_{\alpha\beta}(t', \vec x')
	+\mathcal O (\delta h^{2} )\;,
\end{split}	
\end{align} 
where $\vec x=(x,y)$ and the retarded Green's function $G_{R}^{\mu\nu,\alpha\beta}(t, \vec x; t',  \vec x')$ and the Fourier transform $\tilde G_{R}^{\mu\nu,\alpha\beta}(\omega, \vec k)$ with the frequency $\omega$ and momentum $\vec k$ have the forms 
\begin{align} \label{RetardedGRT}
	\begin{split}
		G_{R}^{\mu\nu,\alpha\beta}(t, \vec x; t', \vec x')&=-i\theta(t-t')
		\langle[T^{\mu\nu}(t, \vec x),T^{\alpha\beta}(t', \vec x')]\rangle \;,  \\
		\tilde G_{R}^{\mu\nu,\alpha\beta}(\omega, \vec k) 
		&= \int dt \int d^{2}x e^{-i k^\mu x_\mu}G_{R}^{\mu\nu,\alpha\beta}(t, \vec x; 0, \vec 0) \;. 
	\end{split}
\end{align} 
Thus one finds
\begin{align}
\begin{split}
	\tilde G_{R}^{xy,xx-yy}(\omega,\vec{0}) &= 
	2i\omega \eta_{H}+\mathcal O (\omega^{2})\;, \quad 
	\tilde G_{R}^{xy,xy}(\omega,\vec{0}) =  
	p-i\omega\eta +\mathcal O (\omega^{2})\;. 
\end{split}	
\end{align}
One can put the expressions in covariant form as 
\begin{align}
	\qquad~~ \eta_H &=\lim_{\omega\to 0}\frac{\epsilon_{ik}\delta_{jl}}{4i\omega}
	\tilde G_{R}^{ij,kl} (\omega,\vec{0})\;, \quad 	
	\eta =\lim_{\omega\to 0}\frac{\delta_{ik}\delta_{jl} 
		-\epsilon_{ik}\epsilon_{jl}}{8i\omega}
	\tilde G_{R}^{ij,kl} (\omega,\vec{0})\;,  \\
	\qquad~~ \sigma_H &=\lim_{\omega\to 0}\frac{\epsilon_{ij}}{2i\omega}
	\tilde G_{R}^{i,j} (\omega,\vec{0})\;, \quad \quad 	
	\sigma =\lim_{\omega\to 0}\frac{\delta_{ij}}{2i\omega}
	\tilde G_{R}^{i,j} (\omega,\vec{0})\;,
\end{align} 
where $\tilde G_{R}^{i,j}$ are the momentum space Green's functions for the currents. Here we also list the Kubo formulas for the shear viscosity, Hall conductivity and conductivity in addition to that of the Hall viscosity for future reference. 

\newpage 
\section{Ward identities \& Skyrmion Hall transport }\label{sec:WI}

In the previous section, we encounter surprising relations among different physical quantities such as the Hall conductivity, the Hall viscosity, and angular momentum in quantum Hall systems. Here we generalize the Ward identities by including topological charge of Skyrmions and apply them to provide various relations among the transport coefficients and thermodynamic quantities which have been reviewed previously in \S \ref{sec:WIBackground}. 

First, we provide a simple geometric picture for the Ward identity in \S \ref{sec:GeometricPictureWI}, that provides insights on its underlying structure. To apply this idea to the Skyrmion systems, the Ward identity is required to generalize to include the topological contributions. To do so, we introduce the topological Skyrmion charge as a central extension of momentum momentum commutation relation in \S \ref{sec:SkyrmionCentralExtension}. Then, we generalize the Ward identities by including the topological Skyrmions charge. This simple and powerful Ward identities are presented in \S \ref{sec:SimpleWI}. The rest of the section is a direct application or further generalization of this identities. 

As discussed in \S \ref{sec:AngularMomentum}, there are two inequivalent Ward identities with two independent sets of symmetries. We consider the Ward identities with translation and rotation symmetries in \S \ref{WIRotAndTrans} that provides a set of interesting relations among the viscosities and conductivities. Based on these results, we propose a simple way to measure the Hall viscosity or confirm its existence in the insulating magnets when the Magnon excitations can be ignored at very low temperature in \S \ref{sec:WIInsulating}. We show that the topological charge $c$ plays important roles. A possible way to measure the Hall viscosity in the presence of Magnon excitations is proposed in \S \ref{sec:Outlook}. In this review, we focus on Ward identities with translation and rotation symmetries without angular momentum (Option B described in \S \ref{sec:AngularMomentum}) for a clearer and simpler presentation. The Ward identities with rotation invariance and angular momentum have been thoroughly discussed in \cite{Kim:2015qsa}.

We generalize our Ward identities with electric charges in the presence of magnetic field and (electric) currents in \S \ref{sec:WIChargeB}. As an application, we explain the implications of the Ward identities for the phenomenology of the Hall conductivity and its connection to experiments in \S \ref{sec:HallConductivity}. We further provide a simple formula for the Hall viscosity in the charged case in terms of Hall conductivity measurements as a function of momentum in \S \ref{sec:HallViscosityConductor}. Finally, we briefly comment on the case  without translation invariance in \S \ref{WIAngularMomentum}  (Option A described in \S \ref{sec:AngularMomentum}).   

This section is based on \cite{Kim:2015qsa}, that includes the generalization of Ward identities with the topological charges, and previous results \cite{Hoyos:2014lla}\cite{Hoyos:2015yna}. When possible, these results have been compared to be consistent with other results in \cite{Read:2010epa}\cite{Hoyos:2011ez}, especially in \cite{Bradlyn:2012ea}. Let us start with the simple geometric picture of the Ward identity!

\subsection{Geometric understanding of Ward identities} \label{sec:GeometricPictureWI}

\begin{figure}[h!]
	\begin{center}
		\includegraphics[width=0.3\textwidth]{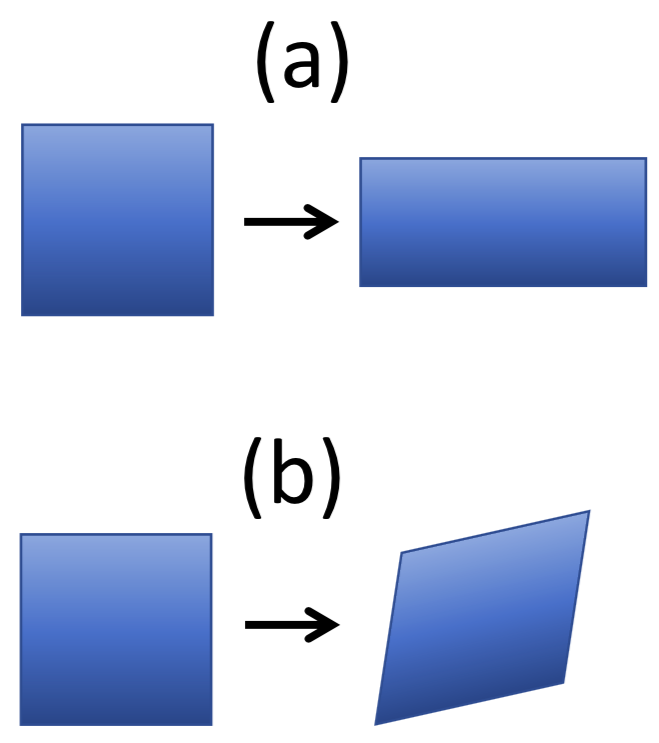} \qquad\qquad 
		\includegraphics[width=0.53\textwidth]{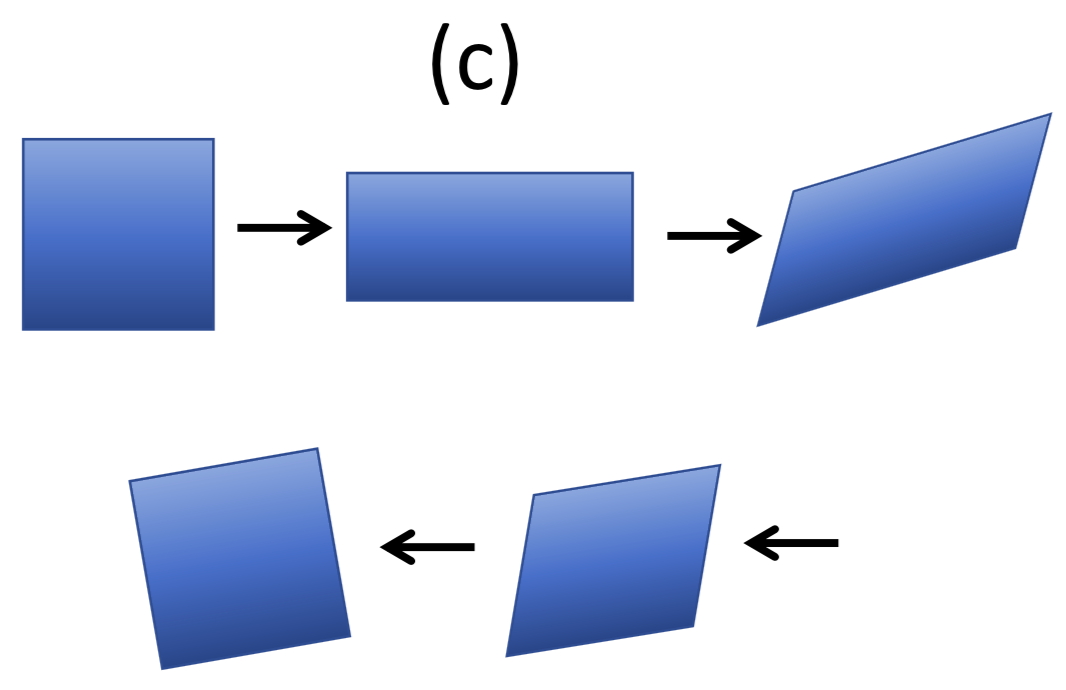} 
		\caption{\footnotesize\small Two in-equivalent area preserving shear transformations in 2 spatial dimensions. (\textbf{a}): Elongation of the square along the horizontal direction. (\textbf{b}): Elongation along the diagonal direction of the square. (\textbf{c}): A combination of these two shear transformations produces a rotation. 
		}
		\label{fig:SjearTR0}
	\end{center}
\end{figure}

Fortunately, there is a simple geometric picture that clearly illustrates the physical meaning of the Ward identities. There are only two independent area preserving shear transformations in 2 spatial dimensions. By performing a series of the shear transformations, one can produce a net rotation. Starting with a square, one can elongate one side and squeeze the other as in the figure \ref{fig:SjearTR0} (\textbf{a}) or stretch along the diagonal direction as in the figure \ref{fig:SjearTR0} (\textbf{b}). These transformations can be written in matrix forms as   
\begin{equation} 
a = \left( \begin{array}{cc}
 1 + \tilde \epsilon & 0  \\
 0 & 1 -  \tilde \epsilon \end{array} \right), \qquad\quad 
 b = \left( \begin{array}{cc}
 1 &  \tilde \epsilon'  \\
  \tilde \epsilon' & 1 \end{array} \right) \;,
\end{equation}
where $ \tilde \epsilon$ and $\tilde \epsilon'$ are the infinitesimal transformation parameters. It turns out that the two matrices, $a$ and $b$, do not commute with each other. Interestingly, the operation $b^{-1} a^{-1} b a$ produces a net rotation. 
\begin{equation} 
 b^{-1} a^{-1} b a =  \left( \begin{array}{cc}
 1 & -2\tilde \epsilon \tilde \epsilon'  \\
 2\tilde \epsilon \tilde \epsilon' & 1 \end{array} \right) + 
 \mathcal O(\tilde \epsilon, {\tilde \epsilon'})^3 \;.
\end{equation}
Thus a certain combination of shear transformations can generate a rotation \cite{Read:2010epa}. This already suggests that viscosities can be related to angular momentum in some cases.   

One can promote this simple geometric picture to the level of a full fledged quantum field theory Ward identity. As reviewed in \S \ref{sec:AngularMomentum} and \S \ref{sec:HallViscosityApplication}, the identity can produce two mutually exclusive relations depending on the choices of the symmetries \cite{Bradlyn:2012ea}\cite{Hoyos:2014lla}\cite{Hoyos:2015yna}. 

If one keeps the rotation symmetry and the angular momentum without translation invariance (Option A described in \S \ref{sec:AngularMomentum}), there exists a rather non-trivial relation between the Hall viscosity and the angular momentum, $\eta_H = \hbar \bar s \bar n/2 $, signifying that the Hall viscosity is a half of the average angular momentum $\bar s = -\ell$ of the system in a suitable unit \cite{Read:2008rn}\cite{Read:2010epa}. The angular momentum can be associated with the topological quantity called shift $\bar {\mathcal S}=2\bar s $. If one keeps both the translation and rotation symmetries without the angular momentum (Option B described in \S \ref{sec:AngularMomentum}), one can obtain a different identity stating that the Hall viscosity $\eta_H$ is related to the Hall conductivity $\sigma_H$ \cite{Hoyos:2011ez}. This happens when one includes a background magnetic field ($B\neq 0$) with the Galilean invariance. These are reviewed in \S \ref{sec:HallViscosityApplication}. 

Here we point out that the previous Ward identities are not suitable for describing the physical system with Skyrmions due to the associated topological nature. In the following section, we introduce the central extension of the topological Skyrmion charge in the momentum momentum commutation relation, which can be directly included in the Ward identities. 

\subsection{Skyrmion charge as a central extension} \label{sec:SkyrmionCentralExtension}

When quantum mechanics came along, the deterministic idea of physics that all the physical quantities can be completely determined shattered. Precise and simultaneous measurements of a particle's position and momentum are not possible at a fundamental level. This is referred to the uncertainty relation $ \Delta x^i \Delta p^j \geq \hbar/2 \delta^{ij} $ that originates from the commutation relations $[x^i, p^j] = i \hbar \delta^{ij} $, while $[x^i,x^j]=0$, and $[p^i,p^j]=0$ where $i,j=1,\ldots,d$ for $d$ - dimensional space. To present the material in a simpler manner, we set $\hbar =1$. 

In some situations, there even exist further limitations for determining the momenta in different coordinate directions. This is realized by the momentum - momentum commutation relation that can be modified by a central extension \cite{WeinbergI}. In quantum field theories, we can formulate this as $ [P^i , P^j] = i C^{ij} $, where $ P^i = \int d^d x ~T^{0i} (\vec x) $ is the momentum operator formed from the local energy momentum tensor $ T^{0i} (\vec x)$. 

There are several obstructions to have a non-zero $C^{ij}$ due to the Jacobi identities, which are necessarily satisfied. Let us start with the definition of the momentum operator that generates translations 
\begin{align}
	[P^i,T^{0j}(t, \vec x)] = -i \partial_i T^{0j} (t, \vec x) \;.
\end{align}
Upon integrating this expression on both sides, we find that the right-hand side is a total derivative. To circumvent this, we need either finite boundary contributions from the boundary or singularities in $T^{0j}$. We provide some concrete realizations of these in the later part of this section. 

When the Jacobi identity involves a boost operator $M^{0i} $, energy and a momentum operators, 
\begin{align}
\begin{split}
	J(M^{0i}, P^0, P^j) &= [M^{0i}, [P^0, P^j]] + [P^0, [P^j, M^{0i}]] + [P^j, [M^{0i}, P^0]]  \\
	&= i C^{ij}  \;,
\end{split}	
\end{align}
where we use $[M^{0j},P^\mu]=i(\delta^{\mu 0}P^j - \delta^{\mu j} P^0)$. Thus the Jacobi identity is violated by the central extension $ C^{ij} $.  In the presence of a boost symmetry, either Lorentz or Galilean, the left-hand side vanishes and thus $C^{ij}=0$. Thus to have the central extension $C_{ij}$ in our interesting system, the boost symmetry is not allowed. 

We may also consider the Jacobi identity involving a rotation operator $M^{ij} $ and two momentum operators, $P^k$ and $P^l$. We also contract the indices $l$ and $j$ to have a simple result.
\begin{align}\label{MPPJacobi}
\begin{split}
	\delta_{lj} J(M^{ij}, P^k, P^l)  &= \delta_{lj} ( [M^{ij}, [P^k, P^l]] + [P^k, [P^l, M^{ij}]] + [P^l, [M^{ij}, P^k]])  \\
	&=i\delta_{lj} ([M^{ij}, C^{kl}] +C^{lj}\delta^{ki} +C^{ki}\delta^{lj} -C^{li}\delta^{kj} - C^{kj}\delta^{li})  \\
	&=i (d- 2) C^{ki} \;, 
\end{split}
\end{align}
where we use $[M^{ij},P^k]=i(\delta^{ki}P^j - \delta^{kj} P^i)$ 
and $d$ is the number of spatial dimensions. The left-hand side vanishes in the presence of rotation symmetry. We have a couple of options to satisfy \eqref{MPPJacobi}. We either discard the rotation symmetry or focus on two spatial dimensions. We choose to consider the physical system with a rotational symmetry, set $d=2$ henceforth, and discard the boost symmetry.   

There is a well known example for this types of constructions, a constant background magnetic field $B$ with a modified energy momentum tensor $T^{0j}_B = T^{0j} - (B/2) \epsilon^j_n x^n J^0 $, where $J^0$ is a charge density operator. This modification corresponds to a minimally coupled momentum operator. 
\begin{align}
	[P^i_B,T^{0j}_B(t, \vec x)] = - i \partial_i T^{0j}_B (t, \vec x) - i q \epsilon^{ijk} B^k J^0 (t, \vec x) \;.
\end{align}
There exists a central extension due to the last term, for $J^0 \neq 0$. Boost symmetry is broken by the background magnetic field. 

Actually there is more interesting physical system that allows a central extension: a spin system with magnetic Skyrmions that is described by a continuous spin configuration $\vec n(t, \vec x)$. 
\begin{align}
	\vec n = (\sin \Theta (\rho) \cos \Phi (\phi), \sin \Theta (\rho)\sin \Phi (\phi), \cos \Theta (\rho) ) \;.
\end{align}
We adapt the coordinate system $\vec x = (\rho \cos \phi, \rho \sin \phi, z)$. The Lagrangian density for $\vec n$ is given by
\begin{align} \label{LagrangianSkyrmion}
	\mathcal L = \dot \Phi (\cos \Theta - 1) - (J/2) \partial_i \vec n \cdot \partial_i \vec n \;,  
\end{align}
where $ \dot \Phi $ is the time derivative of $\Phi $, whose Dirac commutator with conjugate momentum $p_{\Phi} (t, \vec x) = \cos \Theta (t, \vec x) -1 $ is given by  
\begin{align}
[\Phi (t, \vec x), p_{\Phi} (t, \vec x')] = i \delta^2 (\vec x - \vec x') \;.
\end{align}
The energy momentum tensor can be computed as $ T^{0i} (t, \vec x) = p_\Phi (t, \vec x) \partial_i \Phi (t, \vec x) $. 
Then \cite{Watanabe:2014pea}\cite{Toma1991}
\begin{align}\label{CommutationRelationC}
	[P^i,T^{0j}(t, \vec x)] = -i \partial_i T^{0j} (t, \vec x) + i \epsilon^{ij} (\epsilon^{kl} p_\Phi (t, \vec x) \partial_k \partial_l \Phi (t, \vec x) )\;,
\end{align}
This system has a central extension in the commutation relations because the field $\Phi$ has a vortex singularity at $\Theta = \pi$. The singularity also breaks boost symmetry. Note that the conservation equation $ \partial_\mu T^{\mu\nu}=0$, with $\mu, \nu = 0,1,2$, can be shown to be satisfied explicitly. 

\begin{figure}[t!]
	\begin{center}
		\includegraphics[width=0.93\textwidth]{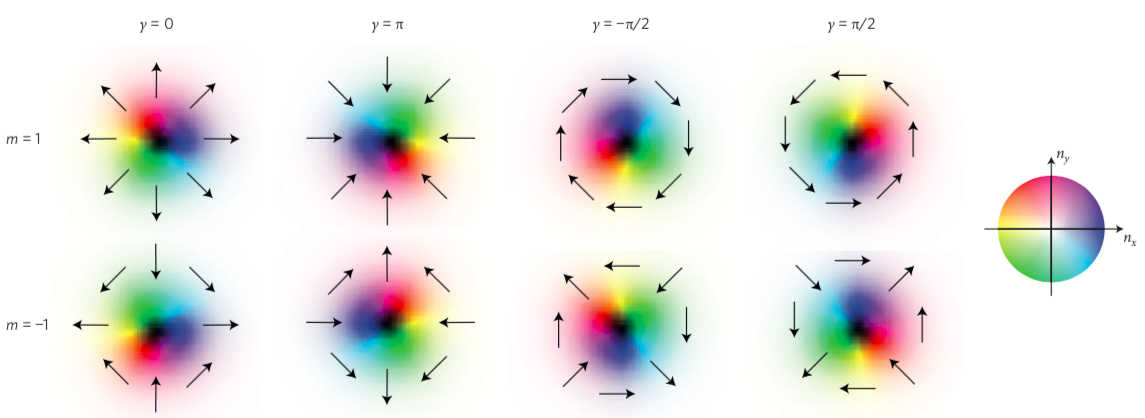} 
		\caption{\footnotesize\small Skyrmion structures with varying $m$ and $\gamma$. The arrows indicate the direction of the in-plane spin component, and the brightness indicates the normal component to the plane, with white denoting the up direction and black the down direction. All the structures of the anti-skyrmions ($m = -1$) are equivalent on rotation in the $xy$ plane. Reproduced with permission from \cite{SkyrmionTopologicalReview}.
		}
		\label{fig:SkyrmionKinds}
	\end{center}
\end{figure}

For our purpose, it is convenient to use a local version of \eqref{CommutationRelationC}
\begin{align}\label{CommutationRelationLocal}
	[T^{0i}(t, \vec x),T^{0j}(t, \vec x')]  = i \left(- \partial_i T^{0j} + \partial_j T^{0i}  + i \epsilon^{ij} c \right) \delta^2 (\vec x - \vec x') \;.
\end{align}
where $c$ is topological charge density. While we have presented this relation in the context of a particular spin model, it holds for Skyrmion systems in general, independent of the details such as the form of Lagrangian given in \eqref{LagrangianSkyrmion}. 

The total Skyrmion charge is the integral of the charge density $c$. To see this, let us consider a particular spin configuration that is composed of up-spins $\uparrow$ at infinity, $\cos \Theta (\rho=\infty) = 1$, and down-spins $\downarrow$ at the center, $\cos \Theta (\rho=0) = -1$. We further specify $\Phi(\phi) = m\phi + \gamma $. $m$ represents the winding number along the $\phi$ direction and $\gamma$ parameterizes the Skyrmions with different phases. See, for example \cite{SkyrmionTopologicalReview} and figure \ref{fig:SkyrmionKinds}. The topological Skyrmions charge is integral of the density $C^{ij} = \epsilon^{ij} C $.
\begin{align} \label{TopologicalChargeDensity}
	C & = \int d^2 x ~\vec n \cdot \left[\frac{\partial \vec n}{\partial x} \times \frac{\partial \vec n}{\partial y}  \right]  
	 = \int_0^\infty  d\rho \int_0^{2\pi} \!\! d\phi \frac{d \Theta}{d\rho} \frac{d \Phi}{d\phi}	\sin \Theta \nonumber \\
	 & = \cos \Theta (\rho)  \Big|_{0}^{\infty} \Phi(\phi) \Big|_{0}^{2\pi} 
	 = 2 \cdot  2\pi m.
\end{align} 
Here $m$ is the number of $2\pi$ rotations of the spin between the infinity and the center. Only $m = \pm 1$ are stable Skyrmions. Magnetic Skyrmions of this sort with $\gamma=1/2$ have been first observed in MnSi using neutron scattering experiment \cite{SkyrmionExp1} and also in thin films with a real space Lorentz transmission electron microscope in $Fe Co Si$ \cite{SkyrmionExp2}. They are named as Bloch Skyrmions. The Skyrmions with $\gamma=0, \pi$ have been also discovered on the $Fe $ monolayer on the $Ir(111)$ by direct image of spin-polarized scanning tunneling microscopy \cite{Heinze2011}.  They are know as Neel Skyrmions. Since then Skyrmions have been observed in many different materials such as conductor $MnSi $, semi-conductor $ Fe_x Co_{1-x} Si$ and insulator $ Cu_2 O Se O_3$ with different geometry including 3 dimensional bulk, 2 dimensional thin film (mono layer) and 1 dimensional nano wires \cite{SkyrmionTopologicalReview}.

\subsection{General Ward identities with a topological charge} \label{sec:SimpleWI}

We are going to present a simple quantum field theory Ward identities that is a consequence of symmetries and conservation laws. Thus we age going to use the conservation equation $\partial_\mu T^{\mu\nu}=0$. In addition to this, we adapt the topological charge given in \eqref{CommutationRelationLocal} in the identity, which is achieved for the first time in \cite{Kim:2015qsa}. 

We consider a commutator of two energy currents, $T^{0j}(x^\mu )$ and $T^{0l}(x'^{\mu}) $, or two momentum densities, multiplied by a step function $ \theta(x^0-x'^0)$, which is called the retarded Green's function introduced in \eqref{RetardedGRT}. 
\begin{align} \label{RetardedGRT2}
	G^{0j0l}(x^\mu ;x'^{\mu} )=  i\theta(x^0-x'^{0}) \langle [ T^{0j}(x^\mu ), T^{0l}(x'^{\mu} )] \rangle \;.
\end{align} 
Let us examine a simple Ward identity by taking two time derivatives on this equation \eqref{RetardedGRT2}. Then there are 4 terms, with the two derivatives acting on the two $T$s, on one $T$ and the step function, or both on the step function. They are organized as   
\begin{align} \label{ComputationWI2}
\begin{split}
	{\partial_0' \partial_0 G^{0j0l}(x^\mu ;x'^{\mu}) }
	& = \partial_n \partial_m' G^{njml}(x^\mu ;x'^{\mu})	\\
	& - [ \delta' (x^0-x'^0) + \delta (x^0-x'^0) (\partial_0 - \partial_{0}') ] 
	C^{0j0l}(x^\mu ;x'^{\mu}) \;, 
\end{split}	
\end{align}  
where we use the conservation equation $\partial_0 T^{0j} =- \partial_m T^{m j}$ to rewrite the first term so that the Green's function turns into $ G^{njml}(x^\mu ;x'^{\mu} )$
\begin{align} \label{ContactTermsT2}
\begin{split}
	G^{njml}(x^\mu ;x'^{\mu} ) &=  i\theta(x^0-x'^{0}) \langle [ T^{nj}(x^\mu ), T^{ml}(x'^{\mu} )] \rangle  \;, \\
	C^{0j0l}(x^\mu ;x'^{\mu} ) &\to C^{0j0l}(x^0,\vec x ,{\vec x}' ) =  i\langle [ T^{0j}(x^0,\vec x ), T^{0l}(x^0,{\vec x}' )] \rangle \;.
\end{split}	
\end{align} 
The second line of \eqref{ComputationWI2} has three terms are defined at the same time due to the delta function. Then the Green's function turn into equal time correlator that is also called contact terms.  

Now the new key ingredient is the fact that the contact terms can be rewritten as one point functions and the topological Skyrmion charge by using \eqref{CommutationRelationLocal}. Here we also focus on the systems with the translation and rotation symmetries without angular momentum (Option B described in \S \ref{sec:AngularMomentum}). Then we perform a Fourier transform in the time direction as $\int d(x^0\- x'^0)e^{i\omega (x^0 - x'^0)} \cdots $ on both sides of \eqref{ComputationWI2}, we arrive at our general result in this simplest case.  
\begin{align}\label{GeneralWI}
	\begin{split}
		\omega^2 G^{0j0l} (\omega,\vec x ,{\vec x}' ) 
		&= \partial_n \partial_m' G^{njml} (\omega,\vec x ,{\vec x}' ) \\
		&+(1/2) [2i\omega \epsilon^{jl} \epsilon^n_{\ m} \partial_n \langle T^{0m}(\vec x ) \rangle 
		- \partial_l \partial_n \langle T^{nj} (\vec x ) \rangle
		-  \partial_j \partial_m \langle T^{ml} (\vec x ) \rangle ] \delta(\vec x -{\vec x}' )  \\
		&~{- i \omega c^{jl} \delta(\vec x -{\vec x}' )} \;.
	\end{split}
\end{align} 
This is abstract, yet general Ward identities with a topological charge (Topological WI). Here $ G^{0j0l} (\omega,\vec x ,{\vec x}' ) $ and $G^{njml} (\omega,\vec x ,{\vec x}' ) $ are two spatial and four spatial index Green's functions, while $ \langle T^{0m}(\vec x ) \rangle $, $ \langle T^{nj} (\vec x ) \rangle $ and $ \langle T^{ml} (\vec x ) \rangle $ are one point functions. These are specified further in the following sections, depending on the situations in the physical systems of interest. We note the presence of the central charge $c^{jl}$. 
 
Ward identities are actually consequences of the conservation equation $\partial_\mu T^{\mu j} = 0$. Here we further generalize to include the central extension, the topological charge of Skyrmions, that arises from the commutation relations among energy momentum operators \cite{Kim:2015qsa} based on \cite{Hoyos:2014lla}\cite{Hoyos:2015yna}. As advertised before, there are two independent and exclusive cases we can consider depending on symmetries. For simplicity, we focus on the case with translation and rotation invariance without the angular momentum, Option B described in \S \ref{sec:AngularMomentum}. 

\subsection{Topological WI with rotation and translation symmetries} \label{WIRotAndTrans}

Starting from the general Ward identity \eqref{GeneralWI}, we focus on a the physical system with the rotation and translation symmetries and examine the consequences of the topological Ward identities. The translation symmetry does not allow any non-trivial spatial dependence for the one point functions. Thus all the contact terms in the second line in \eqref{GeneralWI} vanish. This makes the Ward identity particularly simple. Then, \eqref{GeneralWI} becomes 
\begin{align}\label{TranslationWI}
	&\omega^2  G^{0j0l} (\omega,\vec x ,{\vec x}' ) 
	= \partial_n \partial_m' G^{njml} (\omega,\vec x ,{\vec x}' ) 
	- i \omega c^{jl} \delta(\vec x -{\vec x}' ) \;.
\end{align}  

The presence of the translation symmetry allows us to perform a Fourier transform 
\begin{align}
	G^{\mu\nu\alpha\beta}(\omega,\vec x - {\vec x}') = 1/(4\pi^2) \int d^2 \vec q ~ e^{i\vec q\cdot (\vec x -{\vec x}' )} ~ \tilde G^{\mu\nu\alpha\beta}(\omega,q) \;.
\end{align} 
The rotation symmetry is useful to organize the most general tensor structures for the retarded Green's functions. In the momentum space, the two spatial index Green's function can be written as 
\begin{align}\label{IndexStructures1}
	\begin{split}
		\tilde G^{0i0k} &= -i\omega \big[\delta^{ik} \bm{\kappa}_\delta +\epsilon^{ik} \bm{\kappa}_{\epsilon}
		+ q^i q^k\bm{\kappa}_{q} +(\epsilon^{in}q_n q^k+\epsilon^{kn}q_n q^i) \bm{\kappa}_{q\epsilon} \big]\;,
	\end{split}
\end{align}  
where $\bm{\kappa}_\delta$ and $\bm{\kappa}_{q} $ are the symmetric thermal conductivities. The proper name might be momentum conductivities which are related to the response of the momentum density and current, $T^{0i}$ and $T^{ij}$. The other two, $\bm{\kappa}_{\epsilon}$ and $ \bm{\kappa}_{q\epsilon} $, are the anti-symmetric thermal conductivities that contribute in the absence of the parity symmetry. In the presence of electric charges, there are additional contributions with the electric conductivities.

The four spatial index Green's function has the familiar tensor structure as we already reviewed in \S \ref{sec:HallViscosity} with the rotation symmetry. They are related to the viscosities. 
\begin{align}\label{IndexStructures2}
	\begin{split}
		\tilde G^{njml} 
		&= -i\omega [\eta(\delta^{nm}\delta^{jl}+\delta^{nl}\delta^{mj} -\delta^{nj}\delta^{ml})
		+\zeta\delta^{nj}\delta^{ml} \\
		&\qquad\quad~~+(\eta_H/2) (\epsilon^{nm}\delta^{jl}+\epsilon^{nl}\delta^{jm}+\epsilon^{jm}\delta^{nl}
		+\epsilon^{jl}\delta^{nm}) ] \;.
	\end{split}
\end{align}  
Here $\eta$ and $\zeta$ are the shear and bulk viscosities that are related to the symmetric part of the tensor. In the absence of parity symmetry, one can use the anti-symmetry epsilon tensor. Thus we have also Hall viscosity $\eta_H$.

After a little algebra for \eqref{TranslationWI} with the Fourier transform, adapting the tensor structures \eqref{IndexStructures1} and \eqref{IndexStructures2}, we obtain 
\begin{align}\label{MomWI2}
	&\omega^2  [  \delta^{jl}\bm{\kappa}_\delta + \epsilon^{jl} \bm{\kappa}_{\epsilon}
	+ q^j q^l \bm{\kappa}_{q}
	+ (\epsilon^{jn}q_n q^l+\epsilon^{ln}q_n q^j) \bm{\kappa}_{q\epsilon} ] \\
	&= \delta^{jl} q^2\eta  + \epsilon^{jl} ({c}+ q^2 \eta_H ) + q^j q^l \zeta  \;. 
\end{align} 
This equation seems to give a single Ward identity. Note that there are 4 independent tensor structures, and thus there are actually 4 independent Ward identities depending on these tensor structures. 
\begin{align} \label{MomWI4}
\begin{split}
	&\omega^2 \bm{\kappa}_\delta =q^2 \eta \;,\quad\qquad
	\omega^2 \bm{\kappa}_\epsilon =c+ q^2 \eta_H \;,\quad~~ \\
	&\omega^2 \bm{\kappa}_{q}= \zeta \;, \quad\qquad\quad
	\omega^2 \bm{\kappa}_{q\epsilon} = 0 \;.
\end{split}	
\end{align}
Thus thermal conductivities are directly related to viscosities except the topological charge. This is the general topological Ward identities for the physical systems with translation and rotation symmetry in momentum space. 

\subsection{Skyrmions Hall transports in insulating materials} \label{sec:WIInsulating}

The topological Ward identities \eqref{MomWI2} and \eqref{MomWI4} are already interesting. One can push a little further by isolating the momentum independent parts in \eqref{MomWI4}. Then, we arrive at the simple relations 
\begin{equation}\label{HallWI}
	\begin{split}
		& \omega^2 \bb{\kappa}^{(0)}_\delta = 0\;, \qquad 
		\omega^2 \bb{\kappa}^{(0)}_\epsilon = c\;,
	\end{split}
\end{equation} 
where the superscript $^{(0)}$ denotes the momentum independent contribution only. Intuitively, the reason $ \bb{\kappa}^{(0)}_\delta $ vanishes, while $ \bb{\kappa}^{(0)}_\epsilon $ does not, is the fact that Skyrmions are associated with spontaneously broken translation symmetry along with the broken parity. The imprints of the broken parity symmetry can only enter through the parity odd parts of the conductivity at zero momentum. More precisely, the second identity predicts that the formation of a single Skyrmion results in the creation of a unit of thermal Hall conductivity $\bb{\kappa}^{(0)}_\epsilon$ in units of the quantized topological charge density. The frequency dependence is a consequence of the pole structure of the Goldstone boson that manifests itself in the retarded momentum correlator. 

In the presence of disorder, the behavior $ \bb{\kappa}^{(0)}_\epsilon = c /\omega^2 $ could, in principle, be lifted. However, numerical simulations have confirmed that Skyrmion motions are unaffected by impurities, in contrast to the case of domain walls \cite{Impurity}. Furthermore, recent experiments confirmed that Skyrmions move very efficiently with velocities reaching the predicted for perfect pinning free systems \cite{Litzius2017}. The thermal Hall conductivity $\bb{\kappa}_\epsilon$ is dissipationless and exists even at zero temperature. While our Ward identity relations are valid at finite temperatures as well, measurements will be cleaner at very low temperatures, where additional dissipative contributions are suppressed. Another interpretation of eq. \eqref{HallWI} is that the Skyrmions carrying the thermal current propagate in an effective magnetic field given by the Skyrmion charge density $c_{ij}$, leading to a thermal Hall effect.  

For the momentum dependent terms in \eqref{MomWI2} and \eqref{MomWI4}, we obtain 
\begin{equation}\label{NeutralWIMomen}
	\begin{split}
		\omega^2 \bar{\bb{\kappa}}_\delta  
		= q^2 \eta  \;, \quad\quad
		\omega^2 \bar{\bb{\kappa}}_\epsilon 
		= q^2 \eta_H \;, \quad \quad
		\omega^2 \bb{\kappa}_{q}  
		= \zeta \;,
	\end{split}
\end{equation} 
where the bar $~\bar{}~$ signifies the non-constant momentum dependent part. For example, 
$\bar{\bb{\kappa}}_\epsilon =  \bb{\kappa}_\epsilon - \bb{\kappa}^{(0)}_\epsilon 
= q^2 \bb{\kappa}_\epsilon^{(2)} + q^4 \bb{\kappa}_\epsilon^{(4)} + \cdots $. 
Thus, thermal conductivities are directly connected to the viscosities of the system, which are previously confirmed \cite{Hoyos:2015yna}. 
Furthermore, it follows from \eqref{MomWI4} that $ \bb{\kappa}_{q\epsilon} =0$. 

\begin{figure}[h!]
	\begin{center}
		 \includegraphics[width=0.73\textwidth]{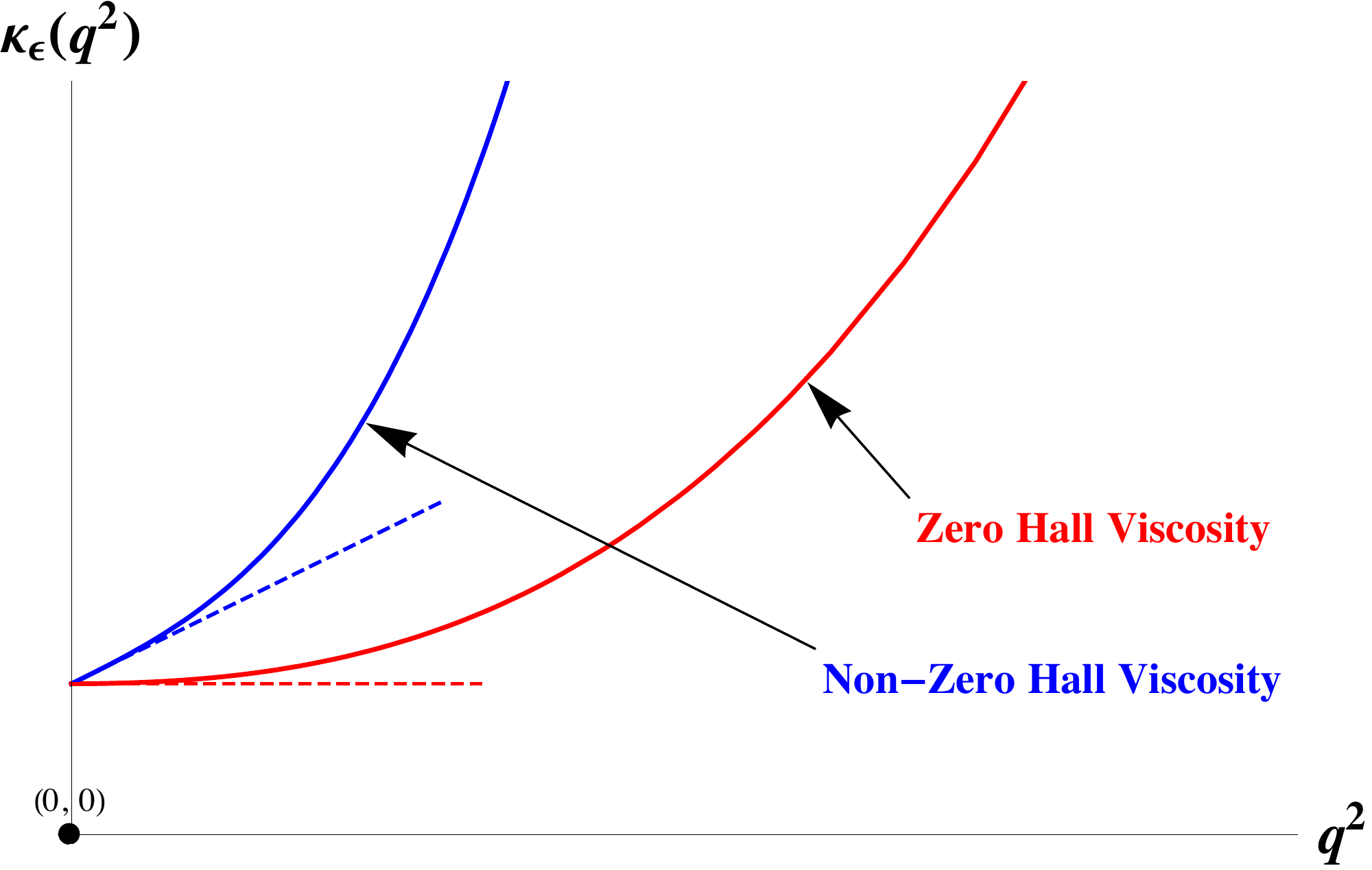} 
		\caption{\footnotesize\small Confirmation of the existence of the Hall viscosity. Intercept of thermal Hall conductivity $\bb{\kappa}^{(0)}_\epsilon $ is non-zero and is proportional to the Skyrmion charge density. Non-vanishing slope $\bb{\kappa}^{(2)}_\epsilon $ as a function of momentum squared $q^2$ confirms the existence of the Hall viscosity. 
		}
		\label{fig:HallViscosityMT1}
	\end{center}
\end{figure}

Finally, we can describe a simple way to measure the Hall viscosity or to confirm its existence. Combining the second equations in \eqref{HallWI} and \eqref{NeutralWIMomen}, we get 
\begin{equation} \label{MeasuringHV}
	\begin{split}
		\eta_H = c \frac{ \bar{\bb{\kappa}}_\epsilon}{q^2 \bb{\kappa}^{(0)}_\epsilon} 
		~\to~  c \frac{ \bb{\kappa}_\epsilon^{(2)} }{\bb{\kappa}^{(0)}_\epsilon}    \;,
	\end{split}
\end{equation} 
where we take the limit $q^2 \to 0$. Once the thermal Hall conductivity $\bb{\kappa}_\epsilon $ is measured as a function of $q^2 $, the Hall viscosity is nothing but the Skyrmion density multiplied by the ratio between the slope and $\bb{\kappa}_\epsilon$-intercept $\bb{\kappa}_\epsilon(q^2=0)$. Note that this is only applicable in the presence of nonzero Skyrmion density. See the figure \ref{fig:HallViscosityMT1}. 

Here we consider the physical system such as insulating magnets in the presence of  Skyrmions without Magnon contributions, such as at very low temperature, where the low energy excitations such as Magnons can be suppressed. In general, the Magnon  excitations can play an important role. We propose some simple ways to measure the Hall viscosity in the presence of the Magnons excitations in \S \ref{sec:Outlook}.  

\newpage
\subsection{Topological WI with magnetic fields and electric currents }\label{sec:WIChargeB}

The topological Ward identities \eqref{GeneralWI} in abstract form and \eqref{MomWI2} with translation and rotation symmetries in momentum space are quite general already. Yet, they only concern about the conservation equations related to the energy momentum tensor. Here we generalize this Ward identity by including both a background magnetic field and an electric current. Electric currents turn out to be very useful to control the Skyrmions because of the interaction between the Skyrmion spins and electron spins. 

In the presence of the magnetic field $B$ and the electric current $J^\mu$, the conservation equations and momentum generators are modified as follows in 2+1 dimensions.  
\begin{align}
	\partial_\mu T^{\mu i}=B\epsilon^i_{\ j} J^j \;, \qquad 
	T^{0j}_B = T^{0j}-(B/2)\epsilon^j_{\ n}x^n J^0 \;. 
\end{align} 
The first equation is from the general expression $ \partial_\mu T^{\mu\nu} = F^{\nu\rho}J_\rho$ for both energy momentum tensor and current, while the  second one can be thought as a minimal coupling substitution in the presence of the electromagnetic potential. 

These modification produce several changes for the Ward identities. When one uses the conservation equation $\partial_0 T^{0i} = - \partial_m T^{mi} + B\epsilon^i_{\ m} J^m $, the momentum-momentum correlator $G^{0j0l} \sim \langle [T^{0j}, T^{0l}] \rangle$ on the left hand side of \eqref{TranslationWI} also includes $G^{0j,m} \sim \langle [T^{0j}, J^m] \rangle$, $G^{n,0l} \sim \langle [J^n, T^{0l}]\rangle$, and $G^{nm} \sim \langle [J^n, J^m]\rangle$. Similar to the thermal conductivities, all these two spatial index Green's functions in the momentum space are the conductivities: thermal conductivities $\bb{\kappa}\sim \langle[T,T]\rangle $, thermoelectric conductivities $\bb{\alpha}\sim \langle[T,J]\rangle$ and $\bb{\alpha}^*\sim \langle[J,T]\rangle$ and also familiar electric conductivities $\bb{\sigma}\sim \langle[J,J]\rangle $. On the right hand side of \eqref{TranslationWI}, we have only one change, the addition of $B \langle J^0\rangle $, which is nothing but the magnetic field times the electric charge density $\langle J^0\rangle = \rho$. 

While the computation is slightly more complicated, it is straightforward. After some algebra, we obtain the topological Ward identities in the presence of the translation and rotation symmetry. 
\begin{align}
	\begin{split}
		& \omega^2 G^{0j0l} -i\omega B\epsilon^j_{\ n}G^{n,0l}+i\omega B\epsilon^l_{\ m}G^{0j,m}
		+B^2\epsilon^j_{\ n}\epsilon^l_{\ m} G^{nm}\\
		&\qquad\quad = \partial_n \partial_m' G^{njml}  - i\omega\epsilon^{jl}\left[ {c} -B\langle J^0\rangle \right]\delta (\vec x - {\vec x}' )  \;.
	\end{split}
\end{align}
After performing the Fourier transform and using the index structures similar to \eqref{IndexStructures1} and \eqref{IndexStructures2}, we re-express the equation using various conductivities.
\begin{align}\label{ChargedWI}
	\begin{split}
		& \delta^{jl}  \big[\omega^2 \bb{\kappa}_\delta 
		+ i\omega B \left(\bb{\alpha}_\epsilon+ {\bb{\alpha}}^*_\epsilon+ q^2 [ \bb{\alpha}_{q\epsilon}\!-\! {\bb{\alpha}}^*_{q\epsilon}]  \right) + B^2 \left( \bb{\sigma}_\delta + q^2 \bb{\sigma}_{q} \right) \big]  \\
		&+{\epsilon^{jl} \big[\omega^2 \bb{\kappa}_\epsilon
			- i\omega B \big(\bb{\alpha}_\delta + {\bb{\alpha}}^*_\delta + q^2 [\bb{\alpha}_{q} + {\bb{\alpha}}^*_{q}]/2  \big) + B^2 \bb{\sigma}_\epsilon \big] } \\
		&+ q^j q^l  \big[ 
		\omega^2 \bb{\kappa}_{q} - 2i\omega B (\bb{\alpha}_{q\epsilon} -  {\bb{\alpha}}^*_{q\epsilon}) - 
		B^2 \bb{\sigma}_{q} \big]   \\
		&+ (\epsilon^{jo}q^l + \epsilon^{lo}q^j) q_o \big[ 
		\omega^2 \bb{\kappa}_{q\epsilon} + i\omega B (\bb{\alpha}_{q} \!-\! {\bb{\alpha}}^*_{q})/2 - B^2 \bb{\sigma}_{q\epsilon} \big]  \\
		&= \epsilon^{jl}  \big[{c} - B\rho + q^2 \eta_H \big] 
		+ \delta^{jl} q^2  \eta + q^j q^l  \zeta   \;. 
	\end{split}
\end{align}
This is a general topological Ward identities formula in the presence of charge density $\langle J^0 \rangle = \rho$ and a magnetic field $B$. While the expressions are a little complicates, the result is quite illuminating. The left hand side has four tensor structures that are nicely packaged along with all possible combinations of the allowed conductivities depending on the tensor structures. Thus there are four independent Ward identities that can be readily extracted from the expression. Applying this to the parity breaking hydrodynamics without the topological extension, ${c}=0$, we find that these results are consistent with those based on the magnetically charged black hole solutions \cite{Hoyos:2015yna}\cite{Hartnoll:2007ip}. 

To make the expression a little simpler, we isolated the momentum independent parts of the Ward identities \eqref{ChargedWI}. Then only terms with the tensor structures $ \delta^{jl} $ and $ \epsilon^{jl}$ contribute. 
\begin{equation}\label{BHallWI}
	\begin{split}
		& \omega^2 \bb{\kappa}^{(0)}_\delta
		+i\omega B (\bb{\alpha}^{(0)}_\epsilon+ {\bb{\alpha}}^{*(0)}_\epsilon) + B^2 \bb{\sigma}^{(0)}_\delta  
		= 0 \;, \\
		& \omega^2 \bb{\kappa}^{(0)}_\epsilon
		-i\omega B (\bb{\alpha}^{(0)}_\delta + {\bb{\alpha}}^{*(0)}_\delta) + B^2 \bb{\sigma}^{(0)}_\epsilon 
		= c - B\rho \;,
	\end{split}
\end{equation}
where the superscript $^{(0)}$ denotes the momentum independent part. One can check these expressions reduce to \eqref{MomWI4} and \eqref{MomWI2} when one simply takes $B\to 0$ (and $\rho \to 0$). In the opposite limit $B \to \infty$ without the topological charge, we have 
\begin{equation} \label{LargeBLimit}
\bb{\sigma}^{(0)}_\delta = 0 \;, \qquad\quad \bb{\sigma}^{(0)}_\epsilon = -\rho/B \;. 
\end{equation}
This result is consistent with the known results of the Hall conductivity from the dyonic black hole with an appropriate identification of the sign convention \cite{Hoyos:2015yna}\cite{Hartnoll:2007ai}. 

At the non-zero momentum, there are four independent relations connecting viscosities and  conductivities as in the neutral case. Here we only consider the parity odd contribution that is related to the Hall transports and the Hall viscosity. 
\begin{align} \label{BWIHallViscosity}
	q^2 \eta_H = \omega^2 \bar{\bb{\kappa}}_\epsilon + B^2 \bar{\bb{\sigma}}_\epsilon 
	-i\omega B \Big[\bar{\bb{\alpha}}_\delta + \bar{{\bb{\alpha}}}^*_\delta
	+\frac{\bb{\alpha}_{q} + {\bb{\alpha}}_{q}^*}{2} \Big].
\end{align}
where the bar $~\bar{} ~$ indicates the non-constant momentum dependent part that is already defined above. We consider this formula to propose a simple way to examine the existence of the Hall viscosity below. 

Before moving on, we comment on the identities with broken translational symmetry. Then one can include the angular momentum $\ell$ according to Option A considered in \S \ref{sec:AngularMomentum}. With the broken translation, one can have contributions from the one point functions. Then the following modifications appear after similar computations. 
\begin{align}
\eta_H \to \eta_H + \ell/2 \;, \qquad \zeta \to \zeta-i(p-BM)/\omega \;,
\end{align}
where $M$ is magnetization that can be defined  as $\langle J^{i} \rangle =\epsilon^{ik}\partial_k M$. More details can be found in \cite{Kim:2015qsa}. 

\subsection{Topological WI for conducting materials at zero momentum} \label{sec:HallConductivity}

Here we review a set of experiments that measure the electric Hall conductivity or Hall resistivity in the presence of the topological Skyrmions. In the seminal papers \cite{HC1}\cite{HC2}, the authors of two different groups have measured the electric Hall conductivity of the MnSi with varying magnetic fields for a wide range of fixed temperature (or varying temperature for a wide range of fixed magnetic field) that cover the so-called A-Phase, which is a small region in the phase space where the Skyrmions exist. What they observe is a unique rectangular-shape function like excess of Hall conductivity only in the region of the A-Phase. Similar experimental results have been confirmed with more systematic measurements later on in \cite{HC4}.  

How does this happen? Skyrmions are electrically neutral objects because they are made from bunch of spins tightly arranged in a beautiful fashion. Then why do they display electric responses? This happens through the interactions between the spins of Skyrmions and those of conduction electrons that are modeled by the ferromagnetic spin coupling \cite{FerroCoupling2}\cite{SkyrmionTopologicalReview}. In the strong coupling limit, the spin wave function of the conduction electrons is identified with that of the localized spin $\vec n(x^\mu)$ of the Skyrmions. This limit is described by a tight binding model with Hund's rule coupling. There is an equivalent way to say this. The Skyrmionic spin configurations create an emergent magnetic field $b$ with magnitude $b=c/2$ for the conduction electrons due to the tight binding interactions between their spins, where $c$ is the topological charge density of Skyrmions. 

This Hunt rule coupling is not built in the Ward identities. One can model this coupling by combining the energy momentum tensors and electric currents through the modification of the parameters existing in the topological Ward identities. This produces the effects of the interaction between the thermal and charge responses. In particular, we note that the Skyrmion charge density $c$ produces an emergent magnetic field $b=c/2$ \cite{SkyrmionTopologicalReview}, which can change the dynamics of conduction electrons. This is also the case for the magnetic field $B$, which is modified in the presence of $b$. For simplicity, we assume that the emergent magnetic field is homogeneous and constant, which is true for all practical measurements. For the flip side, the motion of the conduction electrons will also influence the thermal response of the Skyrmions due to the tight binding. 

After taking these effects into account, the second equation of the Ward identity \eqref{BHallWI} at the vanishing momentum becomes 
\begin{equation}\label{bBHallWI}
	\begin{split}
	\omega^2  \bb{\kappa}^{(0)}_\epsilon
	-i\omega B_b  (\bb{\alpha}^{(0)}_\delta + {\bb{\alpha}}^{*(0)}_\delta) + B_b^2 \bb{\sigma}^{(0)}_\epsilon  
	= c_b - B_b\rho  \;.
	\end{split}
\end{equation}
This Ward identity is of the same form as \eqref{BHallWI}, with the modifications $B\to B_b \equiv B + b$, which contributes to the charge response, and $c \to c_b \equiv c + c_{el} $, which incorporates an additional contribution to the thermal response from the conduction electrons $c_{el}$. Through these modification, the topological charge density $c $ and the background magnetic $B $ do not alter. The quantities $c $ and $b $ are constants and independent of $B $, while $c_{el} $ is expected to be proportional to $B $ and depends on the tight binding strength. We expect that $b, c, c_{el}$ can be readily identifiable experimentally. In particular, $b$ can be identified from a rectangular-shape function like signature in the Hall conductivity $ \bb{\sigma}_\epsilon$ \cite{HC1}\cite{HC2}\cite{HC4}\cite{HC3}\cite{HC5}, as one passes into and out of the A-Phase, in which the Skyrmions develop a finite density $ c $. Such behavior will also confirm the presence of a nonzero density $c$, which will likewise produce a similar rectangular-shape function like contribution in the thermal Hall conductivity $\bb{\kappa}_\epsilon$ (with an additional $B$-dependent $c_{el}$) when experimental setup sweeps the magnetic field $B $ or the temperature $T$ independently.  

If the ferromagnetic binding between the Skyrmion and conduction electron spins would not exist, the electric Hall conductivity would only pick up contributions from the conduction electrons, and $B_b$ would reduce to $B$. On the other hand, the thermal Hall conductivity would include  both contributions, $c$ and $c_{el}$, with the latter being independent of $B$. Thus the tight binding and the corresponding strength are readily verifiable.

\subsection{Hall viscosity in conducting materials} \label{sec:HallViscosityConductor}

Once the experimental setup described in \S \ref{sec:HallConductivity} is well established, one can move on to the momentum dependent Ward identities that is described in \eqref{BWIHallViscosity}. With this equation and proper modification considered in \S \ref{sec:HallConductivity}, we propose a simple way to measure the Hall viscosity in the conducting materials. 

\begin{figure}[h!]
	\begin{center}
		\includegraphics[width=0.73\textwidth]{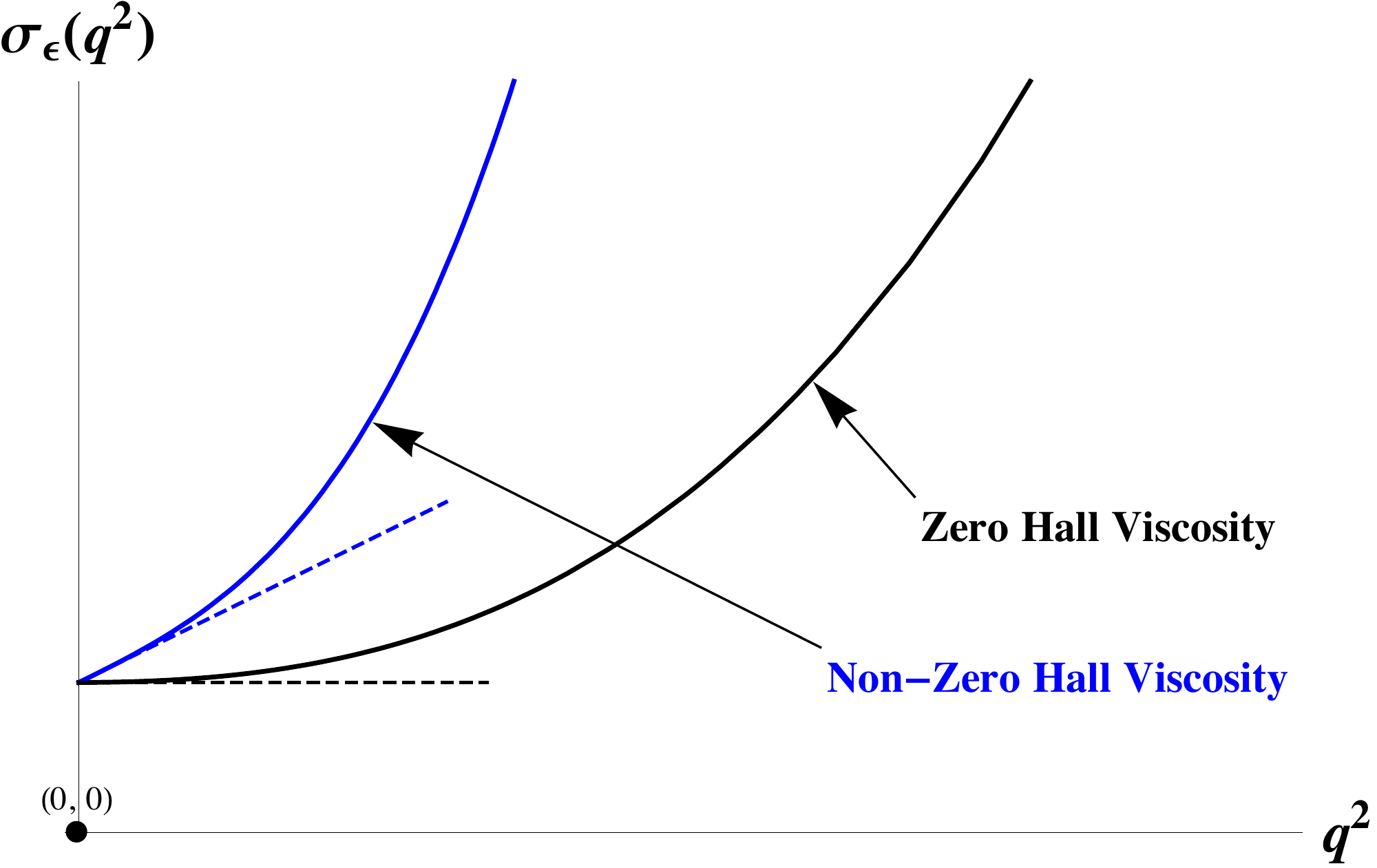} 
		\caption{\footnotesize\small Confirmation of the existence of the Hall viscosity in the conducting magnetic materials. Intercept of electric Hall conductivity $\bb{\sigma}^{(0)}_\epsilon $ is non-zero and is proportional to the Skyrmion charge density. Non-vanishing slope $\bb{\sigma}^{(2)}_\epsilon $ as a function of momentum squared $q^2$ confirms the existence of the Hall viscosity. 
		}
		\label{fig:HallViscosityMT2}
	\end{center}
\end{figure}

Let us divide equation \eqref{BWIHallViscosity} by the second equation of \eqref{BHallWI}, followed by substituting $ B \to B_b, c \to c_b$ as discussed in the previous section. The expression is still complicated. To make things more clear, we take the approximation $\omega/B_b \to 0$ and the zero momentum limit $q^2 \to 0$. We expect these approximations are straightforward and reliable as discussed the results related to \eqref{LargeBLimit}. Then, we obtain an illuminating expression.  
\begin{equation} \label{MeasuringHV2}
	\begin{split}
		\eta_H = (c_b - B_b \rho)~ \frac{ \bb{\sigma}_\epsilon^{(2)} }{\bb{\sigma}^{(0)}_\epsilon} \;.
	\end{split}
\end{equation} 
This is similar to the neutral case described in \eqref{MeasuringHV}. Once the electric Hall conductivity $\bb{\sigma}_\epsilon$ is measured as a function of $q^2$ in the presence of the rotation symmetry, the Hall viscosity is nothing but the modified Skyrmion density $ c_b - B_b \rho$ multiplied by the ratio between the slope and $\bb{\sigma}_\epsilon$-intercept $\bb{\sigma}_\epsilon(q^2=0)$. See the figure \ref{fig:HallViscosityMT2}. We note that one can apply this to the physical systems without topological objects, such as quantum Hall systems, where $c_b=c=0$ and $B_b = B$. This is quite different from the neutral case, which is not applicable in the absence of the topological charge. 

Let us consider the opposite limit $B_b/\omega \to 0 $ after dividing equation \eqref{BWIHallViscosity} by the second equation of \eqref{BHallWI}, followed by substituting $ B \to B_b, c \to c_b$. Then $\eta_H$ reduces to 
\begin{align}	
	\eta_H = (c_b - B_b \rho)~ \frac{ \bb{\kappa}_\epsilon^{(2)} }{\bb{\kappa}^{(0)}_\epsilon} \;.
\end{align}	
Here $\bb{\kappa}_\epsilon$ is the thermal Hall conductivity. This is different from \eqref{MeasuringHV} with the modification $c \to c_b - B_b \rho $ due to the presence of the conduction electrons and the magnetic fields $B$. Note that this identification of $\eta_H$ can also be applied to the systems without Skyrmions.

\subsection{Topological WI with angular momentum \& Hall viscosity} \label{WIAngularMomentum}

Finally, we discuss the topological Ward identities in the presence of the rotation symmetry and the angular momentum without the translation symmetry (Option A discussed in \S \ref{sec:AngularMomentum}) in the context of the insulating magnets. Generalizations for the case with angular momentum for the conducting materials has been done in \cite{Kim:2015qsa}. 

When the system of interest is not translationally invariant, there are additional contributions to the general topological Ward identity \eqref{GeneralWI} because the spatial derivatives of the one point functions no longer vanish. However, the zero momentum identity \eqref{HallWI} will be unmodified because all the contact terms with one point functions are accompanied with spatial derivatives. 
\begin{equation}\label{HallWIOptionB}
\begin{split}
& \omega^2 \bb{\kappa}^{(0)}_\delta = 0\;, \qquad 
\omega^2 \bb{\kappa}^{(0)}_\epsilon = c\;,
\end{split}
\end{equation} 
Here we still use the same notation for the conductivities in the coordinate space that are still constrained with the rotation symmetry as in \eqref{IndexStructures1} and \eqref{IndexStructures2}. Note that we do not Fourier transform the two and four index Green's functions due to the lack of translation invariance, while we keep the time translation symmetry and the corresponding Fourier frequency $\omega$. 

Let us consider some of the contact terms, $\langle T^{0i} \rangle$ and $\langle T^{ij} \rangle$, that appear in \eqref{GeneralWI}. A particularly interesting one arises with the spontaneously generated angular momentum $\ell$ \cite{Toma1991,Liu:2012zm} in the systems with broken parity. Without the spatial translation symmetry, the momentum generator can develop an expectation value 
\begin{align}
	\langle T^{0i} \rangle = (1/2)\epsilon^{ik}\partial_k\ell \;,
\end{align} 
which is described in \eqref{AngularMom}. It turns out that the structure of the term associated with the angular momentum yields a tensor similar to $\eta_H$. As a result, $\eta_H$ in \eqref{NeutralWIMomen} is modified to $\eta_H + \ell/2$ \cite{Hoyos:2015yna}. 

The other contact term actually provides $\langle T^{ij} (\vec x) \rangle = \delta^{ij} p(\vec x^2) $, where $p$ is pressure, another universal contribution to the thermodynamic description. Similar to the angular momentum, due to the particular tensor structure, the contribution would replace $\zeta$ in the general Ward identities by the combination $ \zeta -\frac{i}{\omega} p $.

Recently, Skyrmions have been observed in the insulating material $Cu_2 O Se O_3$ and various experiments regarding the Hall thermal conductivity and angular momentum have been carried out \cite{Insulating1}\cite{ThermalHC1}\cite{Rotation1} not to mention in the conducting materials \cite{SkyrmionExp1}\cite{SkyrmionExp2}\cite{Jonietz2010}\cite{Heinze2011}. For the insulating magnets with Skyrmions, our Ward identity provides a simple relation among parity violating transport coefficients as   
\begin{align}\label{WIInsulator}
	&\omega^2 \bb{\kappa}_\epsilon 
	= c + \partial^2 \Big(\eta_H +\frac{\ell}{2} \Big)  \;,
\end{align} 
which is derived from \eqref{GeneralWI} in the absence of the translation symmetry. Recent experiments have the ability to measure the Skyrmion density, thermal Hall conductivity and angular momentum in Skyrmion materials \cite{ThermalHC1}. Such measurements could in principle be used to confirm the existence of Hall viscosity.

\newpage 
\part{Skyrmion Hall Experiments} \label{sec:PartII}

In Part \ref{sec:PartI}, we have reviewed some universal features of Skyrmion Hall transport using the quantum field theory Ward identities, a first principle method based on symmetries and conservation equations. Various transport and thermodynamics quantities are related to each other. We have had the opportunity to talk about Hall viscosity that has not been discussed in Skyrmion physics before.    

In this second part, we would like to survey various physical quantities that play key roles in understanding the transport phenomena of the Skyrmions from the {\it phenomenological point of view}. We approach this subject with two enveloping themes. First, we review various mechanisms of spin torque, the Landau-Lifshitz-Gilbert equation, and Thiele equation that captures the dynamics of Skyrmions in \S \ref{sec:SpinTorqueBackground}. Second, we review thermo-electromagnetic effects that capture various transport coefficients using the H. B. Callen's thermodynamics of irreversible processes in \S \ref{sec:ThermoEMBackground}. Then, we survey the experimental status of the Skyrmion Hall transport experiments in \S \ref{sec:SkyrmionHallMeasurements}. 

Readers can refer to several review materials are available focusing on different aspects of Skyrmion physics  \cite{SkyrmionTopologicalReview}\cite{SpinCaloritonics,FundApplReview,HallTransportReview,AdvancesSkyrmion,CollectiveSpinReview,JiangReview,QSkyrmionReview}.

\section{Spin torque: Background Materials  } \label{sec:SpinTorqueBackground}

We are interested in the magnetic properties of materials, which are fairly complicated in general. Thus we focus on primarily ferromagnetic materials. Contrast to the electronic crystal structure that is crucially related to the underlying lattice structure, the formation of magnetic structure is mainly due to the exchange interaction of the atom in the materials, which is quite independent of the total magnetic moments relative to the lattice. Of course, there are other, less important, interactions that alter this magnetic structure. They are a direct magnetic interactions between the magnetic moments of the atom and interactions of the magnetic moments and the electric fields of the crystal lattice. These interactions are relativistic effects and are suppressed by $\sim \mathcal O(v^2/c^2)$, where $v$ and $c$ are the atomic velocity and the speed of light. Thus, if we are only concerned with the exchange interaction, the magnetization $\vec M$ can be considered as a conserved quantity and thus an independent variable. The corresponding thermodynamic potential is a function of the magnitude $M$. This has been useful for understanding the Curie point of magnetic materials where $M$ is small and one can expand the thermodynamic potential as a function of $M$. This approach provides a universal and simple understanding for the spontaneous magnetization and susceptibility \cite{LandauLifshitzEMContinuousMedia}. 

It is well known that experimental hysteresis curves of ferromagnetic substances clearly show that beyond certain critical values of the applied magnetic field, the magnetization saturates, becomes uniform and aligns parallel to the magnetic field. In order to incorporate this experimental fact, from phenomenological grounds, Landau and Lifshitz \cite{LandauLifshitz} introduced the basic dynamical equation for magnetization or spin $\vec M(\vec r,t)$ in bulk materials, where the effect of relativistic interactions were also included as a damping term. In 1954, Gilbert \cite{Gilbert} introduced a more convincing form for the damping term, based on a Lagrangian approach, and the combined form is now called the Landau-Lifshitz-Gilbert (LLG) equation, which is a fundamental dynamical system in applied magnetism. LLG equation can be understood from the view of various torque acting on the local magnetization. 

\begin{figure}[h!]
	\begin{center}
		\includegraphics[width=0.42\textwidth]{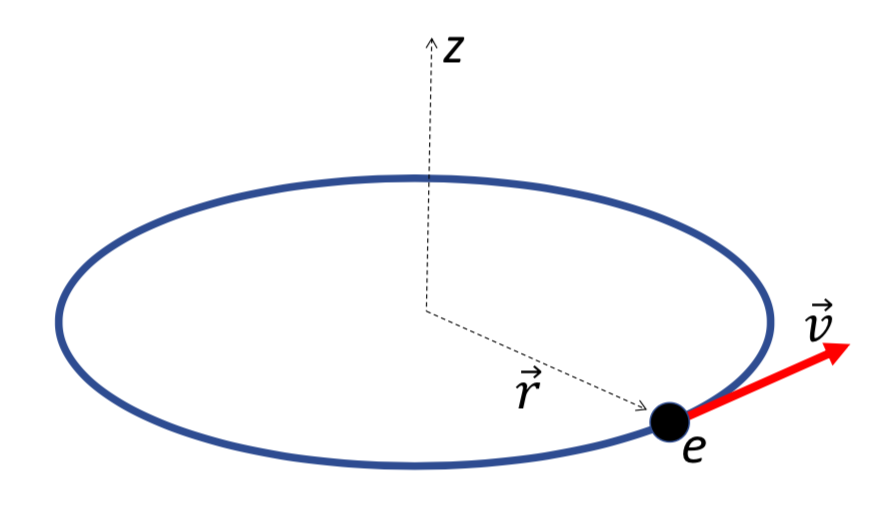} \qquad\qquad 
		\includegraphics[width=0.4\textwidth]{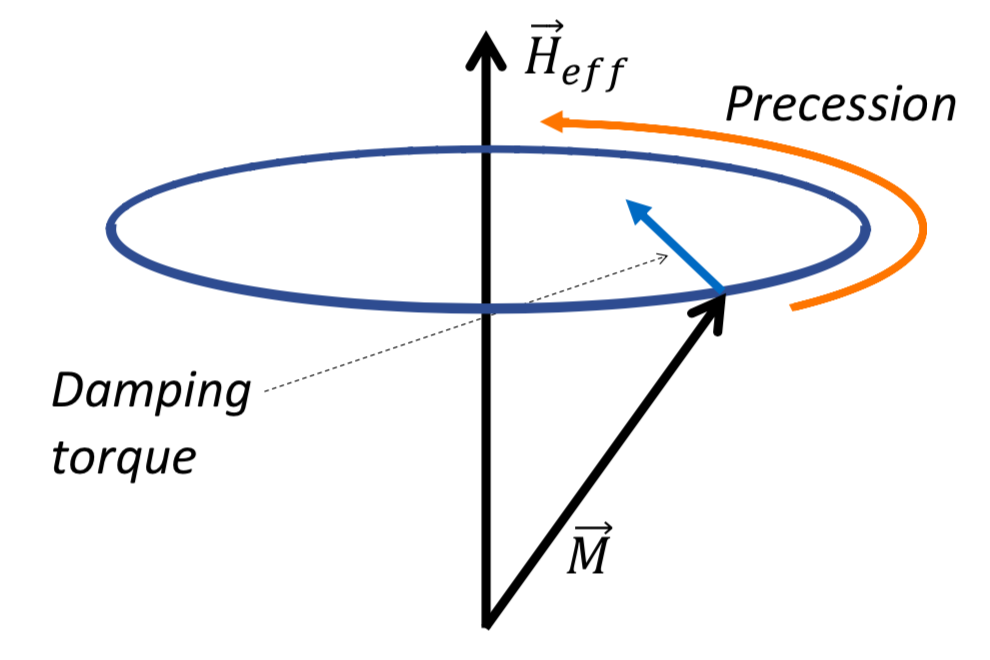} 
		\caption{\footnotesize\small (\textbf{a}) A simple classical model of an atom provides an intuitive picture for gyromagnetic ratio. (\textbf{b}) Spin motion due to the precession and damping terms given in the LLG equation \eqref{GLLGEQ}. }
		\label{fig:GyroscopicRatioAndPrecession}
	\end{center}
\end{figure}

To understand the basic origin of the torque, $\vec T = \frac{d\vec M}{d t}$,  acting on a local magnetization $\vec M = - \< \vec S \>$ in an elementary fashion, we consider a classical model of an atom with an electron with charge $e$ orbiting counterclockwise with radius $r$ in $xy$ plane. See the figure \ref{fig:GyroscopicRatioAndPrecession} (\textbf{a}). The gyromagnetic ratio $\gamma_0$ is given by the magnetic dipole moment $\vec \mu = \mu_z \hat z= \frac{e v r}{2} \hat z$ over the angular momentum $\vec L = \vec r \times \vec p = r m_e v \hat z$ for the atomic orbit. $\gamma_0 = \frac{|\mu_z|}{L_z} = \frac{|e|}{2 m_e} = \frac{\mu_B}{\hbar}$, where $\mu_B$ is the Bohr magneton, $\hbar$ Plank constant, $e$ the electric charge of an electron, and $m_e$ the mass of an electron.  For a quantum mechanical spin, we need an extra factor of 2 for the gyromagnetic ratio. 

The definition of a torque for the angular momentum $\vec L$, and in tun for the dipole moment $\vec \mu$, is given by $\vec T = \frac{\partial \vec L}{\partial t} =\frac{1}{ \gamma_0} \frac{\partial \vec \mu}{\partial t}$. Thus, for a macroscopic volume that contains many atoms, the torque acting on a local magnetization $\vec M$ is given as $\frac{\partial \vec M}{\partial t} = -\gamma_0 \vec T $. There are several different torques action on the magnetization, which are described by the general form of the Landau-Lifshitz-Gilbert (LLG) equation. 
\begin{align}\label{GLLGEQ}
\frac{\partial \vec M}{\partial t}&= - \gamma_0 \vec M \times \vec H_{eff} + \frac{\alpha}{M_s} \vec M \times \frac{\partial \vec M}{\partial t} + \vec T_{STT} + \vec T_{SOT}  + \vec T_{SHT} \;, 
\end{align}
where $\gamma_0$ is the gyromagnetic ratio and $M_s$ is the saturation magnetization. The first term in the right hand side is the precessional torque that describes the precession of the magnetization around the local effective field $\vec H_{eff}$, which includes the external field, exchange interaction, anisotropy and demagnetization. The second term is damping torques that describes the damping effect on the magnetization toward the equilibrium parallel to the effective field $\vec H_{eff}$. $\alpha$ is called the Gilbert damping parameter \cite{LandauLifshitz}\cite{Gilbert}. See the figure  \ref{fig:GyroscopicRatioAndPrecession} (\textbf{b}). 

\begin{figure}[h!]
	\begin{center}
		\includegraphics[width=0.85\textwidth]{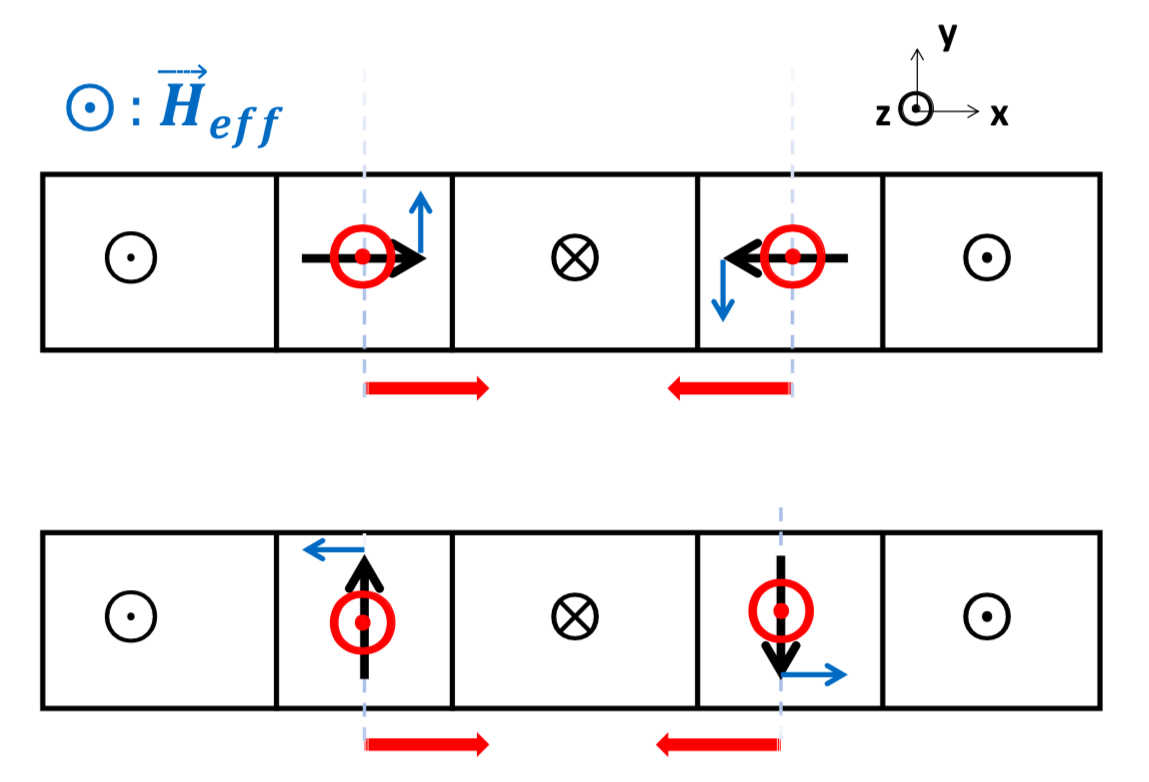} 
		\caption{\footnotesize\small Illustration of the Domain Wall [(\textbf{a}): (left-handed) Neel DW and (\textbf{b}): Bloch DW] motion due to the effective field $\vec H_{eff}$ pointing out of the plane.  Regardless of the DW structure, the domain with the magnetization pointing out of the plane expands due to the precession and damping terms in \eqref{GLLGEQ}. This is explained by the damping term (represented by the red circles) and the associated precessional term (represented by the short solid blue arrow).}
		\label{fig:DWMotion-Field}
	\end{center}
\end{figure}

Before moving forward, we illustrate the application of the precession and damping terms in the context of Domain Wall (DW) motion. We call this as a field driven DW motion. This is depicted in the figure \ref{fig:DWMotion-Field}. In this section, we use similar figures to explain and contrast the other spin torques. See the similar analysis on DWs dynamics using current \cite{Emori2013}. The left, middle, and right boxes are parts of a ferromagnetic layer with the uniform magnetization along the out of plane ($+\hat z$), into the plan ($-\hat z$), and out of plane ($+\hat z$) directions, respectively. There are two DWs with the magnetization with the dark thick black arrows. In the top left DW of the figure \ref{fig:DWMotion-Field}, the magnetization of DW is along the $+\hat x$ direction. Upon applying the magnetic field out of plane (along the $\vec H_{eff} = H_0 \hat z$ direction), the precession term generate the torque $- \gamma_0 \vec M \times \vec H_{eff} $ that is represented as the short solid blue arrow, which is along $+\hat y$ for the magnetization in the top left DW. The damping term can be evaluated by iterating the equation \eqref{GLLGEQ} to get $ \frac{\alpha}{M_s} \vec M \times \frac{\partial \vec M}{\partial t} = \frac{\alpha}{M_s} \vec M \times (- \gamma \vec M \times \vec H_{eff} )$. This is illustrated with the red circle inside the DW region that is $+\hat z$. Thus, the magnetization of the top left DW has the torque that points out of the plane. Thus the Domain with the magnetization out of plane expands. This is illustrated by the thick red arrow outside (below) the ferromagnet layer.   

The third term in \eqref{GLLGEQ} describes the spin transfer torque (STT), whose contributions come from the spin-spin interaction between the Skyrmions and the conduction electron flowing through the Skyrmions. The fourth term describes the spin torque induced by the Rashba effect, whose contributions come from the current-induced spin-orbit interaction (SOT). This Rashba effect can exist in the same layer of the local magnetization $\vec M$ or in a different layer. The last term is a torque due to spin Hall effects (SHE). This Slonczewski-like spin torque is due to the spin Hall effect that is operating on the interface between a ferromagnet layer and a non-magnet layer. We are going to focus on the last three spin torques $\vec T_{STT}, \vec T_{SOT}$ and $ \vec T_{SHT} $  in detail and the corresponding effects on ferromagnetic materials in the following subsections. 

\newpage 
\subsection{Spin transfer torque}\label{sec:STT0}

During the early 2000, surprising giant magneto-resistance effects were reported in magnetic multi-layers \cite{GiantMagExp}.
This stirred up activities to understand them theoretically \cite{Johnson1987}\cite{Valet1993} and to re-evaluate the Landau-Lifshitz-Gilbert equation that had the first two terms on the right hand side of \eqref{GLLGEQ}  
\cite{ZhangLi}\cite{Thaiville}. Understanding this is highly insightful. Here we follow the theoretical understanding of Zhang-Li \cite{ZhangLi}, which has large impacts on the future advancements on the field. 

We considered two different types of electrons: the electrons near the Fermi surface that provide the spin-dependent transport phenomena and the electrons below the Fermi sea that involve the magnetization dynamics. The spin dynamics of the former, itinerant electrons, are described by the full quantum mechanical operators $\vec s$. 
The latter, localized electrons, have much slower dynamics and are described by the classical magnetization vector $\vec S$. Thus we set $\vec S/S = -\vec M(\vec r, t)/M_s$ with $|\vec M| = M_s$, the saturation magnetization. The essential properties of their interactions turn out to be described by the following s-d Hamiltonian 
\begin{align}\label{sdHamiltonian}
H_{sd} = - J_{ex}~ \vec s \cdot \vec S = \frac{S J_{ex}}{M_s}~ \vec s \cdot \vec M(\vec r, t) \;. 
\end{align}	
This Hamiltonian \eqref{sdHamiltonian} can be used to compute the induced spin density for a given $\vec M(\vec r, t)$. In turn, the effects on the magnetization, in the language of the spin-transfer torque, can be obtained by utilizing the induced spin density.  

Using the generalized spin continuity equation by including the Hamiltonian \eqref{sdHamiltonian} and spin relaxation $\Gamma(\vec s)$ due to scattering with impurities and electrons, one can derive the equation of motion for the non-equilibrium conduction electron spin density $ \vec m (\vec r, t) = \< \vec s\>$ and spin current density $\mathcal{J}(\vec r, t) =  \<J\>$ 
\begin{align} \label{ContinuityEQ}
\frac{\partial \vec m}{\partial t} + \vec \nabla \cdot \mathcal{J} = - \frac{S J_{ex}}{\hbar M_s} \vec m \times \vec M(\vec r,t) - \<\vec \Gamma(\vec s)\> \;,  
\end{align}
where $\< \>$ represents the average over all occupied electronic states, and the third term comes from an explicit evaluation of $[\vec s, H_{sd}]/i\hbar$. We note that the spin current $\mathcal{J}(\vec r, t)$ is a tensor with charge current and the spin polarization of the current for the equation to make sense.%
\footnote{To see clearly, one can remind the continuity equation for particle transport given as $ \frac{\partial \rho}{\partial t} + \vec \nabla \cdot \vec J_\rho =0$ with states index $i$ and spin index $\sigma$. The number and corresponding current densities are given by the wave function as $ \rho = \psi^*_{i\sigma} \psi_{i\sigma}$ and $\vec J_\rho = - (i \hbar/m) \psi^*_{i\sigma} \vec \nabla \psi_{i\sigma}$, respectively. Here we assume the summation over the repeated indices and the reality of the current. \\
	For spin degrees of freedom, the spin and spin current densities are $ \vec m = \psi^*_{i\sigma} \vec s_{\sigma,\sigma'} \psi_{i\sigma'}$ and $\mathcal{J} = - (i \hbar/m) \psi^*_{i\sigma}\vec s_{\sigma,\sigma'}  \otimes \vec \nabla \psi_{i\sigma}$, where $\vec s = (\hbar/2) \vec \sigma$ and $\vec \sigma$ is a vector, whose Cartesian components are the three Pauli matrices. The indices of $\mathcal{J}$ are those in spin and real spaces. In general, spin is not conserved. Thus the continuity equation has extra contributions as in \eqref{ContinuityEQ}. 
}  

To solve the equation \eqref{ContinuityEQ} we separate the spin density and current density into two parts. One part, the slower component $\vec M$, satisfies the adiabatic approximation. The dynamics of magnetization is slow compared to that of conduction electrons, and thus the spins of conduction electrons approximately follow the direction of local magnetic moment $\vec M$. The other part $ \delta  \vec m$ captures the deviation from the adiabatic process, which is our main focus.
\begin{align}
\vec m (\vec r, t) &= \vec m_0 (\vec r, t) + \delta  \vec m (\vec r, t)= n_0 \frac{\vec M(\vec r,t)}{M_s} + \delta \vec m (\vec r, t) \;,  \\
\mathcal{J}(\vec r, t) &= \mathcal{J}_0(\vec r, t) + \delta  \mathcal{J}(\vec r, t) = -\frac{\mu_B P}{e} \vec J_e \otimes \frac{\vec M(\vec r,t)}{M_s} + \delta \mathcal{J}(\vec r, t) \;,
\end{align}	 
where $n_0$ is the local equilibrium spin density parallel to the local magnetization, $e$ electron charge, $J_e$ current density, $\mu_B$ Bohr magneton, $P$ spin current polarization of the ferromagnet. 

To solve this analytically in a closed form, we assume $ \<\vec \Gamma(\vec s)\> = \delta \vec m (\vec r, t)/\tau_{sf}$ with a spin-flip relaxation time $ \tau_{sf}$ and only a linear response of $\delta \vec m (\vec r, t)$ to the electric current $J_e$ and time derivative of magnetization $\partial \vec M / \partial t$. Within this semi-classical approximation, we have the following linear relation $\delta  \mathcal{J} = - D_0 \vec \nabla \delta \vec m $ with a diffusion constant $D_0$. Then the non-equilibrium spin density satisfies  
\begin{align}\label{deltamEQ}
D_0 \vec \nabla^2 \delta \vec m -  \frac{1}{\tau_{ex} M_s} \delta \vec m \times \vec M - \frac{1}{\tau_{sf}}  \delta \vec m
=\frac{n_0}{M_s} \partial_t \vec M + \frac{\mu_B P}{ M_s} (\vec v_s \cdot \vec \nabla) \vec M(\vec r,t) \;, 
\end{align}	
where $ \tau_{ex} =  \frac{\hbar}{S J_{ex}}$, the time scale of exchange interaction. Thus $ \delta \vec m$ is determined by two different sources, time variation and spatial variation of magnetization. We discard the first term (with $D_0$) because we consider the slowly varying magnetization in space so that the domain wall width of magnetization is much larger than the transport length scale $\lambda = \sqrt{D_0 (1/\tau_{sf} + i/ \tau_{ex})^{-1}}$. With the observation that the second and third terms in equation \eqref{deltamEQ} are orthogonal each other, the equation can be solved algebraically as
\begin{align}\label{deltamSOL}
\delta \vec m = 
\frac{\tau_{ex}}{1 + \xi_0^2} \left[ -\frac{n_0 \xi_0}{M_s} \partial_t \vec M - \frac{n_0}{M_s^2} \vec M \times \partial_t \vec M - \frac{\mu_B P \xi_0}{ M_s} (v_{s}^i \nabla_i) \vec M - \frac{\mu_B P}{M_s^2} \vec M \times (v_{s}^i \nabla_i) \vec M \right] \;, 
\end{align}	
where $\xi_0=\tau_{ex}/\tau_{sf}$ and $(\vec v_s \cdot \vec \nabla) =(v_{s}^i \nabla_i)$. This induced spin density affects the magnetization $\vec M$.

Now we can compute the spin transfer torque on the magnetization due to the induced conduction spin density $\delta \vec m$ in \eqref{deltamSOL}. The torque action on the slow magnetization $\vec M$ is given by the opposite of the torque listed in \eqref{ContinuityEQ}. Thus, $\vec T_{STT} = - \frac{1}{\tau_{ex} M_s} \vec M \times \vec m = - \frac{1}{\tau_{ex} M_s} \vec M \times \delta \vec m $. 
\begin{align}\label{SpinTorque}
\vec T_{STT} = \frac{\tau_{ex}}{1 \+ \xi_0^2}\! \left[\! -\frac{n_0 }{M_s} \frac{\partial \vec M}{\partial t}+ \frac{n_0 \xi_0}{M_s^2} \vec M \times\! \frac{\partial \vec M}{\partial t} + \frac{\mu_B P \xi_0}{M_s^2} \vec M \!\times\! (v_{s}^i \nabla_i) \vec M + \frac{\mu_B P}{M_s^3} \vec M \!\times\! [\vec M \times\! (v_{s}^i \nabla_i) \vec M] \right] \;. 
\end{align}
The first two terms come from the magnetization variation in time and are independent of the current, while the last two terms are due to the spatial variation of magnetization and are current-driven effect. It is important to know the regime of the validity of the result. This result \eqref{SpinTorque} is valid when the spin transport length scale $\lambda$ is much smaller than DW width or the scale of change of the magnetization. Thus this equation can not be simply applied to the multilayer magnetic materials where the effective domain wall length scale at the interface becomes zero.   

\begin{figure}[h!]
	\begin{center}
		\includegraphics[width=0.85\textwidth]{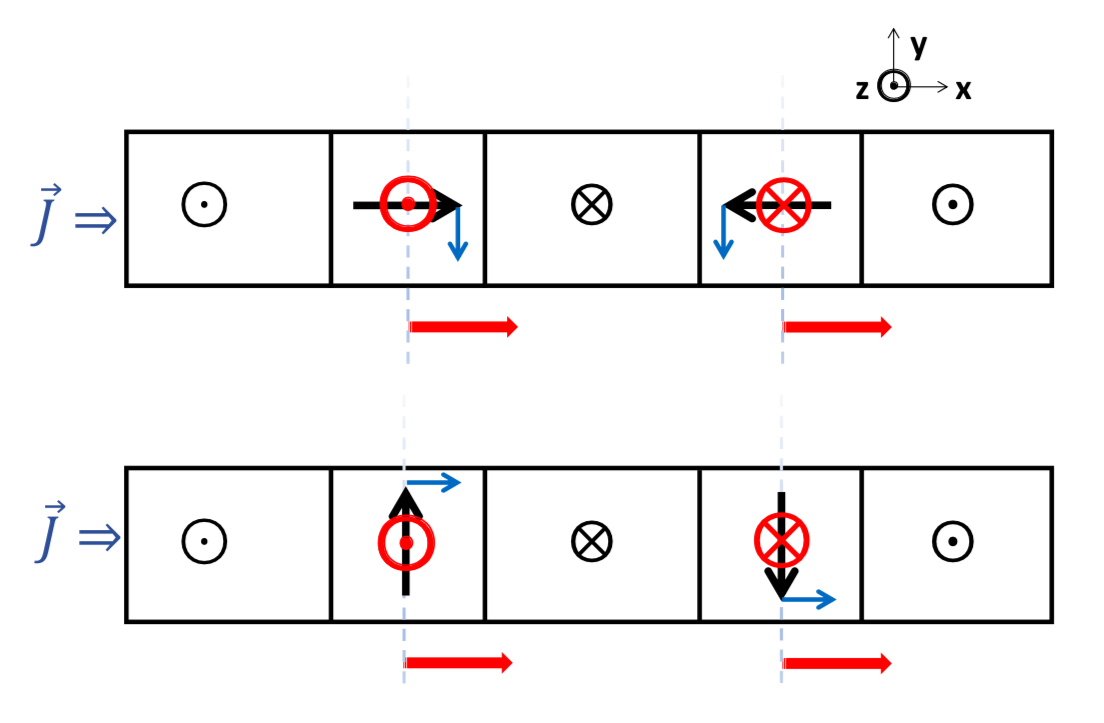}
		\caption{\footnotesize\small Domain Wall (DW) motion due to the spin transfer torque from the spin-spin interaction with conduction electrons. The coefficient of the non-adiabatic term is positive. The net contribution of the spin transfer torque is along the direction of current $\vec J$, opposite to the direction of electron flow. For the spin transfer torque, there is no difference between the Neel and Bloch DWs.}
		\label{fig:STT}
	\end{center}
\end{figure}

The last two terms in \eqref{SpinTorque} are called non-adiabatic and adiabatic spin-transfer torque terms. The coefficients of the both terms are positive. 
In the literature, an alternative form is also used.
\begin{align}\label{SpinTorque20}
\vec T_{STT} \supset \frac{\tau_{ex}}{1 \+ \xi_0^2} \left[ \frac{\mu_B P \xi_0}{M_s^2} \vec M \!\times\! (v_{s}^i \nabla_i) \vec M - \frac{\mu_B P}{M_s^3}  (v_{s}^i \nabla_i) \vec M \right] \;. 
\end{align}
Note that $\vec M$ and its spatial derivatives are orthogonal to each other as $\vec M^2 =constant$. When the magnetizations are uniform, these two contributions vanish due to the derivative. Thus, only non-trivial contribution of this spin transfer torques on the ferromagnets acts on DW region. These terms are useful to move DW as pictured in the figure \ref{fig:STT}. See a recent account for DW motion due to the spin transfer torque \cite{Beach2008}. 

To make the DW motion more transparent, we take the analogy of the effective field: the domain with the direction of the effective magnetic field expands. The {\it non-adiabatic term} can be described as the following effective fields
\begin{align}
\frac{\tau_{ex}}{1 + \xi_0^2}  \frac{\mu_B P \xi_0}{M_s^2} \vec M \times (v_{s}^i \nabla_i) \vec M  = - \gamma_0 \vec M \times \vec H_{\beta} \;, \qquad \vec H_{\beta}  \propto - \frac{\partial \vec M}{\partial x} 
\;,
\end{align}	
where we take $x$ direction as the direction of the current $\vec J = J \hat x$, which is opposite to the direction of the electron flow. Note that we take the $-$ sign in the term with $\vec H_{eff}$ in \eqref{GLLGEQ}. The effective field $\vec H_{\beta}$ is depicted as the red circle. The domains with that effective field expands. Thus the left and middle domains expand to the right. This effectively moves the DW to the right. The non-adiabatic torque term moves the DW against the direction of the flows of electrons, meaning along the direction of the current. The  effective field of the {\it adiabatic term} $\vec H_{\beta'}$ from the last term in \eqref{SpinTorque} is depicted as the blue solid arrow. 

\begin{figure}[h!]
	\begin{center}
		\includegraphics[width=0.35\textwidth]{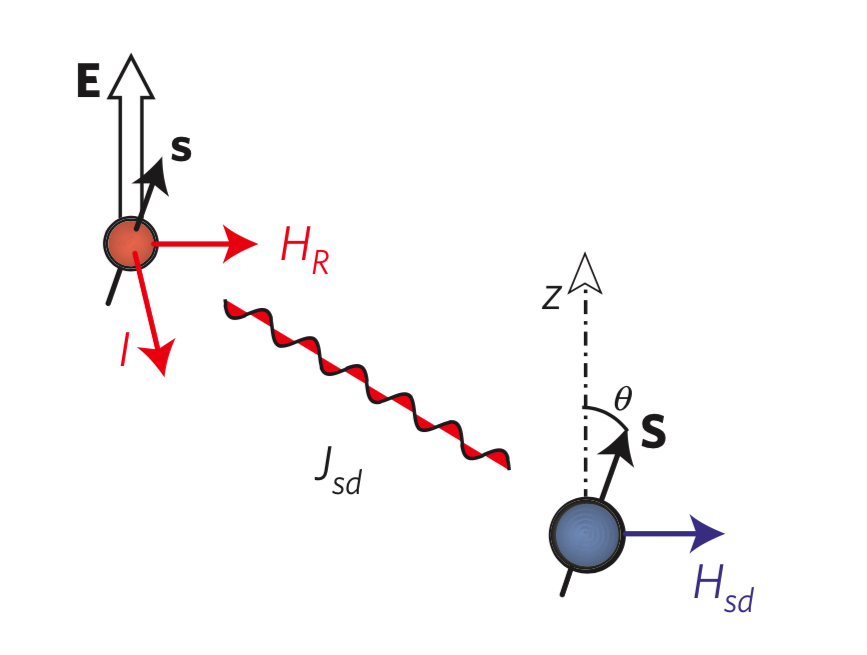} \qquad 
		\includegraphics[width=0.55\textwidth]{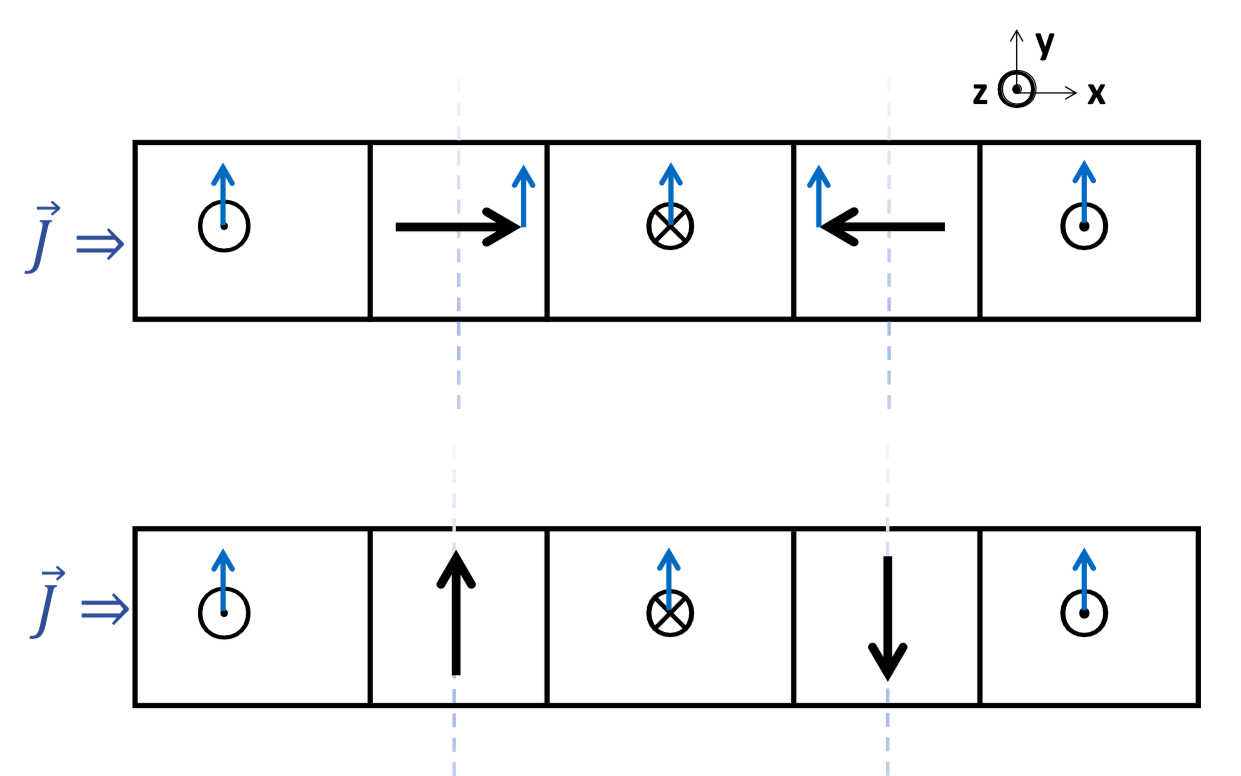} 
		\caption{\footnotesize\small (\textbf{a}): Illustration of the Rashba interaction. Reproduced with permission from \cite{SOT1}. (\textbf{b}): The effect of effective field $\vec H_{sd} $ on the DW structure. The field $\vec H_{sd} $ does not generate a DW motion. }
		\label{fig:SOT}
	\end{center}
\end{figure}

\subsection{Spin-orbit torque} \label{sec:SpinOrbitTorque}

Recently, a different way to produce non-equilibrium polarization of the conduction electron spins has been introduced using the spin-orbit coupling in materials with lacking inversion symmetry \cite{SpinOrbitZhang1}. Electrons moving in asymmetric crystal-field potential experiences a net electric field $\vec E = E \hat z$. This electric field can be transformed to a magnetic field in electron's rest frame as $ \vec B_{eff} = \vec H_{R} = \vec E \times \vec v /(2c^2)= \alpha_R \hat z \times \vec v$, which is called Rashba field \cite{Rashba1}\cite{Rashba2}. This Rashba field makes their spin titled as in the figure \ref{fig:SOT} (\textbf{a}). In ferromagnetic materials, the field $\vec H_R$ combines with the s-d exchange interaction that couples the spins of the conduction electrons to the local magnetization $\vec M$. Thus an electric current passing through a uniformly magnetized ferromagnet exerts an exchange-mediated effective field $\vec H_{sd}$ on the local magnetization $\vec M$ \cite{SOT1}\cite{SOT1A}. See the figure \ref{fig:SOT} (\textbf{a}). 
\begin{align}\label{SpinTorque-SOTDL}
\vec T_{SOT}^{DL} &= -\gamma_0  \vec M \times \vec H_{sd} \;, \qquad \vec H_{sd}  =  \frac{2 \alpha_R m_e P}{\hbar |e| M_s} (\hat z \times \vec J_e) \;, 
\end{align}
where $P$ is the polarization, $\hat z$ is the direction of the local electric field and $\vec J_e$ the electric current. This is referred as damping-like SOT. 

\begin{figure}[h!]
	\begin{center}
		\includegraphics[width=0.85\textwidth]{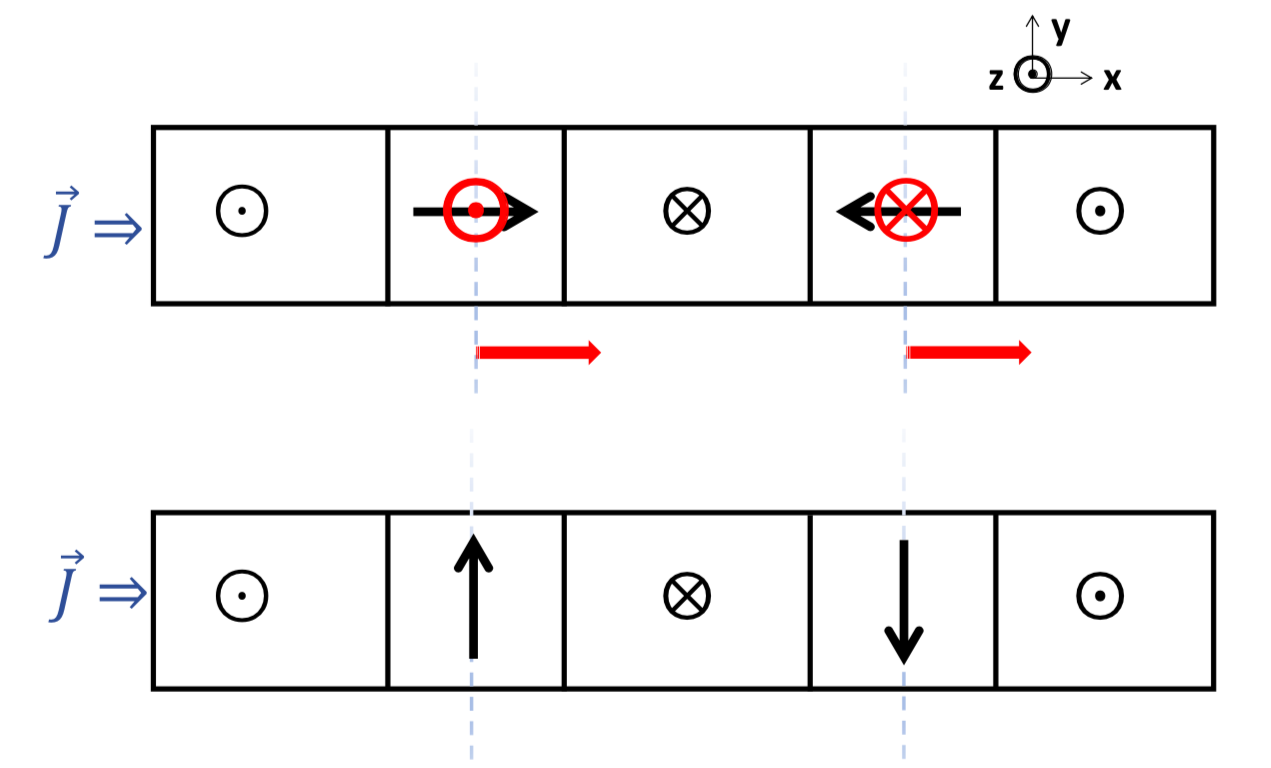} \\
		\includegraphics[width=0.83\textwidth]{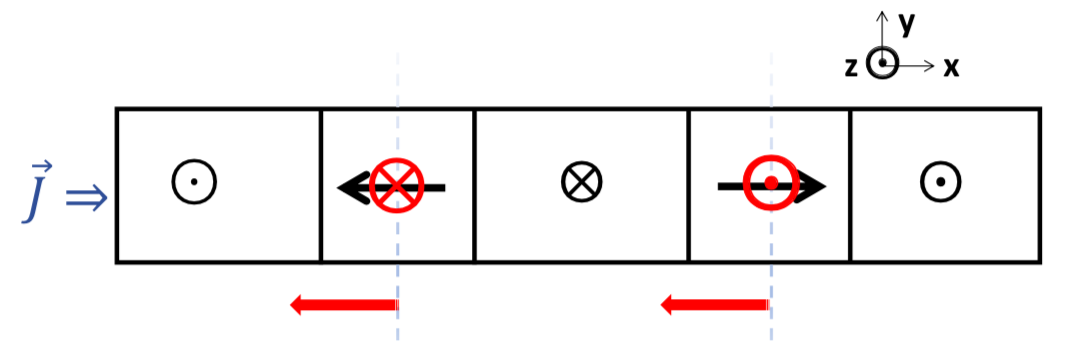}
		\caption{\footnotesize\small (\textbf{a}): Top. Left-handed DW motion under the influence of the spin-orbit torque due to the spin-orbit interaction with the conduction electron. The effective field for the field-like spin-orbit torque (SOT) is $\vec H_{SOT}^{FL} = \vec M \times \vec H_{sd}$, which pushes DW against the direction of electron motion, while there is no DW motion in the presence of damping-like SOT. (\textbf{b}): Middle. Bloch type DW is not influenced by the SOT. (\textbf{c}): Bottom. Right-handed DW moves opposite the direction of current and along with the flow of electrons. Note that this field-like SOT only affects the Neel type DW. }
		\label{fig:SOT-FL}
	\end{center}
\end{figure}

Once the torque \eqref{SpinTorque-SOTDL} is accepted, there can be another contribution called field-like SOT \cite{Litzius2017}. 
\begin{align}\label{SpinTorque-SOTFL}
\vec T_{SOT}^{FL} &=-\gamma_0 \vec M \times \vec H_{SOT} \;,\qquad  \vec H_{SOT}= \frac{\hbar}{2\mu_0 |e|} \frac{\alpha_H}{M_s^2 t}  ( \vec M \times (\hat z \times \vec J_e))  \;,
\end{align} 
where $t$ is the thickness of the ferromagnet layer.
This term can be introduced at the same time with \eqref{SpinTorque-SOTDL}, yet this can be inferred through the Gilbert damping term in the LLG equation \eqref{GLLGEQ} heuristically. It turns out that this field-like SOT term \eqref{SpinTorque-SOTFL} is much more efficient to move Neel type Skyrmions as demonstrated in the figure \ref{fig:SOT-FL}. Thus it is more appropriate to consider this term in terms of effective field 

Combining these two contributions, we have 
\begin{align}\label{SpinTorque-SOT}
\vec T_{SOT} &= -\gamma_0 a_{SOT} \left[ \vec M \times ( \vec M \times (\hat z \times \vec J_e)) + \tilde \eta  \vec M \times (\hat z \times \vec J_e) \right] \;, 
\end{align}
where $a_{SOT}=\frac{\hbar}{2\mu_0 |e|} \frac{\alpha_H}{M_s^2 t} $ and $\tilde \eta$ a parameter to distinguish the strength of field-like SOT compared to that of damping-like SOT. We note that these terms also induce similar terms in the LLG equation through the Gilbert damping term, turning a field-like SOT term into a damping-like SOT term and vice versa. These four combined terms has been used in \cite{Litzius2017} to have a better fit for the Skyrmion motion.

While this spin-orbit effect and the corresponding spin torque seem to be similar to those of the spin transfer torque (STT), the spin-orbit induced torque is fundamentally different from those. STT does not have any effect on uniform magnetization, yet does transfer spin angular momentum directly to DW due to the derivative acting on $\vec M$. On the other hand, SOT transfer orbital angular momentum of the conduction electrons to $\vec M$ through the spin-orbit interaction. SOT can act on homogeneous magnetization and change the magnetization direction. The damping-like SOT does not seem to be useful for deriving a DW motion, while useful for reversible control over the magnetization direction for the direction of the spin-orbit torque can be changed by reversing the current \cite{SOT1}. See the figure \ref{fig:SOT} (\textbf{b}). The structure of this spin torque is useful for understanding spin Hall torque we study in the following section \S \ref{sec:SpinHallTorque}. On the other hand, field-like SOT is efficient to derive only Neel type DW, which make this an effective tool to distinguish the Neel type DW from the Block type DW \cite{Litzius2017}.

\subsection{Spin Hall torque}\label{sec:SpinHallTorque}

Here we briefly introduce the spin Hall effect following the original paper \cite{Dyakonov1971}\cite{Hirsch1999} and Slonczewski mechanism for transferring spin angular momentum into a different layer \cite{Slonczewski}\cite{Berger}. Then we consider the spin Hall torque that is generated on a ferromagnet through the interface with a heavy metal, whose interface produces a large spin-orbit coupling. See a review on Skyrmions in magnetic multilayers \cite{JiangReview}. 

\begin{figure}[h!]
	\begin{center}
		\includegraphics[width=0.42\textwidth]{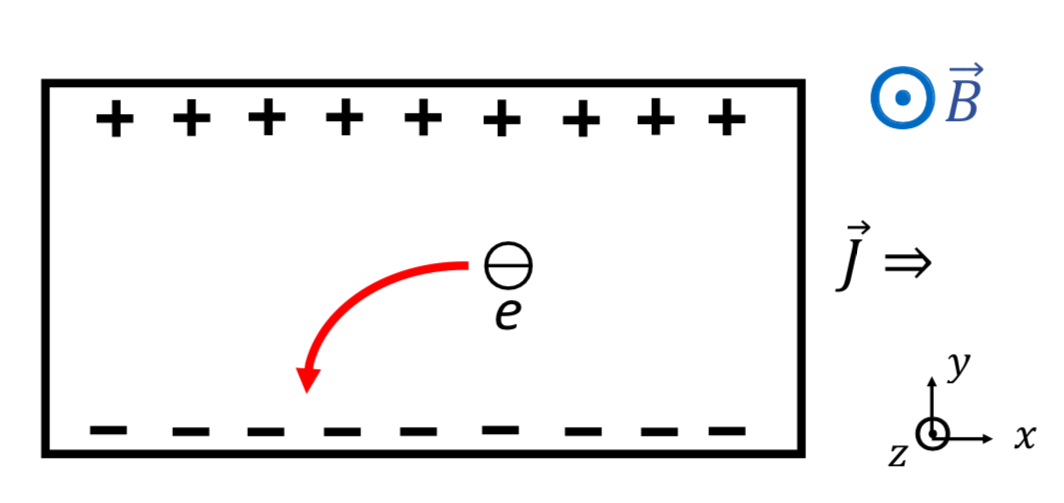} \qquad \qquad
		\includegraphics[width=0.42\textwidth]{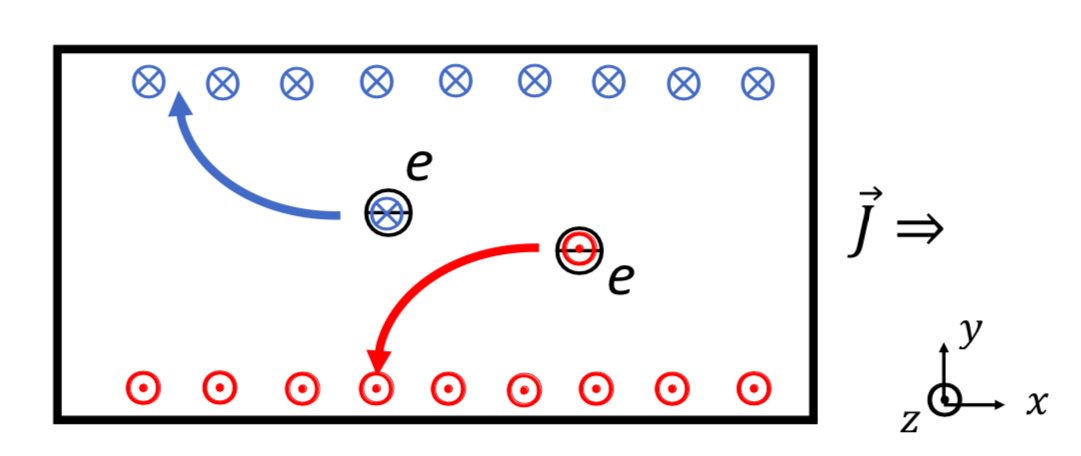} 
		\caption{\footnotesize\small Analogy between the charge Hall effect (\textbf{a}) and the spin Hall effect (\textbf{b}). The charge Hall effects in the presence of magnetic field are from the Lorentz force law and independent of the spin polarization. The spin Hall effects depend on the spins of the electrons without magnetic field. Spin polarization of an electron along $\hat z$ behaves similar to the electron with magnetic field in the same direction. Thus spin up and down electrons move towards the opposite directions. The red circle with a dot represents an electron with an up polarization, while the blue circle with a cross represents an electron with a down polarization. }
		\label{fig:SHE2}
	\end{center}
\end{figure}

The well known Hall effect is the result of magnetic field acting on a moving charged particle through the Lorentz force law $ \vec F = q \vec v \times \vec B$ with an electric charge $q=e$, velocity of the charged particle $\vec v$ and the magnetic field $\vec B$ as illustrated in the figure \ref{fig:SHE2} (\textbf{a}). 

The spin Hall effect is the result of the motion of the intrinsic spin polarization of an electron, depicted as red circle with a dot or a cross depending on the spin directions in the figure \ref{fig:SHE2} (\textbf{b}) briefly \cite{Hirsch1999}. These spin polarizations effectively play the similar role of the magnetic field. Of course, the spin polarization is random unless one polarizes the spin of the electrons. In a thin film, spin polarizations parallel to the film ($ \pm \hat y$) produce the force out of the plane resulting in no visible effects. On the other hand, electrons with spin polarization perpendicular to the film ($\pm \hat z $) produces net spin accumulations, down spins along the $+y$ side and up spins along the $-y$ side of the thin material. Thus electron flow produces the net spin excess and thus net magnetization excess. 

\begin{figure}[h!]
	\begin{center}
		\includegraphics[width=0.62\textwidth]{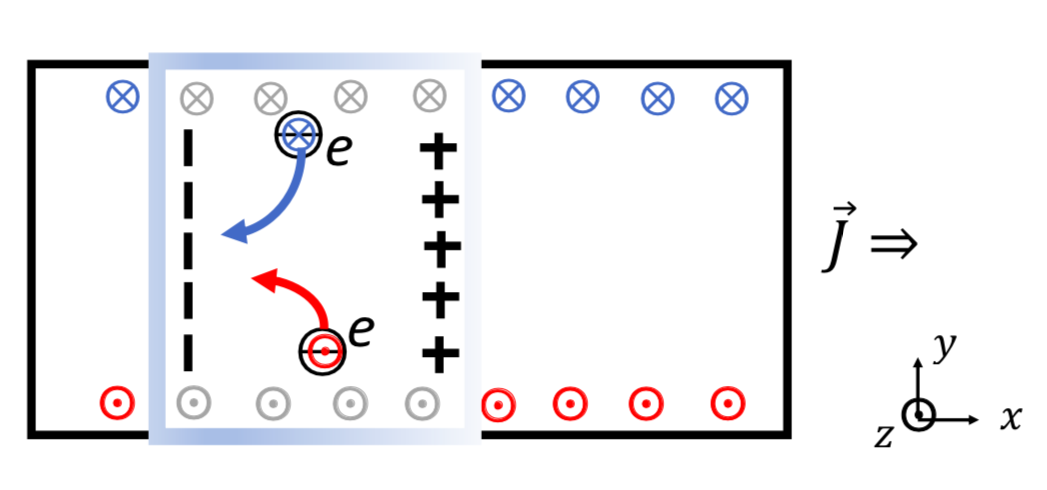}
		\caption{\footnotesize\small A metal such as $Pt$ is overlaid to provide a pathway for the conduction electrons to move to reduce the spin imbalance. Inverse spin Hall effect that results in measurable electric potential.  }
		\label{fig:InverseSpinHallEffect}
	\end{center}
\end{figure}

The polarized spins accumulated in opposite ends move toward each other to remove the spin imbalance and produce spin current when there is a pathway. In the figure \ref{fig:InverseSpinHallEffect}, a metal layer is overlaid on top of the setup of figure \ref{fig:SHE2} (\textbf{b}). The electrons with opposite spins that move opposite direction move toward the same direction, producing a charge imbalance. The resulting electric potential or electric field can be measured. This is the inverse spin Hall effect (ISHE) \cite{Hirsch1999}.

In the literature the notion of the electromotive force $\vec E_{SHE}$ is used to explain the spin Hall effects. Due to the motion of an electron the following electromotive force $\vec E_{SHE}$ is generated depending on its polarization.  
\begin{align}\label{EMFSHE}
\vec E_{SHE} = D_{ISHE} \vec J_N \times \vec \sigma \;,
\end{align}	
where $\vec J_N$ is the direction of actual flow of the particles and $\vec \sigma$ is the polarization vector of the particle. In the figure \ref{fig:InverseSpinHallEffect}, the direction of $\vec E_{SHE} \propto + \hat x$, and thus electrons move to the opposite direction $- \hat x $.

\begin{figure}[h!]
	\begin{center}
		\includegraphics[width=0.75\textwidth]{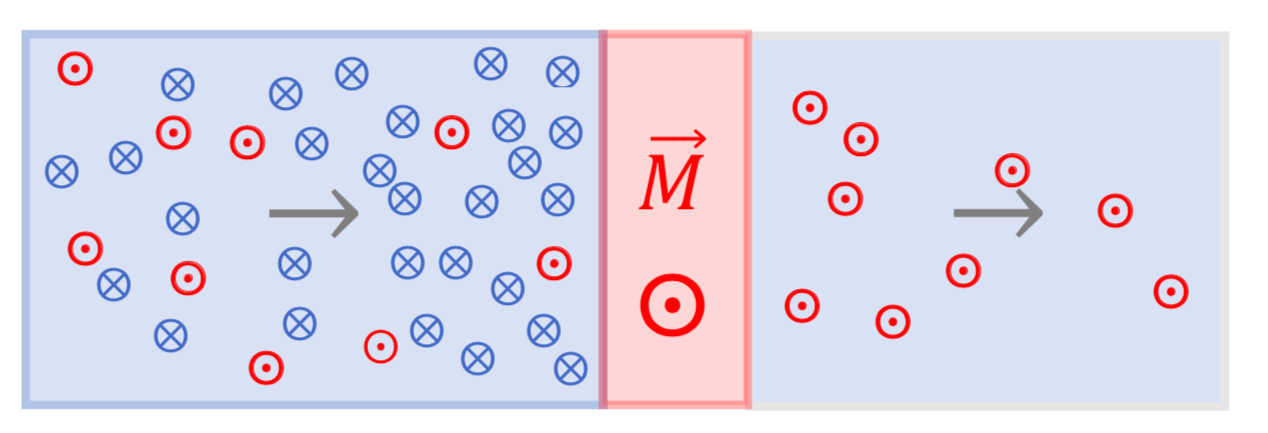} 
		\caption{\footnotesize\small The middle conducting ferromagnetic layer, with up direction magnetization, serves as a spin polarizer by the spin transport of the majority spins, up spins with arrow head. Accumulation of the minority spins, down spins with arrow tail, on the left side of the ferromagnetic layer and that of the majority spins on the right side.   }
		\label{fig:SHE3}
	\end{center}
\end{figure}

The spin Hall effect can be used to produce spin torque when combined with the so-called Slonczewski mechanism \cite{Slonczewski}\cite{Berger} of transferring spin into a different layer. In ferromagnetic conductors, the majority spins (electrons with the same polarization as the polarization of the ferromagnet layer) have large conductivity, while the minority spins (electrons with the opposite polarization compared to the polarization of the ferromagnet layer) have poor conductivity. In the interface between a normal metal and a ferromagnetic metal, majority spins are preferentially transmitted. Thus ferromagnet sitting in the middle of the figure \ref{fig:SHE3} acts as a spin polarizer. With the electrons injected from left, there are accumulated spins on both sides of the ferromagnet layer, the minority (down) spins on the left side and the majority (up) spins on the right side of the ferromagnet with the up direction magnetization. These collective magnetization due to the accumulated spins can provide Slonczewski torque with appropriate arrangements of ferromagnet and non-magnet layers.   
Slonczewski computed the spin torques using 5 different layers that are composed of three normal metal layers and two ferromagnetic layers (one with a fixed magnetization direction $\vec m_f$ and another with flexible magnetization $\vec M$). The spin torque acting on $\vec M$ is described as 
\begin{align}
\vec T_{Slonczewski} = \frac{J_{e} \hbar \epsilon}{2 e M_s^2 t_f} \vec M \times (\vec M \times \vec m_f) \;,
\end{align}       
where $\vec J_e$ is the current, $\bar \epsilon$ is efficiency, or spin Hall angle, of transferring the spin torque, $t_F$ the thickness of ferromagnetic layer with flexible magnetization $\vec M$, and $\vec m_f$ the magnetization of ferromagnetic layer with fixed magnetization direction. This is quantum mechanical phenomena and does not really depends on whether the spins really penetrate to the other layer or not.  

\begin{figure}[h!]
	\begin{center}
		\includegraphics[width=0.85\textwidth]{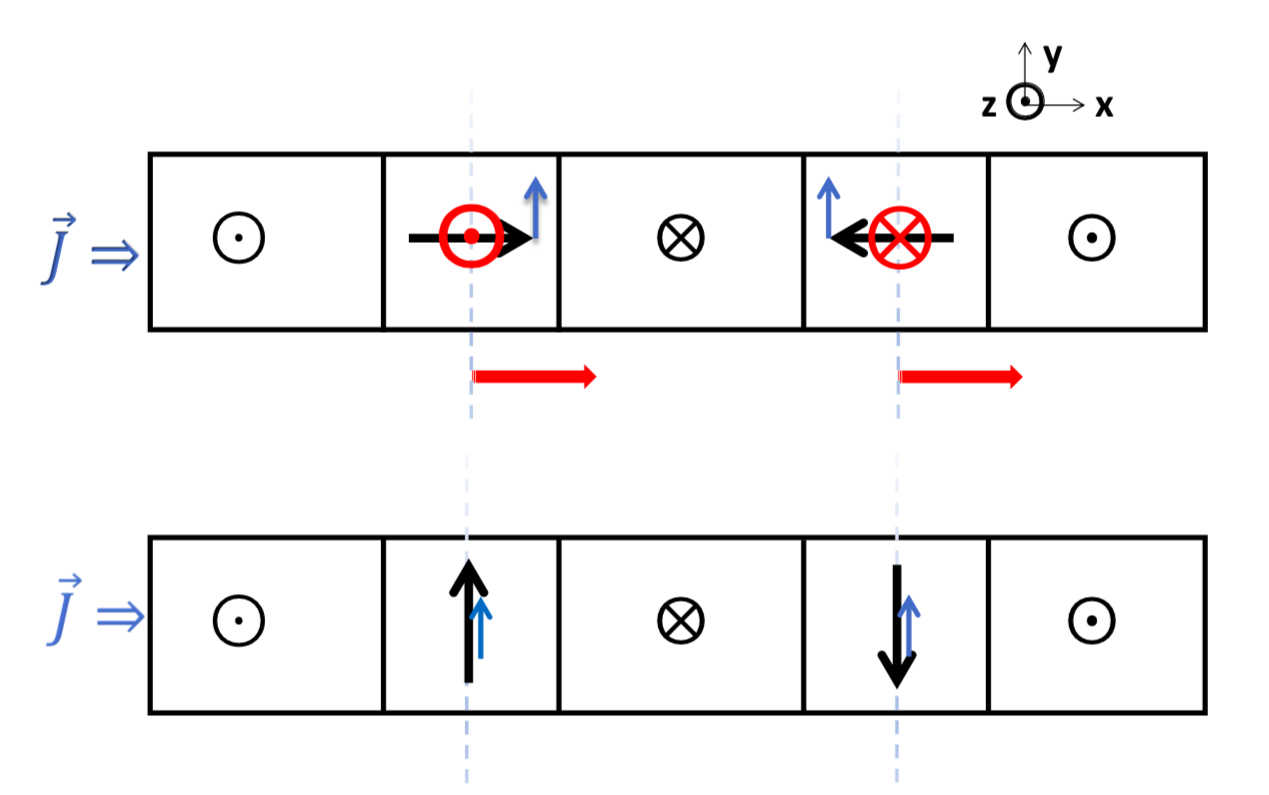}
		\caption{\small DW motion under the influence of the spin Hall torque \eqref{SHTFieldHSL}. The current $\vec J_{HM}$ flows in the other layer (below the ferromagnet layer) and the blue arrows are the direction of $\hat z \times \vec J_{HM} $. The red circles are the direction of $\vec H_{SL}$, which is similar to $\vec H_{eff}$ considered before, yet crucially different due to its dependence on the magnetization direction. (\textbf{a}): Neel DW. The left and middle domain expands to the right and DWs move to the right. (\textbf{b}): Bloch DW. There is no spin Hall torque acting on DW. The spin Hall torque vanishes for the directions of $\vec H_{SL}$ and the magnetization are parallel each other. Unlike the STT, SHT acts differently on Neel and Bloch DWs.   }
		\label{fig:SHE10}
	\end{center}
\end{figure}

This Slonczewski torque can be applied for the case with heavy metal (bottom layer) layered with a ferromagnetic material (top layer). The current injected in the heavy metal experiences the spin Hall effect so that the electrons with spin along $+\hat y$ rise near the boundary between the metal and ferromagnet layers. This accumulated spin polarization takes the role of $\vec m_f = \hat z \times \vec J_{HM}$. The efficiency $\epsilon$ is replaced by the spin Hall angle $\theta_{SH}$. The corresponding Slonczewski torque has the form 
\begin{align} \label{SHTFieldHSL}
\vec T_{SHT} = - \gamma_0 \vec M \times \vec H_{SL} \;, \qquad \vec H_{SL} = \frac{\hbar \theta_{SH}}{2 e M_s^2 t_F} (\vec M \times (\hat z \times \vec J_{HM})) \;,
\end{align} 
Here it is convenient to consider the field $\vec H_{SL}$. Similar to the $\vec H_{eff}$ that makes the magnetization field $\vec M$ to precess and the corresponding damping force direct the magnetization $\vec M$ along the direction $\vec H_{eff}$. The $\vec H_{SL}$ field pointing up as in figure \ref{fig:SHE10} expands the domain with up magnetization, while the $\vec H_{SL}$ field pointing down as in figure \ref{fig:SHE10} expands the domain with down magnetization. It turns out that $\vec H_{SL}$ depends on the magnetization in the ferromagnetic layer that makes the domain walls move. 

\subsection{Landau-Lifshitz-Gilbert equation }

The Landau-Lifshitz-Gilbert (LLG) equation we started to consider in \eqref{GLLGEQ} consists of a precessional term due to an effective field and a phenomenological damping term as described in the beginning of this section. We introduced three more torques, $\vec T_{STT}, \vec T_{SOT}$ and $ \vec T_{SHT} $, one by one in the previous sections. We include them to the LLG equation. 

Let us include the spin transfer torque $\vec T_{STT}$ we derived in \eqref{SpinTorque} to the LLG equation \eqref{GLLGEQ}. Note that the first term in \eqref{SpinTorque} simply renormalizes the gyroscopic ratio $\gamma_0$, while the second term renormalizes the Gilbert damping parameter $\alpha$. Thus the first two temporal spin terms in \eqref{SpinTorque} can be absorbed by the redefinition $ \gamma_0' = \gamma_0 (1+\eta')^{-1}$  and $ \gamma_0'\alpha'  = \gamma_0 (\alpha+\xi_0 \eta')$ with $\eta'= (n_0/M_s)/(1+\xi_0^2)$. To make the notation uncluttered, we omit the prime for these parameters. Thus, we have two new contributions to the LLG equation, the last two spatially non-uniform magnetization vector terms in \eqref{SpinTorque}, in the presence of conduction electrons. Then the LLG equation with the spin transfer torque has the form. 
\begin{align}\label{LLGEQ}
\partial_t \vec M &= - \gamma_0 \vec M \times \vec H_{eff} + \frac{\alpha}{M_s} \vec M \times \partial_t \vec M
+ \frac{\beta}{M_s} \vec M \times (v_{s}^i \nabla_i) \vec M + \frac{\beta}{\xi_0 M_s^2} \vec M \times [ \vec M \times  (v_{s}^i \nabla_i)  \vec M ] \;,
\end{align}
where $\beta = P \mu_B \xi_0 / M_s (1 + \xi_0^2)$. 

When we consider the Skyrmions, we use the approximation $|\vec M|=constant$. This is justified because $|\vec M|$ varies only weakly in the Skyrmion phase. The last term in \eqref{LLGEQ} is simplified as $ \frac{\beta}{\xi_0 M_s^2} \vec M \times [ \vec M \times  (v_{s}^i \nabla_i)  \vec M ] = - \frac{\beta}{\xi_0}  (v_{s}^i \nabla_i) \vec M $ for $\vec M$ and its spatial derivatives are orthogonal each other. After normalizing the space scale compared to the time scale so that $\beta/\xi_0 \to 1$, the LLG equation becomes 
\begin{align}\label{LLGEQ5}
\big(\partial_t +  v_{s}^i \nabla_i \big)  \vec M&= - \gamma \vec M \times \vec H_{eff} + \frac{\alpha}{M_s} \vec M \times \big( \partial_t + \frac{\beta}{\alpha} v_{s}^i \nabla_i  \big)\vec M  \;. 
\end{align}
Here the time and spatial derivatives come along together. The time derivative becomes the drift velocity $\vec v_d$ of the Skyrmions when we consider their motion later, and $\vec v_s$ is the velocity of the conduction electrons passing through the Skyrmions. The combination $\vec v_d - \vec v_s$ is the relative velocity between the center of Skyrmion and conduction electrons.  

We also generalize the LLG equation \eqref{GLLGEQ} by including the spin-orbit torque $\vec T_{SOT} $ and also the Hall torque $\vec T_{SHT} $ in a unified fashion. 
\begin{align} \label{GLLGEQ-Field}
\frac{\partial \vec M}{\partial t}&= - \gamma_0 \vec M \times (\vec H_{eff} +\vec H_{sd} +\vec H_{SOT} + \vec H_{SL})+ \frac{\alpha}{M_s} \vec M \times \frac{\partial \vec M}{\partial t} \;, 
\end{align}
where $\vec H_{sd}$, $\vec H_{SOT}$ and $\vec H_{eff}$ are given in \eqref{SpinTorque-SOTDL}, \eqref{SpinTorque-SOTFL} and \eqref{SHTFieldHSL}, respectively. Combining $\vec H_{sd}$, $\vec H_{SOT}$ and $\vec H_{SL}$ with $\vec H_{eff}$ is convenient and actually reasonable because all these acts on the ferromagnetic layer uniformly. All the spin torques can be combined to control the motion of DW or Skyrmions. The spin transfer torque $\vec T_{STT}$ which has the form \eqref{GLLGEQ-Field} is rather distinct from those of field like spin torques $\vec T_{SOT} $ and $\vec T_{SHT} $ given in \eqref{LLGEQ5}. 

Before moving on, we would like to introduce the effect of the topological emergent electro-magnetic fields to the spin torque and the LLG equation. This has been considered in \cite{FerroCoupling2}\cite{DMinteractionLGModel}. 
\begin{align}\label{LLG-EMField}
\vec T_{EM} = -\frac{\alpha'}{M_s^3} \big[ \vec M \cdot \big( \partial_i \vec M \times (\partial_t + \frac{\beta'}{\alpha'} v_s^k \nabla_k) \vec M \big) \big] \partial_i \vec M 
= -\big(\alpha' E_i^e + \beta' (\vec v_s \times \vec B^e)_i \big) \partial_i \vec M  \;,
\end{align}
where $ e_i = E_i^e = \vec n \cdot (\partial_i \vec n \times \partial_t \vec n)$ and $b_i = B_i^e = \frac{1}{2} \epsilon_{ijk} \vec n \cdot (\partial_j \vec n \times \partial_k \vec n)$ with $\vec n = \vec M/M_s $. The spin structure of the Skyrmions provides the emergent electro-magnetic fields (EEMF). This EEMF interact with the conduction electrons due to the Hunt coupling, strong ferromagnetic coupling, between the electron spins and the spins of the Skyrmions. They are higher order terms in derivatives. Nevertheless, these contributions can be significant for the metallic materials.  

\subsection{Thiele equation}

On 1973, Thiele published a paper that describes the steady state motion of magnetic domains, which can be also directly applied to the topological magnetic Skyrmions as well. His paper is beautifully written and captures deep insights. Here, we would like to follow his original paper \cite{Thiele}. 
The $i$-th component of the LLG equation \eqref{GLLGEQ} without the three torque terms (or \eqref{LLGEQ} without the last two terms) has the form 
\begin{align}\label{LLGEQi}
-\frac{1}{\gamma_0} \partial_t M_i + \frac{\alpha}{\gamma_0 M_s} \epsilon_{ijk} M_j \partial_t M_k - \epsilon_{ijk} M_j H^{eff}_k = 0 \;.
\end{align}
We consider the effects of the omitted torque terms $\vec T_{STT}, \vec T_{SOT}$ and $ \vec T_{SHT} $ later. 

We focus on studying the spatially constant saturation magnetization, $M_s^2 = M_i M_i=constant$. This implies $ M_i \frac{\partial M_i}{\partial x^j}=0$, meaning that the magnetization vector and its spatial derivatives are orthogonal to each other. Furthermore, contracting $M_i$ with \eqref{LLGEQi} followed by summing over the index $i$ reveals that $ M_i \partial_t M_i=0 $. They are summarized as 
\begin{align}\label{LLGEQiConstraint}
M_i \partial_t M_i =0 \;, \quad M_i \frac{\partial M_i}{\partial x^j}=0\;, \quad  M_s^2 = M_i M_i=constant \;.
\end{align}
These relations tells that $ \vec M$ is orthogonal to $ \partial_t \vec M$ and also to $ \vec \nabla \vec M $. 

Thiele realized that the following equation is equivalent to \eqref{LLGEQi} with the conditions \eqref{LLGEQiConstraint}. 
\begin{align}\label{LLGEQiEquivalent}
\tilde \beta M_j + \frac{-1}{\gamma_0 M_s^2} \epsilon_{jkl} M_k \partial_t M_k +  \frac{-\alpha}{\gamma_0 M_s} \partial_t M_j + H^{eff}_j = 0 \;.
\end{align}
One can verify the equivalence by multiplying $ -\epsilon_{jik} M_k$ to \eqref{LLGEQiEquivalent}, summing over $j$ and renaming indices. 
Multiplying $M_j$ to \eqref{LLGEQiEquivalent} seems to produce a new result $\tilde \beta M_s^2+ M_j H_j=0$, yet it just fixes the value $\tilde \beta$, which does not enter into the physical quantities we consider. Note that the first three terms are mutually orthogonal. With the identification of $\tilde \beta$, one can derive \eqref{LLGEQiEquivalent} directly from \eqref{LLGEQi} by acting $ \epsilon_{ijk} M_k$ on it. 

To derive the Thiele's equation, we consider the steady motion of spin structure such as a domain wall or Skyrmion lattice with the center position $X_i$ that moves with a drift velocity $v^d_i$, while field position is denoted by $x_i$. Thus $M_i = M_i (x_j - X_j), X_j = v^d_j t $ and $ \partial_t M_i  = -v^d_j \partial M_i/\partial x_j$.
We multiply $-\partial M_j/ \partial x_i $ on \eqref{LLGEQiEquivalent} to produce the force density for each term. The first term in \eqref{LLGEQiEquivalent} vanishes. The second term gives $\hat g _{ij} v_j = \frac{1}{M_s^2 \gamma_0'} \epsilon_{lmn} M_l \frac{\partial M_m}{\partial x_i} \frac{\partial M_n}{\partial x_j} v_j $ that can be rewritten in a more convenient form. 
\begin{align}
f^g_i =\epsilon_{ijk} g_j v_k 	\;, \qquad g_j
=  \frac{-1}{2 M_s^2 \gamma_0'} \epsilon_{jpq}\epsilon_{lmn} M_l \frac{\partial M_m}{\partial x_p} \frac{\partial M_n}{\partial x_q}   \;.   
\end{align}
The integral of this $f^g_i$ over a total volume is the total Magnus force with the gyromagnetic coupling vector $\vec g$. Even without considering the topological object such as Skyrmions, this gyromagnetic term contains the topological charge density \eqref{TopologicalChargeDensity} that can be identified with $\vec n = \vec M/M_s$ in two spatial dimensions. Thus we can readily apply for the systems with Skyrmions.     

The third term proportional to $\alpha$ gives 
\begin{align}
f^\alpha_i = \alpha' d_{ij} v_j 
\;, \qquad d_{ij}= \frac{-1}{M_s \gamma_0'} \frac{\partial M_k}{\partial x_i} \frac{\partial M_k}{\partial x_j}  \;. 
\end{align}
Here $d_{ij}$ is a second rank tensor. 
The integral of this $f^\alpha_i$ over a total volume is the total dissipative drag force. This term contains a second rank tensor $d_{ij}$ that is similar to the non-uniform energy term considered in Landau-Lifshitz \cite{LandauLifshitzEMContinuousMedia}. Multiplying $-\partial M_j/ \partial x_i $ to the last term in \eqref{LLGEQiEquivalent} followed by integrating over the volume captures the total externally applied force $\vec F$. Thus we have 
\begin{align}\label{ThieleEQ}
\vec G \times \vec v_d + \alpha \mathcal {D} \cdot \vec v_d + \vec F' = 0 \;,
\end{align} 
where $\vec G = \int_V \vec g d $, $\mathcal {D}  = \int_V d  ~dV $ with $\mathcal {D} $ and $d$ as second rank tensors. 
This is the original form developed by Thiele \cite{Thiele}. The Thiele equation \eqref{ThieleEQ} describes the translational motion of the center of a domain wall or Skyrmions without macroscopic deformation of their structure. While the Skyrmion is extended object and the spin configuration is complex, the quantity $\vec g$ is proportional to the topological charge of the Skyrmions. $\vec g$ is also directly related to the emergent magnetic field that we discuss below.

There are several different generalization of the Thiele equation. We introduce them one by one here. First, we can generalize the Thiele by including the contribution of the current with the velocity $\vec v_s$ due to conduction electrons in conducting materials. The equation \eqref{LLGEQ} has the combination $ (\partial_t + v_{s}^i \nabla_i ) \vec M$ in two different places, whose dependences on $\vec M$ are the same as the first two terms in \eqref{LLGEQi}. This combination has more suggestive form, $ (\partial_t + v_{s}^i \nabla_i ) \vec M = (v^i_s -  v^i_d) \nabla_i \vec M $, when we use $ \partial_t M_i  = -v^d_j \partial M_i/\partial x_j$. Then the Thiele equation without external force or pinning effects becomes 
\begin{align}\label{ThieleEQ2CC}
\vec G  \times (\vec v_s - \vec v_d) + \mathcal {D}  \cdot (\beta \vec v_s -\alpha \vec v_d) = 0 \;.
\end{align} 
This can be useful for the Skyrmions without pinning or impurity effect \cite{Impurity}\cite{Litzius2017}.

If one includes the rotation of an extended object such as Skyrmions, one needs to go beyond the rigid limit of the Skyrmions. This can be treated approximately by including a term as $\vec n (t, \vec x) = \vec n_0 (\vec x - \vec R(t)) + \vec n_1 + \delta \vec n$, where $\vec n_0 $ describes the motion of the center of Skyrmion with a coordinate $R(t)$ and $ \delta \vec n$ the motion of conduction electrons. $ \vec n_1$ is proportional to the time derivative of $R(t)$. This give rise to the mass term of the Skyrmion motion described in the following equation \cite{ThermalHC1}. 
\begin{align}\label{ThieleEQ2CC01}
\mathcal{M} \frac{d\vec v_d}{dt} +  \vec G  \times (\vec v_s - \vec v_d) + \mathcal {D}  \cdot (\beta \vec v_s -\alpha \vec v_d) = \vec F \;.
\end{align} 
Here we also include the acceleration term with $\mathcal M = \frac{1}{2} \int d^2 x \vec n_0 \cdot \left[ {\partial \vec n_1}/{\partial \dot {\vec R}_i} \times {\partial \vec n_0}/{\partial \dot x_i} \right]$, the mass of the domain wall or Skyrmion along with the generalization of the force terms such as pinning force. The terms involved with the current $\vec v_s$ describe the spin transfer torque due to the conduction electrons on the Skyrmions through the interactions between the spins of the Skyrmions and spins of the conduction electrons. 

Here we also generalize the Thiele equation with the spin-orbit torque terms given in \eqref{SpinTorque-SOT}. We multiply $-1/\gamma_0$ and $-\epsilon_{jlk} M_k$ to the $l$ the component of \eqref{SpinTorque-SOT}, followed by contracting the index $l$, and finally multiply $- \partial M_j/\partial x_i$, moving to the other side with $-$ sign, and summing over the index $j$. After some algebra, we get 
\begin{align}\label{LLGEQ-SOT-Thiele0}
f^{SOT}_i = a_{SOT} ~ \vec M \cdot \left( \frac{\partial \vec M}{ \partial x_i} \times \left[ (-\hat z \times \vec J_e) + \eta \vec M \times  (\hat z \times \vec J_e)   \right] \right) \;, 
\end{align}
which can be rewritten as $f^{SOT}_i = \mathcal{T}_{ij} J_{e,j}= (\mathcal{T}_{ij}^{1} + \mathcal{T}_{ij}^{2} ) J_{e,j}$.
\begin{align}\label{LLGEQ-SOT-Thiele}
\mathcal{T}_{ij}^{1} = a_{SOT} \epsilon_{zlj} \epsilon_{lmn} \frac{\partial M_m}{ \partial x_i} M_n \;, \qquad 
\mathcal{T}_{ij}^{2} = -a_{SOT} \eta \left[ \epsilon_{zmn} M_m \frac{\partial M_n}{ \partial x_i} M_j -  \epsilon_{jmn} M_m \frac{\partial M_n}{ \partial x_i} M_z  \right] \;. 
\end{align}
Here $\mathcal{T}_{ij}$ is a second rank tensor. 
Thus the Thiele equation of the magnetization $\vec M$ in the presence of an electric current $\vec J_e$ has multiple contributions. We take into account the spin-orbit torque (SOT) as well as the spin transfer torque (STT). 
\begin{align}\label{ThieleSOT}
\vec G  \times (\vec v_s - \vec v_d) + \mathcal {D}  \cdot (\beta \vec v_s -\alpha \vec v_d) + \mathcal{T} \cdot \vec J_e = 0 \;.
\end{align} 

We present the generalization of the Thiele equation in the context of controlling the Skyrmions using the spin Hall current, instead of spin transfer torque, using the electric current flowing in another layer. One prominent example is putting together the ferromagnetic material with a heavy metal layer. By driving polarized currents along the heavy metal, one can pump the spin into the ferromagnetic material which becomes a standard experimental technique. From the LLG equation by including the spin Hall effect term in  \eqref{GLLGEQ-Field}, we can derive the corresponding Thiele contribution following the procedure in this section. 
\begin{align}\label{LLGEQ-SHE-Thiele0}
f^{SHT}_i = -\frac{1}{\gamma_0} \frac{g \mu_B \theta_{SH}}{2 e M_s^2 t_F}  [ \vec M \times ( \hat z \times \vec J_{HM}) ]_j  \left(-\frac{\partial M_j}{ \partial x_i} \right) \;, 
\end{align}
which can be evaluated to give 
\begin{align}\label{LLGEQ-SHE-Thiele}
f^{SHT}_i = \mathcal{B}_{ij} J_{HM,j} \;, 
\qquad \mathcal{B}_{ij} = -\frac{1}{\gamma_0} \frac{g \mu_B \theta_{SH}}{2 e M_s^2 t_F} \epsilon_{zlj} \epsilon_{lmn} \frac{\partial M_m}{ \partial x_i} M_n  \;. 
\end{align}
Here $\mathcal{B}_{ij}$ is a second rank tensor. 
The corresponding Thiele equation comes as 
\begin{align}\label{ThieleEQ2CC4}
\vec G  \times \vec v_d + \alpha \mathcal {D} \cdot \vec v_d + 4\pi  {\mathcal B} \cdot \vec J_{HM} = 0 \;.
\end{align} 
This form of Thiele equation can be combined with the \eqref{ThieleEQ2CC} that has the contribution of spin transfer torque terms due to the conduction electrons.

Lastly, we find the contribution of the topological emergent electro-magnetic fields \eqref{LLG-EMField} to the Thiele equation \eqref{ThieleEQ}. The analysis is straightforward as described in this section. We apply $(\vec M/M_s^2)  \times $ to \eqref{LLG-EMField}, take $j$ th component of it, contract with $ \partial M_j/\partial x_i$, and perform the integral over the volume. The results are 
\begin{align} \label{ThieleEQ-Added}
\int_V M_s B_e^2 dV \Big( \beta \vec v_s - \alpha \vec v_d\Big) -  \int_V M_s \vec B_e B_e^i dV  \Big( \beta v^s_i - \alpha v_i^d\Big) \;.
\end{align}
The first term can be directly added to the contribution to the third term in \eqref{ThieleEQ2CC}. 
The second term vanishes when the magnetic field is perpendicular to the velocities.

\newpage 
\section{Thermo-electromagnetic effects: Background Materials } \label{sec:ThermoEMBackground}

We review thermo-electromagnetic effects that capture various transport coefficients using H. B. Callen's thermodynamics of irreversible processes \cite{Callen1948}\cite{CallenThermodynamics}, which is considered in \S \ref{sec:IrrThermodynamics}. We consider the thermoelectric effects in detail such as Seebeck, Peltier and Thompson effects to explain the basic idea in \S \ref{sec:SeebeckPeltierThompson}. We extend this to thermo-electromagnetic effects by including magnetic field in \S \ref{sec:ThermoElectroMagnetic}. Recently, spin transport has been newly developed. We include the discussion of the spin transport coefficients by incorporating different chemical potential for up and down spins in \S \ref{sec:SpinTransport}. As an application, we apply this idea to the Spin Seebeck and Peltier effect in \S \ref{sec:SpinSeebeck}. Finally, we review the low energy collective excitations of the spin, Magnon, and its Hall effects in \S \ref{sec:MagnonHallEffect}.

\subsection{Irreversible thermodynamics} \label{sec:IrrThermodynamics}

Here we briefly consider the theory of irreversible thermodynamics developed by Callen \cite{Callen1948}\cite{CallenThermodynamics}. The equilibrium statistical mechanics assume the existence of an enormous number of quantum states among which continual and spontaneous transitions occur in the time frame of a macroscopic observation with equal probability of each of the quantum states. The theory of non-equilibrium statistical mechanics further assume time symmetry of physical laws, meaning that all laws of physics remain unchanged if the time is everywhere replaced by its negative and if simultaneously the magnetic field is replaced by its negative. Onsager's reciprocity theorem is a canonical example \cite{OnsagarI}\cite{OnsagarII}. 

As an application of Onsager's reciprocity relation, we consider the thermo-electromagnetic effects that happen when simultaneous flow of electric current and heat in a system under the influence of magnetic field. We first consider thermo-electric effects following \cite{Callen1948}\cite{CallenThermodynamics}. We focus on a solid with electrons as charge carriers. If $s$ is the local entropy density, we have 
\begin{align}
ds  = \frac{1}{T} du - \sum_{k} \frac{\mu_k}{T} dn_k \;,  
\end{align}  
where $u$ is the local energy density, $\mu$ the electrochemical potential per electron (not to be confused with the vector $\vec \mu$ or a component $\mu_z$ that we use as magnetic dipole moment in \S \ref{sec:SpinTorqueBackground}), $n$ the number density of electrons, and $k$ refers to the electrons and other contributions such as atomic nuclei. This equation can be rewritten as that of current densities as 
\begin{align}
\vec J_{entropy}   = \frac{1}{T} \vec J_U - \frac{\mu}{T} \vec J_N \;,  
\end{align}  
where $\vec J_{entropy} , \vec J_U,$ and  $\vec J_N $ are the current densities of entropy, energy, and number of electrons, respectively. Here we assume the other components are immobile except electrons. 

To get a crucial formula relating the currents and the corresponding forces that produce the currents, we consider the entropy density per volume as an extensive variable densities. $s= s(u, n) $. Then $ ds= \frac{1}{T}  du - \frac{\mu}{T} dn$. Then the corresponding current densities satisfy $ \vec J_{entropy}   = \frac{1}{T} \vec J_U - \frac{\mu}{T} \vec J_N$. Now we are ready to compute the entropy production rate per volume by using the continuity equation as 
\begin{align}\label{EntropyEQ0}
\frac{ds}{dt} =  \frac{\partial s}{\partial t} + \vec \nabla \cdot \vec J_{entropy} = \frac{1}{T}  \frac{\partial u}{\partial t} - \frac{\mu}{T} \frac{\partial n}{\partial t} 
+ \vec \nabla \cdot \big( \frac{1}{T} \vec J_U - \frac{\mu}{T} \vec J_N \big) =  \vec \nabla \big( \frac{1}{T} \big) \cdot  \vec J_U - \vec \nabla \big( \frac{\mu}{T} \big) \cdot  \vec J_N \;, 
\end{align}
where we use the continuity equation for the energy density $ \frac{\partial u}{\partial t} + \vec \nabla \cdot \vec J_U=0$ and number density $\frac{\partial n}{\partial t} + \vec \nabla \cdot \vec J_N=0 $. A little more convenient form is heat current instead of the energy current. Thus we have 
\begin{align}\label{EntropyEQ}
\frac{ds}{dt} =  \vec \nabla \big( \frac{1}{T} \big) \cdot  \vec J_Q - \frac{1}{T}  \vec \nabla  \mu \cdot  \vec J_N \;, 
\end{align}
where $\vec J_Q = \vec J_U - \mu \vec J_N $. The heat current in a steady-state flow shows something we usually see in thermodynamics. By taking the divergence of heat current and using the divergencelessness of $\vec J_U$ and $ \vec J_N $, we get $\vec \nabla \cdot \vec J_Q =  -\vec \nabla \mu \cdot \vec J_N $. Thus the rate of heat current increase is equal to the rate of decrease in the potential energy current. By plugging in \eqref{EntropyEQ}, we get $\frac{ds}{dt} =  \vec \nabla \big( \frac{1}{T} \big) \cdot  \vec J_Q +  \frac{1}{T} \vec \nabla \cdot  \vec J_Q =  \vec \nabla \cdot \big( \frac{ \vec J_Q }{T} \big)  $. Thus in steady state flow, the increase in entropy comes from the flow of heat and also appearance of heat current.

The significance of this equation \eqref{EntropyEQ} is the relation between the currents $ \vec J_Q, \vec J_N$ and the corresponding generalized forces, called affinities, $\vec \nabla \big( \frac{1}{T} \big), \frac{1}{T}  \vec \nabla \mu $, respectively. The dynamical equations that connect them are (in 1 dimension)
\begin{align} \label{TransportFormula1D}
\begin{split}
- \vec J_N &= L_{11} (\vec \nabla \mu) / T +  L_{12} \vec \nabla (1/T) \;, \\
\vec J_Q &=   L_{12} (\vec \nabla \mu) / T +  L_{22} \vec \nabla (1/T) \;, 
\end{split}
\end{align}
where we use the Onsager relation $ L_{12} (\vec H)= L_{21} (- \vec H)$. Here we comment on $\mu$. The chemical potential can include different contributions such as electric potential $\mu_e$, the spin contribution, and other chemical portions. The charge of an electron is $e$, $\mu_e=e\phi$ with $\phi$ as an electrostatic potential. Thus, $ \vec \nabla \mu_e =- e\vec E$. We will be back to the spin-up and spin-down chemical potentials $\mu_\pm$ later in this section.  

These coefficients $L$ are related to transport coefficients with no applied magnetic fields. For example, electric conductivity is the electric current density $ \vec J_e = e \vec J_N$ per electric field $\vec E$ in an isothermal system $\vec \nabla T=0$ (without controlling over the heat current). Thus 
\begin{align}\label{ElectricConductivity}
\sigma = \vec J_e/\vec E = e \vec J_N/ \vec E = -e \vec J_N/ (\vec \nabla \mu_e/e) = e^2 L_{11} /T \;.
\end{align} 
Similarly, the heat conductivity $\kappa$ is defined as the heat current density per unit temperature gradient when there is no electric current $\vec J_N=0$ (without controlling the change of the chemical potential). To enforce the condition $\vec J_N=0$, we solve the first equation in \eqref{TransportFormula1D}. Thus $\vec \nabla \mu = - (L_{12}/L_{11})T \vec \nabla (1/T) $. Plugging this into the second equation in \eqref{TransportFormula1D},
\begin{align}\label{ThermalConductivity}
\kappa = - \vec J_Q/ \vec \nabla T = \frac{L_{11} L_{22} - L_{12}^2 }{L_{11} T^2} \;.
\end{align} 
There are three independent transport coefficients $L_{11}, L_{22}$ and $L_{12}$. In addition to the electric and thermal conductivities, one can define the thermoelectric power as the potential gradient (electric field) per unit temperature gradient when there is no electric current $\vec J_N=0$ (without controlling the heat current).
\begin{align}\label{ThermoelectricPower}
\epsilon = \vec E / \vec \nabla T = - \frac{1}{e} \vec \nabla \mu/ \vec \nabla T = - \frac{ L_{12}}{eT L_{11}} \;.
\end{align} 
Thus we have $ L_{11} = T\sigma / e^2, L_{12} = -\epsilon \sigma T^2/e$ and $L_{22} =T^2 \kappa + T^3 \epsilon^2 \sigma $. Thermoelectric power $\epsilon$ is another name of the Seebeck coefficient, that is sometimes denoted as $S$. ($\epsilon$ is not to be confused with $\varepsilon$ that is used for energy density in \S \ref{sec:Hydrodynamics} and $\epsilon_{\mu\nu}$ that is used as totally antisymmetric tensor with indices throughout this review.)  

Now the dynamic equation \eqref{TransportFormula1D} can be rewritten in terms of the electric, thermal conductivities $\sigma, \kappa$ and thermoelectric power $\epsilon$
\begin{align} \label{TransportFormula1D2}
\begin{split}
- \vec J_N &= (\sigma / e^2) \vec \nabla \mu +  (\epsilon \sigma /e)  \vec \nabla T  \;, \\
\vec J_Q &=  -(T \epsilon \sigma /e) \vec \nabla \mu  - (\kappa + T \epsilon^2 \sigma)  \vec \nabla T \;, 
\end{split}
\end{align}
In terms of electric current, electric field, heat current, and temperature difference, we get 
\begin{align} \label{TransportFormula1D4}
\left( \begin{array}{c}
\vec J_e  \\
\vec Q \end{array} \right) 
=
\left(  \begin{array}{cc} 
\sigma  & \epsilon \sigma  \\
T \epsilon \sigma &~ \kappa + T \epsilon^2 \sigma \end{array} \right)
\left( \begin{array}{c}
\vec E\\
- \vec \nabla T \end{array} \right) \;.
\end{align}
Thus we express all the transport coefficients in terms of the conductivities and thermoelectric power. Frequently, alternative form is used in the literature that change the relations as $\vec E (\vec J)$. By converting the \eqref{TransportFormula1D4}, we get 
\begin{align} \label{TransportFormula1D7}
\left( \begin{array}{c}
\vec E  \\
\vec Q \end{array} \right) 
=
\left(  \begin{array}{cc} 
\rho  &  \epsilon   \\
T \epsilon  &~ -\kappa  \end{array} \right)
\left( \begin{array}{c}
\vec J_e\\
\vec \nabla T \end{array} \right) \;,
\end{align}
where the resistivity is inverse of the conductivity, $\rho=1/\sigma$. As mentioned before, the thermoelectric power is also Seebeck coefficient $\epsilon = S $. In the coming section, we explorer the physical properties of the transport coefficients in more detail.

\subsection{Seebeck, Peltier, Thompson effects} \label{sec:SeebeckPeltierThompson}

In this subsection, we continue to discuss one dimensional thermo-electric transport phenomena given in \eqref{TransportFormula1D}, \eqref{TransportFormula1D4}, or \eqref{TransportFormula1D7} without magnetic fields. We briefly review the experimental setups for the Seebeck effect and its coefficient thermo-electric power, the Peltier effect, and the Thompson effect because they frequently mentioned in the literature.  

\begin{figure}[h!]
	\begin{center}
		\includegraphics[width=0.4\textwidth]{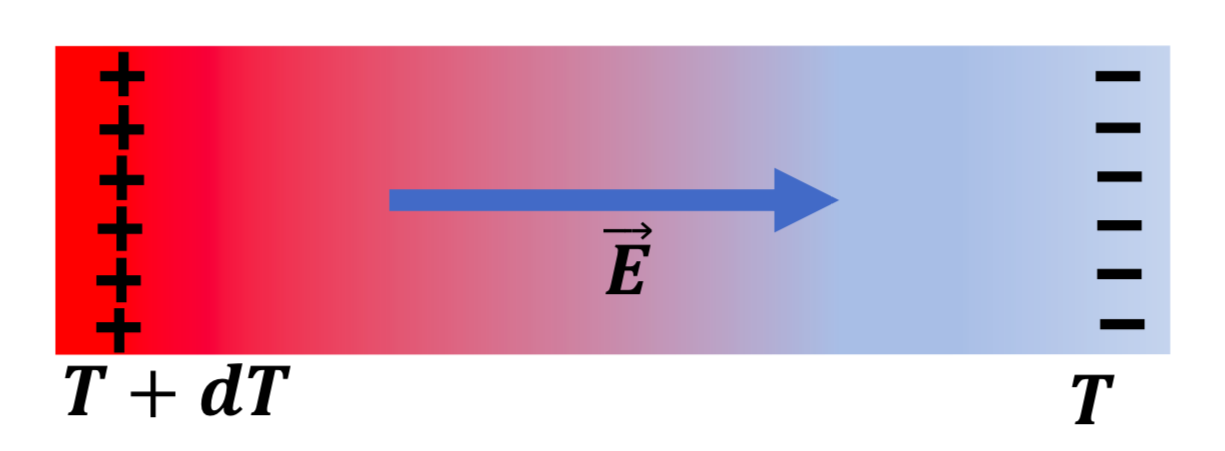}\qquad\qquad  
		\includegraphics[width=0.4\textwidth]{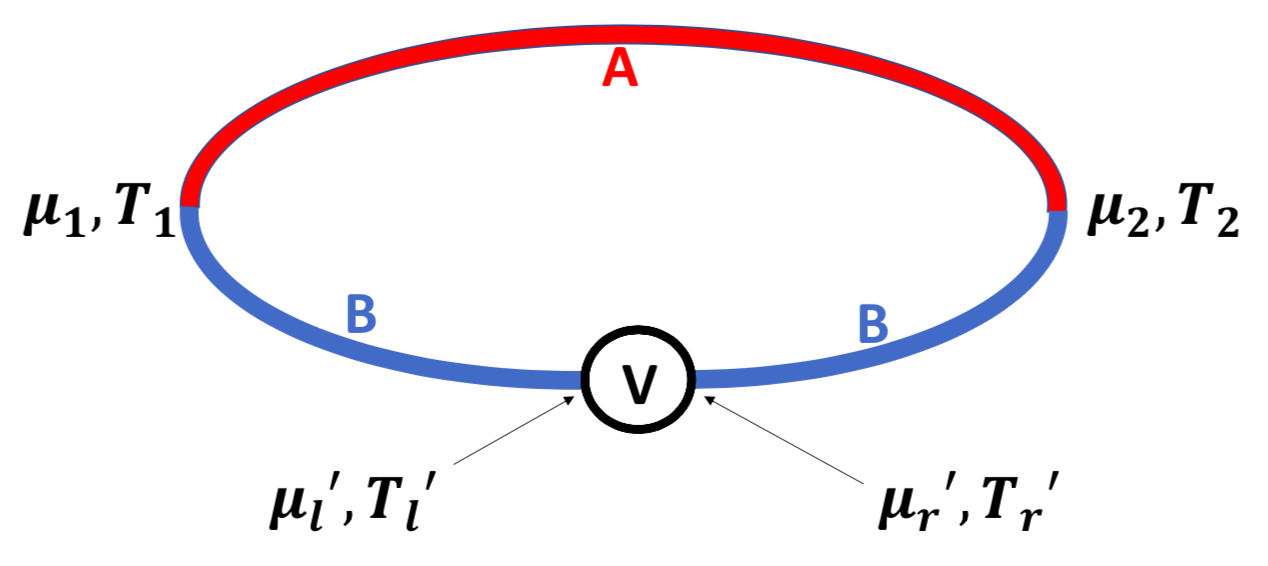}
		\caption{\small Seebeck effect. Electric field is generated by a temperature gradient in (\textbf{a}). Illustration of the thermocouple in (\textbf{b}). Two different materials $A$ and $B$ with different thermoelectric powers, $\epsilon_A$ and $\epsilon_B$ are connected at two different points with different temperatures and generate different electric potential difference. The potential difference can be measured by the voltmeter $V$. 
		}
		\label{fig:Seebeck}
	\end{center}
\end{figure}

In general, Seebeck effect is the phenomenon that temperature gradient produces electric field in conducting materials. When one side of the conductor is hotter than the other, the carriers, that are mobile, in the hotter side are excited, diffused and accumulated in the colder side.%
\footnote{This may not be true for electrons with a particular spin polarization in the in plane magnetic field. As we see below in \S \ref{sec:SpinSeebeck}, electrons with two opposite polarization show opposite behavior, spin up electrons accumulate in low temperature side, while the spin down ones in the high temperature side. Overall, more electrons accumulate in the hotter side. } 
If the carriers have electric charges, electric field can be generated as illustrated in the figure \ref{fig:Seebeck}(\textbf{a}). When the material is connected to a wire, electric current can be generated. 

In particular, the Seebeck effect refers to the production of an electromotive force in a thermocouple with vanishing electric current $\vec J_N=0$. This is directly related to the definition of the thermoelectric power $\epsilon$ given in \eqref{ThermoelectricPower}. Let us consider a thermocouple with junctions for two different materials $A$ and $B$ at temperature $T_1$ and $T_2 (T_2>T_1)$ as in the figure \ref{fig:Seebeck}(\textbf{b}). A voltmeter $V$ is attached in the middle of the material $B$ with a temperature $T'$. The voltmeter has a large resistance and does not allow electric current, while there is no resistance to the heat flow.  

From the first equation in \eqref{TransportFormula1D2}, we get $\vec \nabla \mu = -e \epsilon \vec \nabla T $. 
By integrating this function, one can compute $\mu_2 - \mu_1 = -e \int_1^2  \epsilon_A dT $ with $\epsilon_A$ as the thermoelectric power for the material $A$. The potential difference, Voltage, at the voltmeter can be computed as 
$V = (\mu_r' - \mu_l')/e =  [(\mu_2 - \mu_1) - (\mu_2 - \mu_r')  - (\mu_l' - \mu_1)]/e =  -\int_1^2  (\epsilon_A - \epsilon_B) dT $, 
where we use the fact that the temperature difference at the left and right side of the voltmeter vanishes. Thus the voltage on the voltmeter is integral of the difference of the thermoelectric power of the two materials. 

The meaning of the thermoelectric power $\epsilon$, or the Seebeck coefficient $S$,  can be understood more clearly if we eliminate the terms proportional to $\vec \nabla \mu$ in \eqref{TransportFormula1D2}, we get $\vec J_Q = T\epsilon e \vec J_N - \kappa \vec \nabla T$. The entropy current $\vec J_{entropy}  = \vec J_Q/T = \epsilon e \vec J_N - \kappa \vec \nabla T/T$. Thus the thermoelectric power can be viewed as the entropy that is transported per coulomb by the electron flow. 

\begin{figure}[h!]
	\begin{center}
		\includegraphics[width=0.45\textwidth]{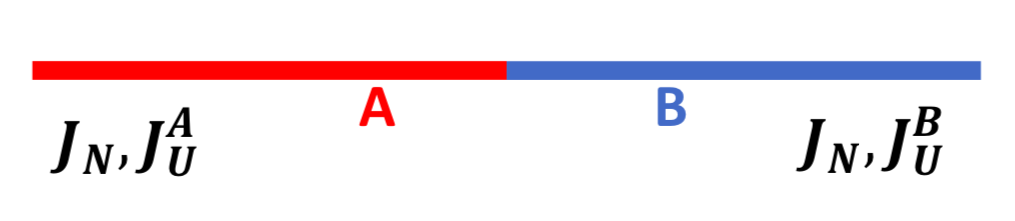}
		\caption{\small Peltier effect is generated when two conducting materials are connected with isothermal junction with electric current.  
		}
		\label{fig:Peltier}
	\end{center}
\end{figure}

The Peltier effect is inverse of the Seebeck effect. Temperature gradient can be generated by the electric field due to the accumulation of the charge carriers. This happens because charge carriers are also heat carriers. In particular, the Peltier effect is the evolution of heat accompanying the flow of electric current $e\vec J_N $ across an isothermal junction (the middle point in the figure \ref{fig:Peltier}) of two conducting materials $A$ and $B$. The isothermal junction prevents heat exchange and thus creates the discontinuity of $\vec J_Q$. The total energy current, $\vec J_U = \vec J_Q + \mu \vec J_N  $, is also discontinuous. Thus,  
\begin{align} 
\vec J^B_U - \vec J^A_U = \vec J^B_Q - \vec J^A_Q  = T(\epsilon_B -\epsilon_A) (e\vec J_N) \equiv \pi_{AB} (e\vec J_N) \;,
\end{align}
where we use that the electric current and the chemical potential are continuous and use $ \vec J_Q = T\epsilon e \vec J_N$ that is evaluated from \eqref{TransportFormula1D2} or \eqref{TransportFormula1D7} with the isothermal condition $\vec \nabla T=0$. The Peltier coefficient $\pi_{AB}=T(\epsilon_B -\epsilon_A)$ is the heat that needs to be supplied to the junction when unit electric current passes from $A$ to $B$. As we see the Peltier effect is directly related to the Seebeck effect and thermoelectric power, which is called as second Kelvin relation.  

\begin{figure}[h!]
	\begin{center}
		\includegraphics[width=0.35\textwidth]{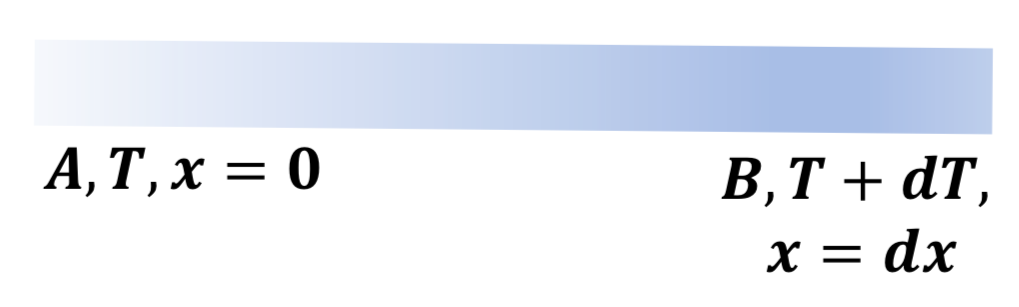} \qquad
		\includegraphics[width=0.35\textwidth]{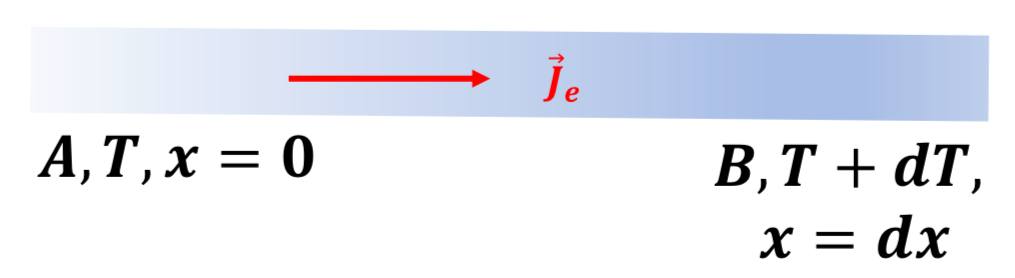}
		\caption{\small Thompson effect. (\textbf{a}): heat current without electric current. (\textbf{b}): heat and electric currents. 
		}
		\label{fig:Thompson}
	\end{center}
\end{figure}

The Thompson effect concerns the heat absorbed per unit electric current and per unit temperature gradient. Let us consider a setup more clear \cite{ThompsonEffect}. Consider the figure \ref{fig:Thompson} (\textbf{a}) that there is a heat current without electric current due to the temperature difference $dT= \delta T$ at two ends $A$ and $B$ of a conducting material. The conductor are placed in heat reservoir so that there would be no heat interchange between the conductor and reservoir. 

Now let the electric current flows as in the figure \ref{fig:Thompson}(\textbf{b}). Then the heat interchange will take place between the conductor and heat reservoir. The heat exchange has two parts: Joule heat and Thompson heat. The total energy flow can be computed by $\vec \nabla \cdot \vec  J_U = \vec \nabla \cdot (\vec J_Q + \mu \vec J_N) =  \vec \nabla \cdot \vec J_Q + \vec \nabla \mu \cdot \vec J_N  $, where we use $\vec \nabla \cdot  \vec J_N=0$ due to current conservation. We are interested in the change of energy as a function of current $\vec J_N$ without changing the temperature profile, we replace $\vec \nabla \mu$ using the first equation of \eqref{TransportFormula1D2} and $\vec J_Q $ using $\vec J_Q = T\epsilon e \vec J_N - \kappa \vec \nabla T$ that is from the combination of the two equations of \eqref{TransportFormula1D2}. Then
\begin{align}
\vec \nabla \cdot \vec  J_U 
&= T \vec \nabla \epsilon \cdot (e \vec J_N) - \vec \nabla \cdot (\kappa \vec \nabla T) - e^2/\sigma \vec J_N^2 \;. 
\end{align}
The first term on the right hand side is the desired Thompson term that represent the heat absorbed from the thermal reservoir when the current $e\vec J$ travels through the temperature gradient. Using $\vec \nabla \epsilon= \frac{d\epsilon}{dT} \vec \nabla T$, the Thompson coefficient (defined as the Thompson heat absorbed per unit electric current and per unit temperature gradient) reads 
\begin{align}
\tau = \frac{\text{Thompson heat}}{\vec \nabla T \cdot (e \vec J_N)} = T  \frac{d\epsilon}{dT}  \;. 
\end{align}
The second term on the right hand side vanishes according to our setup. This can be checked with the condition $\vec \nabla \cdot \vec  J_U = 0$ if $  \vec J_N=0 $, meaning that the change of the energy vanished when there is no electric current. The last term is nothing but the Joule heating term. 

\begin{figure}[h!]
	\begin{center}
		\includegraphics[width=0.45\textwidth]{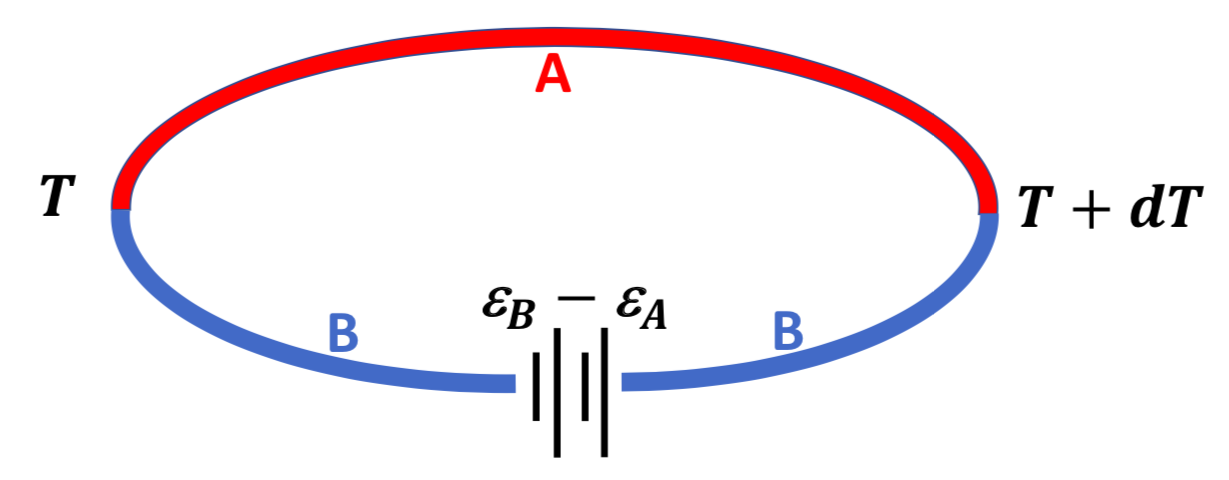} 
		\caption{\small The first Kelvin relation among Seebeck, Peltier, and Thompson effects.  
		}
		\label{fig:SeebeckThompson}
	\end{center}
\end{figure}

There exist an interesting relation between the three coefficients, Seebeck, Peltier and Thompson coefficients. By taking a derivative of the Peltier coefficient, we get $ \frac{d\pi_{AB}}{dT} + \tau_A - \tau_B= \epsilon_A - \epsilon_B $. This can be demonstrated by the thermocouple with the Voltmeter replaced by a battery that negates the Seebeck voltage so that there is no electric current in the thermocouple as in the figure \ref{fig:SeebeckThompson}. Here the thermocouple is placed in the thermal bath that was considered for Thompson heating with the two junctions to the end points of the conductor. This is known as the first Kelvin relation.

\subsection{Thermo-electromagnetic effects } \label{sec:ThermoElectroMagnetic}

\begin{figure}[h!]
	\begin{center}
		\includegraphics[width=0.45\textwidth]{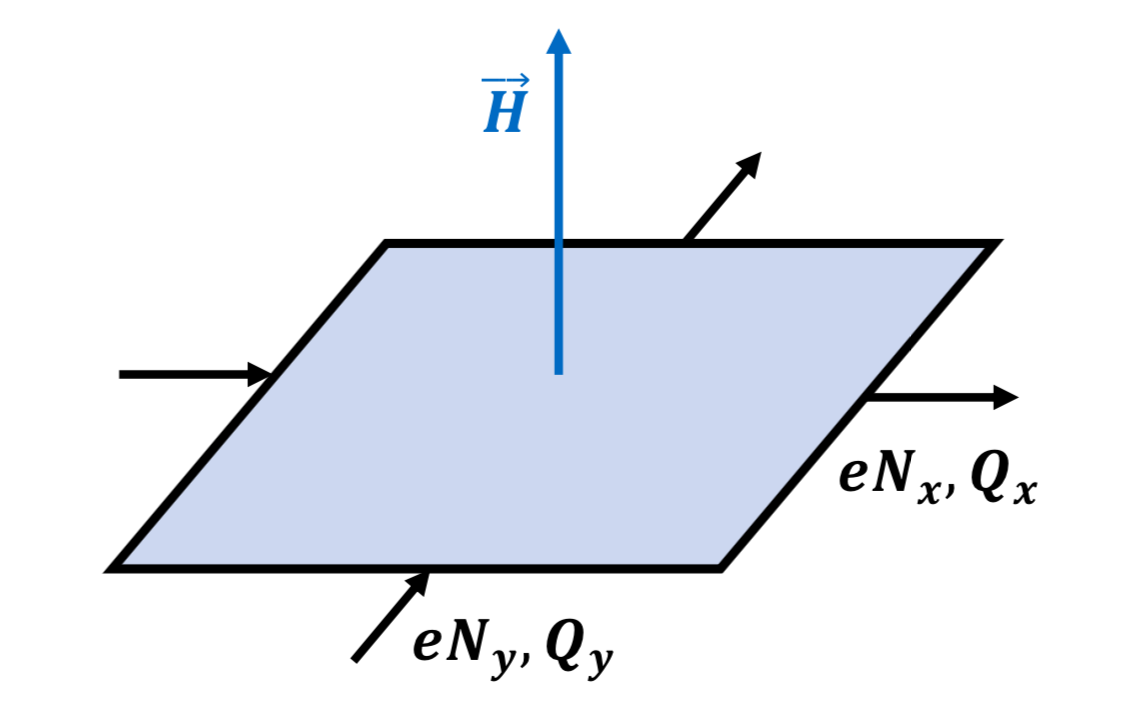} 
		\caption{\small Setup for the thermo-electromagnetic transport coefficients in the presence of the Magnetic Field $\vec H$ as well as the charge $\vec J_e = e\vec J_N$ and heat $\vec Q = \vec J_Q$ currents.  }
		\label{fig:WithMagneticField}
	\end{center}
\end{figure}

We generalize the discussion of the thermoelectric transport coefficients to the two spatial dimensions with magnetic field $\vec H$ perpendicular to the two dimensional $xy$ plane. This is illustrated in the figure \ref{fig:WithMagneticField}. Following the discussion in \S \ref{sec:IrrThermodynamics}, we consider the entropy change in terms of the currents and the corresponding generalized forces. 
\begin{align}
\frac{ds}{dt} =  \vec \nabla \big( \frac{1}{T} \big) \cdot  \vec J_Q - \frac{1}{T}  \vec \nabla  \mu \cdot  \vec J_N 
=  - \frac{1}{T^2} \vec \nabla T \cdot  \vec Q - \frac{1}{T}  \vec E  \cdot  \vec J_e \;. 
\end{align}
Hereafter on we adapt the notation $ J_{e,x} = e J_{N,x}, J_{e,y} = e J_{N,y}, Q_x = J_{Q,x}, Q_y = J_{Q,y}$ and $\vec \nabla \mu_e = -e \vec E $. 

The dynamical equation becomes 
\begin{align} \label{TransportFormula2D}
\begin{split}
-J_{e,x} &= L_{11}' \frac{e^2}{T} E_x + L_{12}' \frac{e^2}{T} E_y + L_{13}' \frac{e}{T^2} (- \nabla_x T) + L_{14}' \frac{e}{T^2} (-\nabla_y T)  \;, \\
-J_{e,y} &= -L_{12}' \frac{e^2}{T} E_x + L_{11}' \frac{e^2}{T} E_y - L_{14}' \frac{e}{T^2} (-\nabla_x T) + L_{13}' \frac{e}{T^2} (-\nabla_y T)  \;, \\
Q_x &= L_{13}' \frac{e}{T} E_x + L_{14}' \frac{e}{T} E_y + L_{33}' \frac{1}{T^2} (-\nabla_x T) + L_{34}' \frac{1}{T^2} (-\nabla_y T)  \;, \\
Q_y &= -L_{14}' \frac{e}{T} E_x + L_{13}' \frac{e}{T} E_y - L_{34}' \frac{1}{T^2} (-\nabla_x T) + L_{33}' \frac{1}{T^2} (-\nabla_y T)  \;.
\end{split}
\end{align}
Isotropy in $x$ and $y$ coordinates puts the `diagonal components' to be the same, $L_{11}' = L_{22}', L_{13}' = L_{24}', L_{31}' = L_{42}', L_{33}' = L_{44}' $. The presence of magnetic field breaks the parity symmetry and allows the off-diagonal components, $L_{21}' = -L_{12}', L_{23}' = -L_{14}', L_{41}' = -L_{32}', L_{43}' = -L_{34}'$. The diagonal components are identified as even functions of the magnetic field, while the off-diagonal components are odd functions of the magnetic field. Moreover, Onsager relations $L_{ij}' (H) = L_{ji}'(-H) $ put further constraints as $ L_{13}' (H) = L_{31}' (-H) = L_{31}' (H)$ and $ L_{14}'(H)=L_{41}'(-H)= -L_{32}'(-H)=L_{32}'(H)$. Thus $ L_{13}' = L_{31}', L_{14}'=L_{32}'$. We identify all these related transport coefficients in \eqref{TransportFormula2D}. 

This set of algebraic equations \eqref{TransportFormula2D} can be converted into another set of equations with the electric currents and the temperature gradients on the right hand side because these are experimentally controlled. 
\begin{align} \label{TransportFormula2DAlternative}
\begin{split}
E_x &= ~~L_{11} J_{e,x} + L_{12} J_{e,y} - L_{13}  \nabla_x T - L_{14} \nabla_y T  \;, \\
E_y &= -L_{12} J_{e,x} + L_{11} J_{e,y} + L_{14} \nabla_x T - L_{13} \nabla_y T  \;, \\
Q_x &= - T L_{13}  E_x - T L_{14} E_y - L_{33} \nabla_x T - L_{34} \nabla_y T  \;, \\
Q_y &= ~~TL_{14}  E_x - TL_{13}  E_y + L_{34} \nabla_x T - L_{33} \nabla_y T  \;.
\end{split}
\end{align}
There is a definite and clear relations between $L$s and $L'$s, even though they are a little complicated. All these relations are explicitly worked out in the appendix \S \ref{appsec:TEME} following the illuminating work \cite{CallenThermodynamics}. We present the results for the transport coefficients and focus on summarizing some interesting parts.   
\begin{align} \label{TransportFormula2DForm}
\left( \begin{array}{c}
E_x \\ E_y \\Q_x\\ Q_y
\end{array} \right) =
\left( \begin{array}{cccc}
1/\sigma_i & H R_i & -\epsilon & - H \eta_N \\ 
-H R_i & 1/\sigma_i & H \eta_N&  -\epsilon\\ 
-T\epsilon & - T H \eta_N & -\kappa_i & - H \kappa_i \mathcal{L}_R \\ 
T H \eta_N & - T\epsilon& H \kappa_i \mathcal{L}_R  & - \kappa_i 
\end{array} \right)
\left( \begin{array}{c}
J_{e,x} \\ J_{e,y} \\ \nabla_x T \\ \nabla_y T
\end{array} \right) 
\end{align}
where $H, T$ are magnetic field and temperature, while $\sigma, R_H, \epsilon, \eta_N, \kappa, \mathcal{L}_R$ are isothermal electric conductivity $\sigma=\sigma_i$, isothermal Hall coefficient $R_H=R_i$, absolute thermoelectric power (Seebeck coefficient), isothermal Nernst effect $\eta_N=\eta_i$, isothermal heat conductivity $\kappa=\kappa_i$, and Leduk-Righi coefficient $\mathcal{L}_R $, respectively. (There are several different usages for $\eta$. This $\eta_N=\eta_i$ is used for the Nernst coefficient, while $\eta_H$, $\eta_{ijkl}$, $\eta_{\mu\nu}$ and $\eta$ are used for the Hall viscosity, shear tensor, metric tensor and shear viscosity, respectively, throughout the review.) In addition to these isothermal coefficients, there are corresponding transport coefficients for the adiabatic process explained in appendix \S \ref{appsec:TEME}.  

\begin{figure}[h!]
	\begin{center}
		\includegraphics[width=0.55\textwidth]{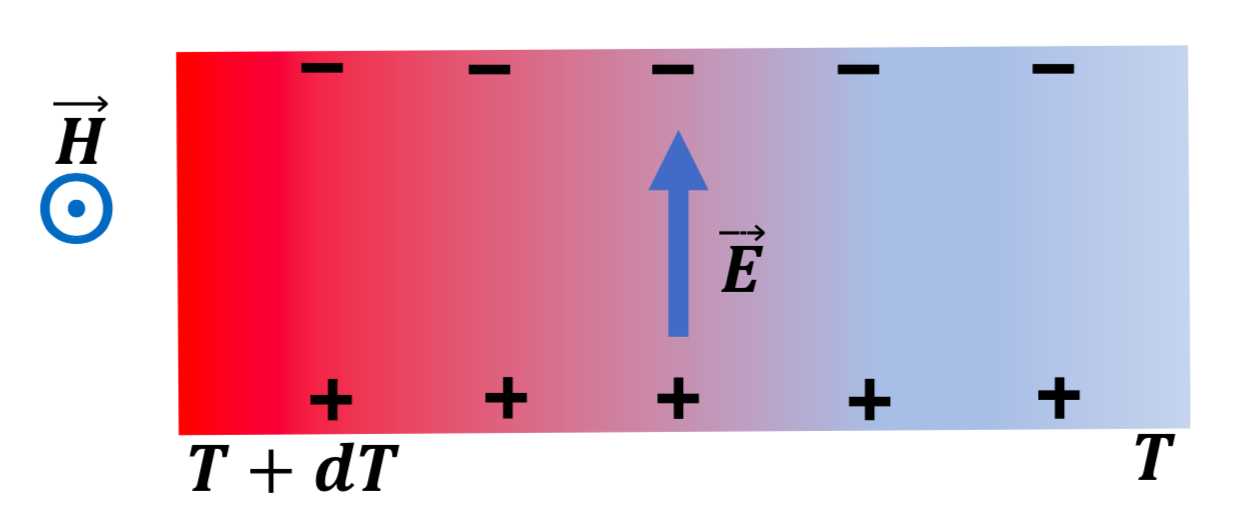} 
		\caption{\small Illustration of the Nernet effect. Temperature gradient produce the electric field and current when connected to a closed circuit. In the presence of magnetic field, there is a transverse electric field and the corresponding transverse current when connected to a circuit. This can be understood as the Seebeck effect in the presence of magnetic field.  
		}
		\label{fig:Nernst0}
	\end{center}
\end{figure}

The form \eqref{TransportFormula2DAlternative} and \eqref{TransportFormula2DForm} suggest the following vector form for the electric field in terms of the temperature gradient in the presence of magnetic field $\vec H$. 
\begin{align} \label{TransportFormula2DVectorFormE}
\vec E &= \rho \vec J_e + R_H \vec H \times \vec J_e + \epsilon (- \vec \nabla T) + \eta_N \vec H \times (- \vec \nabla T )   \;.
\end{align}
In addition to the resistivity $\rho$ and thermoelectric power $\epsilon$, we have additional two transport coefficients, $R_H$ and $\eta_N$. In the presence of magnetic field, the Lorentz force change the direction of the electric current producing the transverse component of the current, which is measured by the Hall coefficient $R_H$. 
More specifically, the Isothermal Hall Effect is defined as $R_H = R_i = \frac{E_y}{H J_{e,x}} $ with $\nabla_x T= \nabla_y T=  J_{e,y} =0$. 
\begin{align}
R_H = \frac{E_y}{H J_{e,x}}  = \frac{L_{12}}{H} \;. 
\end{align}
Similarly, the electric field due to the temperature gradient (this is Seebeck effect with the thermoelectric power) develops transverse component in the presence of the magnetic field. This is called Nernst effect with the Nernst coefficient $\eta_N$. The isothermal Nernst effect, $\eta_N = \eta_i = \frac{E_y}{H \nabla_x T} $ with $\nabla_y T= J_{e,x} =  J_{e,y} =0$, measures the generation of transverse electric field upon applying a temperature gradient.  
\begin{align}
\eta_N = \frac{E_y}{H \nabla_x T} = \frac{L_{14}}{H}  \;. 
\end{align}
This is illustrated in the figure \ref{fig:Nernst0}.   

We can think about the inverse of the Nernst effect, which is the generation of the temperature gradient due to the electric field in the presence of magnetic field. This is similar to the Peltier effect being inverse of the Seebeck effect. The Ettingshausen Effect is given by $E_{tth} = - \nabla_y T/H J_{e,x} $ with $\nabla_x T= Q_y =  J_{e,y} =0$. This coefficient is not in \eqref{TransportFormula2DForm}. It turns out that this coefficient can be rewritten in terms of other transport coefficients. 
\begin{align}\label{EtthRelation0}
E_{tth}  = - \frac{\nabla_y T}{H J_{e,x}} = \frac{T L_{14}}{H L_{33}} = \frac{T \eta_N}{ \kappa}  \;. 
\end{align}
Other similar relations are listed in the appendix \S \ref{appsec:TEME}. We comment one more transport coefficient $\mathcal {L}$ that is in \eqref{TransportFormula2DForm}.
The Leduk-Righi Effect is the generation of the transverse temperature gradient in the presence of a temperature gradient in the presence of magnetic field. It is defined as $\mathcal {L} = \frac{\nabla_y T}{H \nabla_x T} $ with $Q_y = J_{e,x} =  J_{e,y} =0$.  Thus 
\begin{align}
\mathcal {L}_R = \frac{L_{34}}{H L_{33}} = \frac{L_{34}}{H \kappa}  \;,
\end{align}
which is the part of $L_{34}$ in \eqref{TransportFormula2DForm}. The details of exhaustive thermo-electromagnetic transport coefficients are defined and listed in \S \ref{appsec:TEME}. For example, the transport coefficients of the adiabatic process are listed there in detail along with the relations between $L_{ij}$ and $L_{ij}'$.

\subsection{Spin effects on transports} \label{sec:SpinTransport}

In this section we include the effects of spin in the theory of thermodynamics of irreversible process following \cite{Gravier2006}. See also \cite{SpinCaloritonics}.
In the beginning of this section, we mentioned the chemical potential $\mu$ can include more than the electric scalar potential. In \cite{Gravier2006}, this idea was implemented to treat the electrons with different spins are treated as two different charge carriers with two different chemical potential $\mu_\pm$ following the idea that the chemical potential and electric potential adds in Electrochemistry \cite{AtkinEC}. Thus we consider three different currents in 2 or 3 spatial dimensions. We use $\vec Q, \vec J_+, \vec J_-$ as vectors in the corresponding dimensions. $\vec J_+$ and $ \vec J_-$ are the currents with corresponding electrochemical potentials $ \mu_+ = \mu_e + \mu_0 + \delta \mu$ and $ \mu_- = \mu_e + \mu_0 - \delta \mu$, respectively. Here $\mu_e = e \phi$ and $ \phi$ is electrostatic potential. The symmetric combination of these currents is the electric current $ \vec J_e = \vec J_+ + \vec J_-$, while the antisymmetric combination gives the so-called spin current $ \vec J_S = \vec J_+ - \vec J_-$ which is the flow of the spin angular momentum.  

The entropy current \eqref{EntropyEQ0} can be generalized to include the spin contribution. 
\begin{align}\label{EntropyEQ0G}
\frac{ds}{dt} =  \vec \nabla \big( \frac{1}{T} \big) \cdot  \vec J_U - \vec \nabla \big( \frac{\mu_+}{T} \big) \cdot  \vec J_+  - \vec \nabla \big( \frac{\mu_-}{T} \big) \cdot  \vec J_- \;. 
\end{align}
This can be recast into more convenient form 
\begin{align}\label{EntropyEQG}
\frac{ds}{dt} = -\frac{1}{T^2} \vec Q \cdot (\vec \nabla T)  - \frac{1}{T} \vec \nabla  \mu_+ \cdot  \vec J_+ - \frac{1}{T} \vec \nabla  \mu_- \cdot  \vec J_- \;. 
\end{align}
where $\vec Q =  \vec J_U - \mu_+ \vec J_+ - \mu_- \vec J_-$. Now the dynamical equation connecting the currents and generalized forces are 
\begin{align} \label{TransportFormulaSpin}
\left( \begin{array}{c}
\vec Q  \\
\vec J_+ \\
\vec J_- \end{array} \right) 
=
\left(  \begin{array}{ccc} 
L_{qq}  & L_{q+} & L_{q-}  \\
L_{q-}  & L_{++} & L_{+-}  \\
L_{q+}  & L_{-+} & L_{--} \end{array} \right)
\left( \begin{array}{c}
- \vec \nabla T /T^2\\
- \vec \nabla \mu_+ /T \\
- \vec \nabla \mu_- /T \end{array} \right) \;,
\end{align}
where $\vec \nabla \mu_\pm  =\vec \nabla \mu_0 +e \vec E \pm \vec \nabla (\delta \mu) = \vec \nabla \bar{\mu} \pm \vec \nabla (\delta \mu)$ with a generalized potential $V$. The dynamical coefficients $L$ are $d \times d$ matrices for $d$ spatial dimensions, while the vectors are $d$ dimensional column vectors. Note that we already reduced the number of parameters using the Onsager relations along with the symmetry that a positive spin that is a majority spin in a field $H$ is a minority spin in a field $-H$ and vice versa. Thus $ L_{+q} (H) = L_{q+} (-H) = L_{q-} (H)$ and $ L_{-q} (H) = L_{q-} (-H) = L_{q+} (H)$ \cite{Gravier2006}. It is interesting to realize that there is a spin mixing contribution $L_{+-}$ an $L_{-+}$ that are built in this formulation. In the spin mixing process, the charge carriers retains their momenta. 
It turns out that $L_{+-}$ and $L_{-+}$ are independent because $ L_{+-} (H) = L_{-+} (-H) = L_{+-} (H)$.

It is useful to separate the electric charge transport and spin transport. To separate the spin potential $\delta \mu$ from the electrochemical potential $ \mu_0 + \mu_e$, we add and subtract the second and third columns in the $3 \times 3$ matrix in \eqref{TransportFormulaSpin}. Then we add and subtract the second and third row of \eqref{TransportFormulaSpin}. Then 
\begin{align} \label{TransportFormulaSpinComp}
\left(\!\! \begin{array}{c}
\vec Q  \\
\vec J_e \\
\vec J_S \end{array} \!\!\right) 
=
\left(  \begin{array}{ccc} 
L_{qq}  & L_{q+}+L_{q-} & L_{q+}-L_{q-}  \\
L_{q-}+L_{q+}~~  & L_{++}+L_{ss}+L_{--} & ~~L_{++}-L_{aa}-L_{--}  \\
L_{q-}-L_{q+}~~  & L_{++}+L_{aa}-L_{--} & ~~L_{++}-L_{ss}+L_{--} \end{array} \right)
\left(\!\!\! \begin{array}{c}
- \vec \nabla T /T^2 \\
- \vec \nabla \bar \mu/T \\
- \vec \nabla \delta \mu/T \end{array} \!\!\! \right) \;,
\end{align}
where $\vec J_e = \vec J_+ + \vec J_-$, $\vec J_S = \vec J_+ - \vec J_- $, $L_{ss} = L_{+-}+L_{-+} $ and $L_{aa} = L_{+-}-L_{-+} $. 
Let us connect the coefficients to the known transport coefficients. To do so, we assume that there is no spin mixing $ L_{+-} = L_{-+} = 0$. 
Then the conductivities with condition $\vec \nabla T =0$ gives 
\begin{align}
\sigma &= \sigma_+ + \sigma_- =-e \vec J_e/ (\vec \nabla \bar{\mu}_e/e) = e^2 (L_{++} + L_{--}) /T  \;, 
\end{align} 
which determine the coefficients as $L_{++} = \sigma_+ T/e^2 $ and $L_{--} = \sigma_- T/e^2 $. The absolute thermoelectric power can be obtained with $  \vec J_e =0$ as 
\begin{align}
\epsilon_{tot} &= \epsilon_+ + \epsilon_-= - (\vec \nabla \bar{\mu}/e)/ \vec \nabla T = - ( L_{q-}/L_{++} + L_{q+}/L_{--})/(eT) \;, 
\end{align} 
which fixes $L_{q-} = eT L_{++} \epsilon_+ = T^2 \sigma_+ \epsilon_+ /e$ and $L_{q+} = eT L_{--} \epsilon_- = T^2 \sigma_- \epsilon_- /e$. Similarly, the heat conductivity can be evaluated for $\vec J_e =0 $.
\begin{align}
\kappa &= - \vec J_Q/ \vec \nabla T = \frac{1}{T^2} \Big( L_{qq} - \frac{L_{q+} L_{q-}}{L_{++}} - \frac{L_{q-} L_{q+}}{L_{--}}  \Big) \;. 
\end{align} 
This fix the $L_{qq}$ as $L_{qq} =T^2 \kappa + T^3 \epsilon_+ \epsilon_- ( \sigma_+ + \sigma_-) $. 
Thus the coefficient matrix becomes 
\begin{align} \label{SpinTransCoeffs}
\left(  \begin{array}{ccc} 
T^2 \kappa + T^3 \epsilon_+ \epsilon_- ( \sigma_+ + \sigma_-) & -T^2 (\sigma_+ \epsilon_+ + \sigma_- \epsilon_-) /e &  -T^2 (\sigma_+ \epsilon_+ - \sigma_- \epsilon_-) /e \\
-T^2 (\sigma_+ \epsilon_+ + \sigma_- \epsilon_-) /e ~~ & (\sigma_+ + \sigma_-) T/e^2  & ~~(\sigma_+ - \sigma_-) T/e^2- L _{aa}  \\
T^2 (\sigma_+ \epsilon_+ - \sigma_- \epsilon_-) /e ~~  & (\sigma_+ - \sigma_-) T/e^2+ L_{aa} & ~~ (\sigma_+ + \sigma_-) T/e^2 \end{array} \right)
\end{align} 
Where we assume that the conductivities due to the spin mixing is small compared to the direct conductivities, then the spin mixing contribution can be captured by their difference $ L_{+-} - L_{-+} = \delta L ~T/e^2 = L_{aa}$.  

\subsection{Spin Seebeck and Peltier effect} \label{sec:SpinSeebeck}

In this section we consider the newly developed spin Seebeck effect \cite{Uchida2008}\cite{Uchida2010}\cite{Jaworski2010}. 
In ferromagnetic metal, all the three currents exist, while the charge current portion is suppressed in the magnetic insulator. Along the way, we mention the importance of Magnon contribution to the spin Seebeck effect. Magnons can be treated as particle-like excitations similar to electrons, there are similarities between the electron contributions to the electronic transports in metal and the Magnonic contributions to the heat transports in insulator. Thus, the physical understanding developed in previous sections can be directly applied with minimal modifications.  

\begin{figure}[h!]
	\begin{center}
		\includegraphics[width=0.47\textwidth]{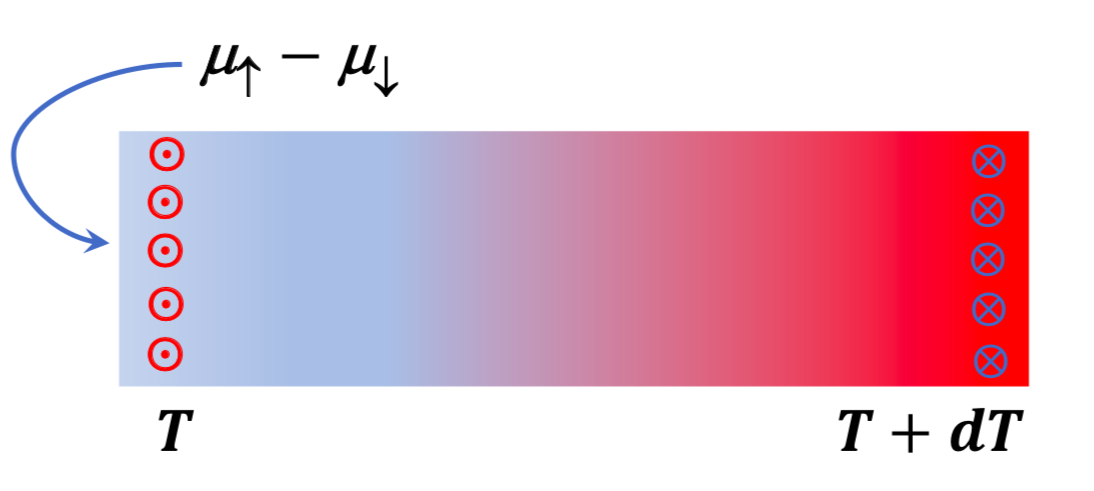} \qquad 
		\includegraphics[width=0.26\textwidth]{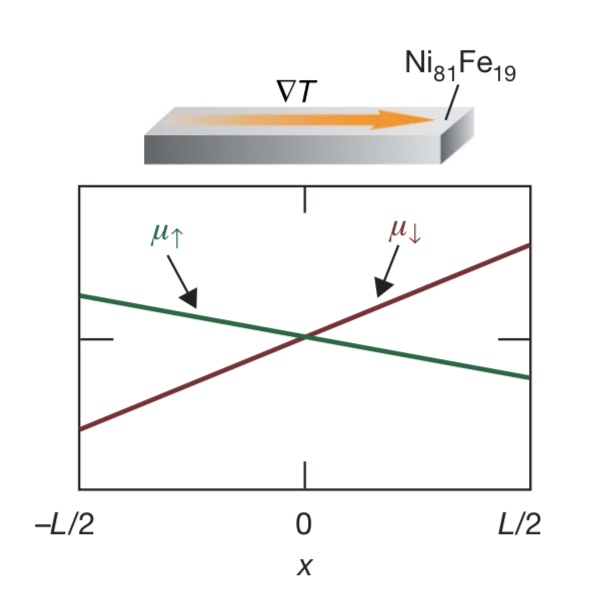}
		\caption{\small Spin Seebeck effect. (\textbf{a}): Electrons with spin up (red arrow head) and spin down (blue arrow tail) components have different Seebeck coefficients in a metallic magnet under a temperature gradient. (\textbf{b}): Illustration of the calculated distributions of electrochemical potentials for spin-up and spin-down electrons. Reproduced from \cite{Uchida2008}. 
		}
		\label{fig:SpinSeebeck0}
	\end{center}
\end{figure}
Similar to the Seebeck effect that two different conducting materials have two different Seebeck coefficients, the electrons with two different spin polarizations have different Seebeck coefficients. When a metallic magnet is experiencing a temperature gradient, these two spin polarizations generate different amount of flow, resulting in spin voltage $ \mu_\uparrow - \mu_\downarrow$ that is proportional to the applied temperature gradient as in the figure \ref{fig:SpinSeebeck0}. Thus the magnet acts as a thermocouple in the spin sector. 

Quantitatively, the electrochemical potential $\mu_\uparrow =\mu_0 + \delta \mu + \mu_e = \mu_\uparrow^c + e\phi  $ has the spin dependent chemical potential $ \mu_\uparrow^c$ that depends on the temperature and its density $n_\uparrow$. The gradient of $ \mu_\uparrow$ has three different contributions. The electrochemical potential for the spin down component $\mu_\downarrow =\mu_0 - \delta \mu + \mu_e = \mu_\downarrow^c + e\phi  $ can be described similarly.
\begin{align}
\vec \nabla \mu_{\uparrow, \downarrow} &= \left(\frac{\partial \mu_{\uparrow, \downarrow}^c}{\partial T}\right) \vec \nabla T + \left(\frac{\partial \mu_{\uparrow, \downarrow}^c}{\partial n_{\uparrow, \downarrow}}\right) \vec \nabla n_{\uparrow, \downarrow} + e \vec \nabla \phi  \;,
\end{align}
where the first term on the right hand side come from the density accumulation and the second term measuring the entropy contribution from the temperature gradient. Thus 
\begin{align}
\vec \nabla \delta \mu = - e S_s \vec \nabla T \;, \qquad S_s = \epsilon_s = - \frac{1}{e} \left(\frac{\partial \mu_{\uparrow}^c}{\partial T} - \frac{\partial \mu_{\downarrow}^c}{\partial T}\right) =  \epsilon_\uparrow - \epsilon_\downarrow \;,
\end{align}
where $\delta \mu = \mu_\uparrow -  \mu_\downarrow$ and $S_s = \epsilon_s $ is the spin Seebeck coefficient. The definition of the Seebeck coefficient (thermoelectric power) is given in \eqref{ThermoelectricPower}. Here we assume that the $ \vec \nabla n_\uparrow $ and $ \vec \nabla n_\downarrow $  decay within the spin diffusion length and do not contribute significantly, which were relevant for the study done in \cite{Uchida2008}.

\begin{figure}[h!]
	\begin{center}
		\includegraphics[width=0.5\textwidth]{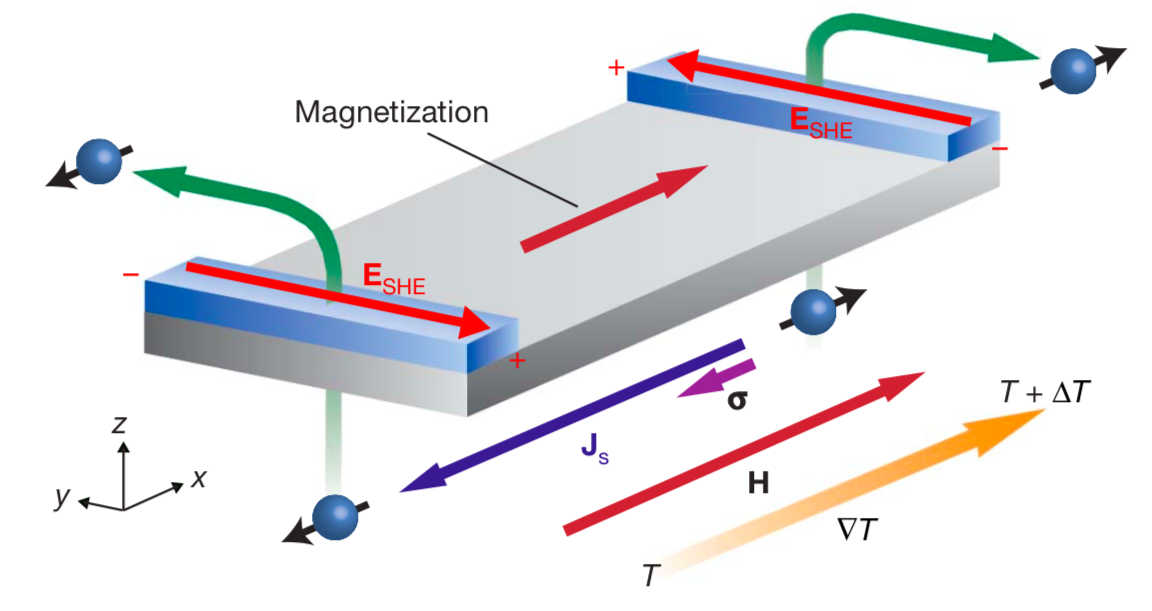} \quad 
		\includegraphics[width=0.4\textwidth]{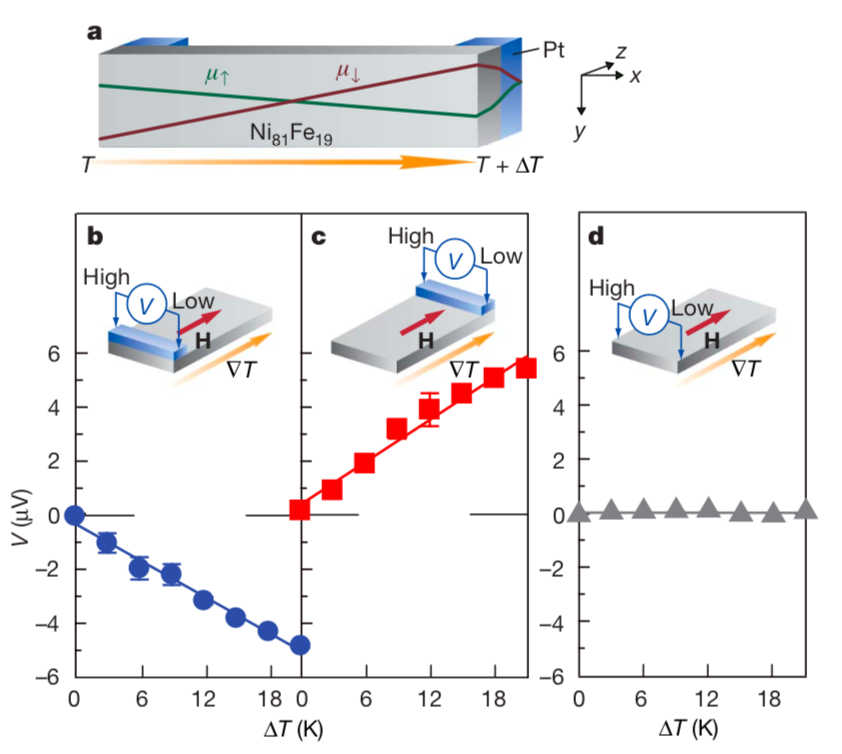} 
		\caption{\small Measurement of the spin Seebeck effect using inverse spin Hall effect. (\textbf{a}): The $Ni_{81}Fe_{19}$ film has $Pt$ wires attached to the ends of the film. 
			(\textbf{b}): Demonstration of spin Seebeck effect by the electromotive forces in the $Pt$ wires. Reproduced from \cite{Uchida2008}. 
		}
		\label{fig:SpinSeebeckInverseSH}
	\end{center}
\end{figure}

The measurement of the spin Seebeck effect utilize the inverse spin Hall effect that we discuss in \S \ref{sec:SpinHallTorque} \cite{Uchida2008}. The setup is depicted in the figure \ref{fig:SpinSeebeckInverseSH} (\textbf{a}). The temperature gradient $\Delta T$ and the in-plane magnetic field $\vec H$ are set in the $+\hat x$ direction. The electron spins are polarized along the magnetic field direction, and the Nernst effects that play roles with transverse magnetic field are suppressed. Electrons with spin ($+ \hat x$) polarized parallel to the magnetic field accumulate in the far end of the long side of the film, while those with spins anti-parallel ($- \hat x$) in the near end.   
Focusing on this near end, the in-film electromotive force \eqref{EMFSHE} due to the spin Hall effect vanishes because the current $\vec J_S$ is parallel to the spin polarization. This is demonstrated by the measurement of the potential without the $Pt$ wire in the inset {\bf d} of the figure \ref{fig:SpinSeebeckInverseSH} (\textbf{b}). Once the $Pt$ wire is placed  on top of the film, the accumulated electrons move up to the wire, produce the current $ \vec J_N \propto + \hat z$, and thus bring about the electromotive force $\vec E_{SHE} \propto \vec J_N \times \vec \sigma = \hat z \times (- \hat x) = -\hat y $, that confirms the figure \ref{fig:SpinSeebeckInverseSH} (\textbf{a}). The negatively charged electrons move against the direction of the electromotive force. As the temperature gradient is increased, there are more electrons accumulated and bigger current to produce more electric potential difference, which confirms the inset {\bf b} of the figure \ref{fig:SpinSeebeckInverseSH} (\textbf{b}). The sign of the electric voltage is opposite in the far end as demonstrated in the inset {\bf c} of the figure \ref{fig:SpinSeebeckInverseSH} (\textbf{b}).

One can evaluate the spin current $\vec J_S = \vec J_\uparrow - \vec J_\downarrow $ using the transport coefficients described in the previous section \S \ref{sec:SpinTransport}. For this purpose, we consider a non-zero heat current $\vec Q$ without electric current $\vec J_e=0$ along with no spin mixing $L_{ \uparrow \downarrow} = L_{ \downarrow \uparrow} =0$. Then, \eqref{TransportFormulaSpinComp}, for general $\vec Q$, has the form 
\begin{align} \label{SpinCurrentQ}
\left(\!\! \begin{array}{c}
0 \\
\vec J_S \end{array} \!\!\right) 
=
\left(  \begin{array}{ccc} 
L_{q \downarrow}+L_{q \uparrow}~~  & L_{ \uparrow \uparrow}+L_{ \downarrow \downarrow} & ~~L_{ \uparrow \uparrow}-L_{ \downarrow \downarrow}  \\
L_{q \downarrow}-L_{q \uparrow}~~  & L_{ \uparrow \uparrow} -L_{ \downarrow \downarrow} & ~~L_{ \uparrow \uparrow}+L_{ \downarrow \downarrow} \end{array} \right)
\left(\!\!\! \begin{array}{c}
- \vec \nabla T /T^2 \\
- \vec \nabla \bar \mu/T \\
- \vec \nabla \delta \mu/T \end{array} \!\!\! \right) \;,
\end{align}
There are two different cases that simplifies: (A) imposing a condition $ \vec \nabla \delta \mu =0$ or (B)  imposing a condition $ \vec \nabla \bar \mu =0$. First we consider the case (B) with $ \vec \nabla \delta \mu =0$. From the first equation of \eqref{SpinCurrentQ}, $\vec \nabla \bar \mu= - \frac{L_{q \downarrow} + L_{q \uparrow}}{L_{ \uparrow \uparrow} + L_{ \downarrow \downarrow}}  \frac{\vec \nabla T}{T} $. Plugging this into the second equation of \eqref{SpinCurrentQ}, we get
\begin{align}
\vec J_S^A = -2 \frac{L_{ \downarrow \downarrow} L_{q \downarrow} - L_{ \uparrow \uparrow} L_{q \uparrow}}{L_{ \uparrow \uparrow} + L_{ \downarrow \downarrow}}  \frac{\vec \nabla T}{T^2} =- \frac{2}{e} \frac{\sigma_ \uparrow \sigma_ \downarrow}{\sigma_ \uparrow + \sigma_ \downarrow} (\epsilon_ \uparrow - \epsilon_ \downarrow)  \vec \nabla T
=- \frac{\sigma}{2e} (1 -P_\sigma^2) S_s \vec \nabla T  \;. 
\end{align}	 
Here we use the notation used in \S \ref{sec:SpinTransport}, and $P_\sigma= \frac{\sigma_ \uparrow - \sigma_ \downarrow}{\sigma_ \uparrow + \sigma_ \downarrow} $ and $ \frac{2\sigma_ \uparrow \sigma_ \downarrow}{\sigma_ \uparrow + \sigma_ \downarrow} = \sigma ( 1-P_\sigma^2)/2$ with $\sigma = \sigma_ \uparrow + \sigma_ \downarrow$. This has been considered in \cite{Slachter2010}. The case (B) with the condition $ \vec \nabla \bar \mu =0$ can be similarly evaluated as 
\begin{align}
\vec J_S^B = -2 \frac{L_{ \downarrow \downarrow} L_{q \downarrow} + L_{ \uparrow \uparrow} L_{q \uparrow}}{L_{ \uparrow \uparrow} - L_{ \downarrow \downarrow}}  \frac{\vec \nabla T}{T^2} = \frac{2}{e} \frac{\sigma_ \uparrow \sigma_ \downarrow}{\sigma_ \uparrow - \sigma_ \downarrow} (\epsilon_ \uparrow + \epsilon_ \downarrow)  \vec \nabla T
= \frac{\sigma_s}{2e} \Big(\frac{1}{P_\sigma^2} - 1\Big) S_{tot} \vec \nabla T  \;,
\end{align}	 
where $ \sigma_s =\sigma_ \uparrow - \sigma_ \downarrow$ and $S_{tot}= \epsilon_{tot} = \epsilon_ \uparrow + \epsilon_ \downarrow$. One can also consider the isothermal case $\vec Q=0$ without restricting the electric current.   

\begin{figure}[h]
	\begin{center}
		\includegraphics[width=0.7\textwidth]{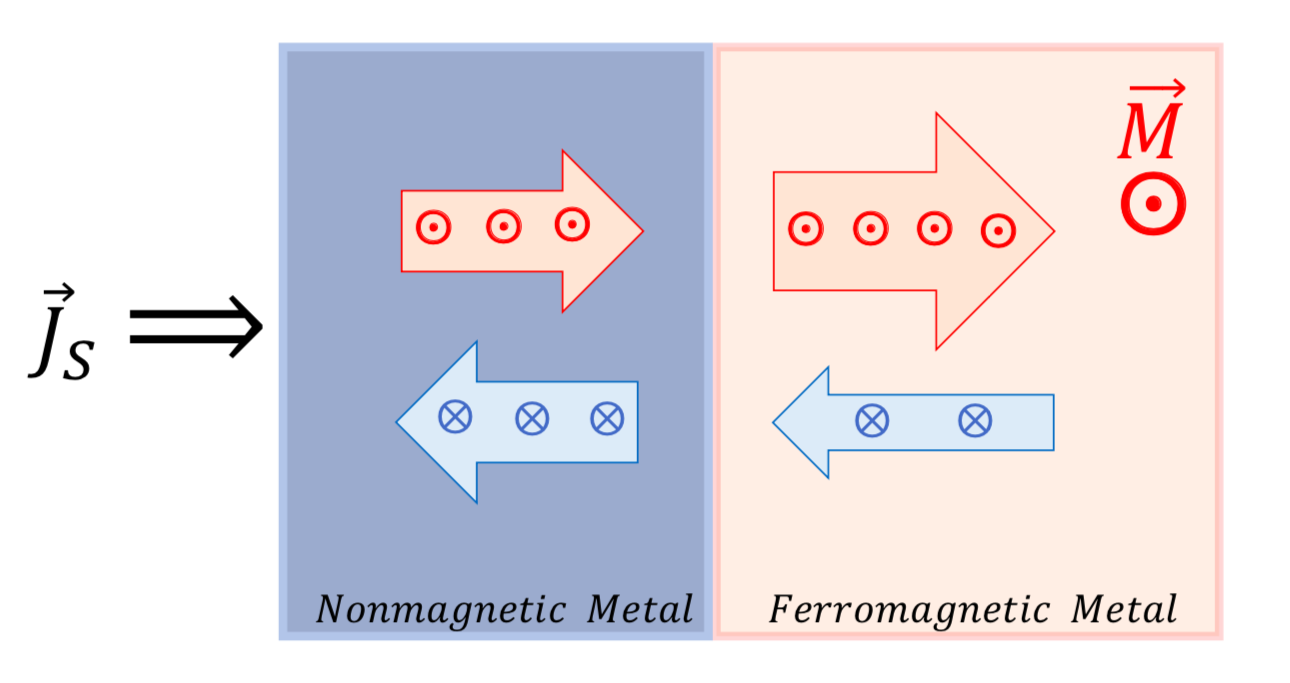}
		\caption{\footnotesize\small  Illustration of the spin Peltier effect. A spin current is push through a nonmagnetic metal and ferromagnetic metal interface. The Peltier heat current vanishes In the nonmagnetic metal while it is non-zero in the ferromagnetic metal because the Peltier coefficients are different for the majority and minority spins.  }
		\label{fig:SpinPeltier0}
	\end{center}
\end{figure}

The Peltier coefficient describes the amount of heat that is carried by an electrical current when it passes through a material. In the context of spin current, this can be made clear by using a pure spin current in a nonmagnetic material as illustrated in the left side of the figure \ref{fig:SpinPeltier0} \cite{Flipse2012}\cite{Gravier2006}. The electric current as well as the heat current vanish because up-spin electrons move toward right while the down-spin electrons move toward left.
The ferromagnetic metal serves as spin polarizer and provides better conductivity for the majority spin (up-spin depicted as a red arrow head). Thus the heat current for the majority spin is larger than that of the minority spin (down-spin depicted as a blue arrow tail), leading in a net heat current. 

More quantitatively, we can we derive the temperature gradient that develops in the ferromagnetic region when a spin current is accompanied by a charge current as in the figure \ref{fig:SpinPeltier0}. For simplicity, we assume that no heat can enter or leave the ferromagnetic metal ($\vec Q = 0$) and disregard Joule heating. Then the first line of \eqref{TransportFormulaSpinComp} gives 
\begin{align} \label{SpinPeltier11}
\vec \nabla T = -\frac{T}{e[\kappa + T \epsilon_\uparrow \epsilon_\downarrow \sigma] } \left(  [ \sigma_\uparrow \epsilon_\uparrow + \sigma_\downarrow \epsilon_\downarrow] \vec \nabla \mu_c - [\sigma_\uparrow \epsilon_\uparrow - \sigma_\downarrow \epsilon_\downarrow ] \vec \nabla \mu_s   \right) \;,
\end{align}
where $\sigma = \sigma_\uparrow + \sigma_\downarrow$, $\bar \mu = \mu_c = \mu_\uparrow + \mu_\downarrow$ is the charge part of the electrochemical potential, and $\delta \mu = \mu_s = \mu_\uparrow - \mu_\downarrow$ is the spin part of the electrochemical potential. With slightly different notation compared to the previous section, we set $  \sigma \epsilon =  \sigma_\uparrow \epsilon_\uparrow + \sigma_\downarrow \epsilon_\downarrow$ and thus $ \epsilon_{\uparrow, \downarrow} = \epsilon - (P_\sigma \mp 1) \epsilon_s/2 $. Then the spin term in \eqref{SpinPeltier11} gives the result 
\begin{align} \label{SpinPeltier22}
\vec \nabla T_s = \frac{T}{e[\kappa + T \epsilon_\uparrow \epsilon_\downarrow \sigma] } \left(\epsilon \sigma_s +  \frac{\sigma}{2} (1- P_\sigma^2) \epsilon_s \right) \vec \nabla \mu_s  \;,
\end{align}
where $\sigma_s = \sigma_\uparrow - \sigma_\downarrow$ and $\vec \nabla T_s$ is the temperature gradient due to the spin current. This result is apparently different from that of \cite{Flipse2012}. There are two equivalent, yet different looking, definitions we employed in the previous sections for the transport coefficients, for example \eqref{TransportFormula1D4} and \eqref{TransportFormula1D7}. Here we use \eqref{TransportFormula1D4} which is different from those in \cite{Flipse2012}.

\subsection{Magnon current \& Magnon Hall effect} \label{sec:MagnonHallEffect}

The spin Seebeck effect has been also observed in the magnetic insulator $ La Y_2 Fe_5 O_{12} $ \cite{Uchida2010} in the absence of conduction electrons and also in the semi-conductor $Ga_{1-s} Mn_s As $ \cite{Jaworski2010}. It turns out that the magnitude of the spin Seebeck effect is independent of the longitudinal charge current due to the absence of conduction electrons in the magnetic insulator \cite{Uchida2010}. This is also demonstrated by revealing that the measured electric potential in $Pt$ wire utilizing the inverse spin Hall effect is the same before and after scratching the $GaMnAs$ sample \cite{Jaworski2010}. These studies indicate that a thermally induced spin voltage, the spin Seebeck effect, is related to the magnetization dynamics through the magnons, quantized spin waves. 

\begin{figure}[h]
	\begin{center}
		\includegraphics[width=0.7\textwidth,height=0.23\textheight]{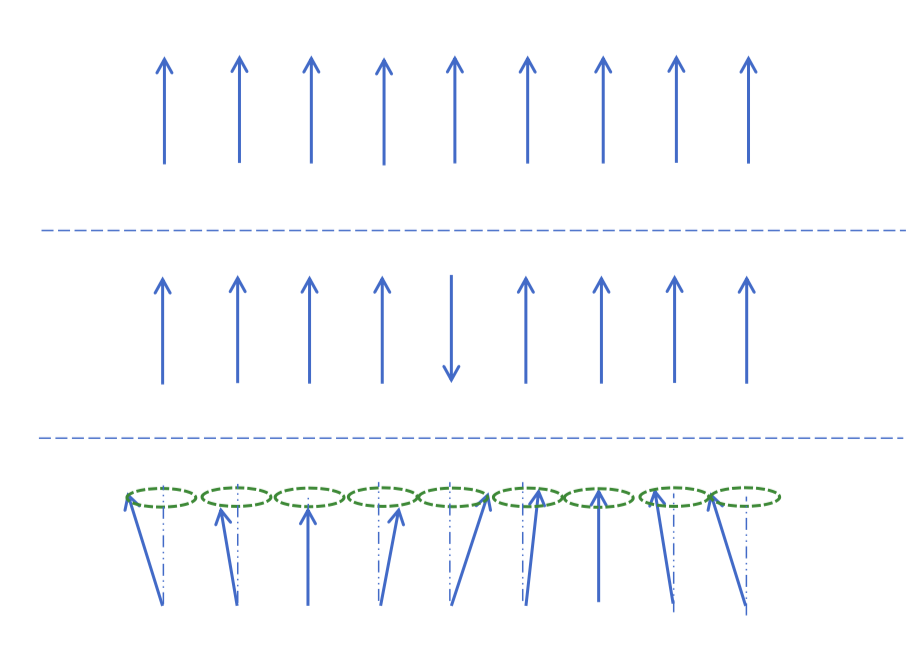} 
		\caption{\footnotesize\small Excitations of ferromagnet. (\textbf{a}): Top configuration representing the ground state of a magnet. (\textbf{b}): Middle spin configuration with a spin flip that represents a high energy excitation. (\textbf{c}): Bottom spins with a spin wave or quantized Magnon that has a lower energy than (\textbf{b}). The spins are precessing around their equilibrium positions. }
		\label{fig:Magnon2}
	\end{center}
\end{figure}
Magnons are low energy excited states in magnetic materials, and can be thought of as a wave of neighboring electron spins precessing at the same frequency but with a different phase, as shown in figure \ref{fig:Magnon2}. They arise because neighbouring electron spins interact strongly in a magnet, making it energetically favorable to excite the collective magnon mode as in figure \ref{fig:Magnon2} (\textbf{b}), rather than flipping a single spin as in \ref{fig:Magnon2} (\textbf{c}).

\begin{figure}[h]
	\begin{center}
		\includegraphics[width=0.57\textwidth]{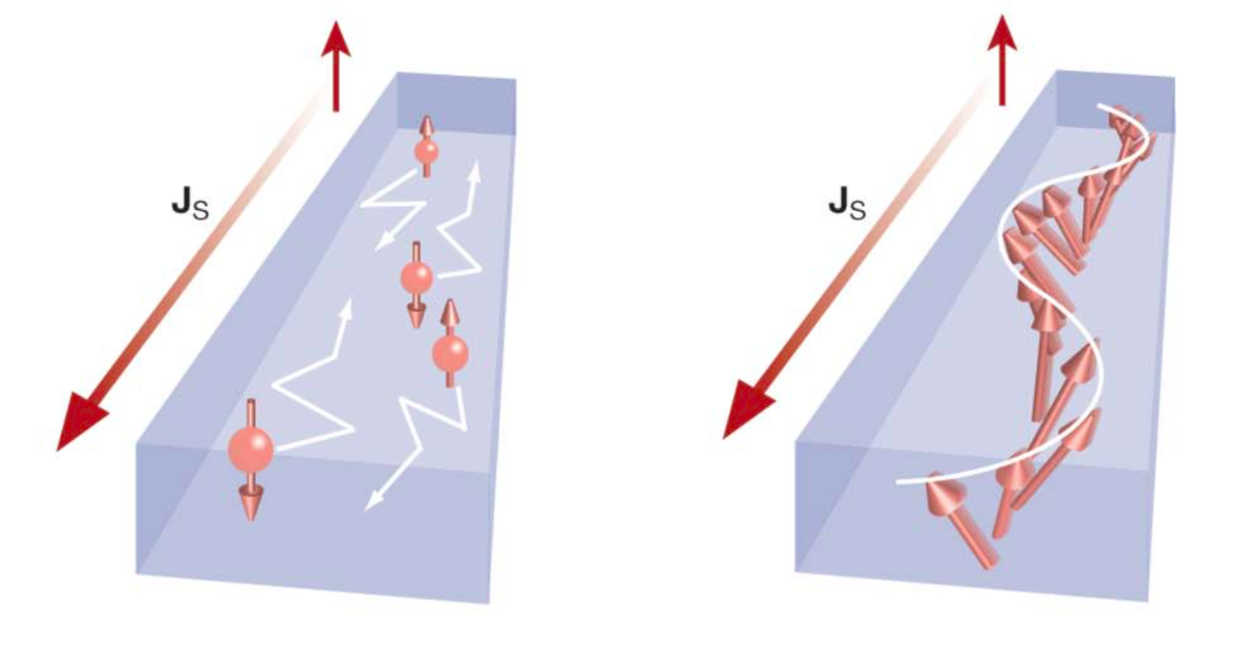} \quad 
		\includegraphics[width=0.37\textwidth]{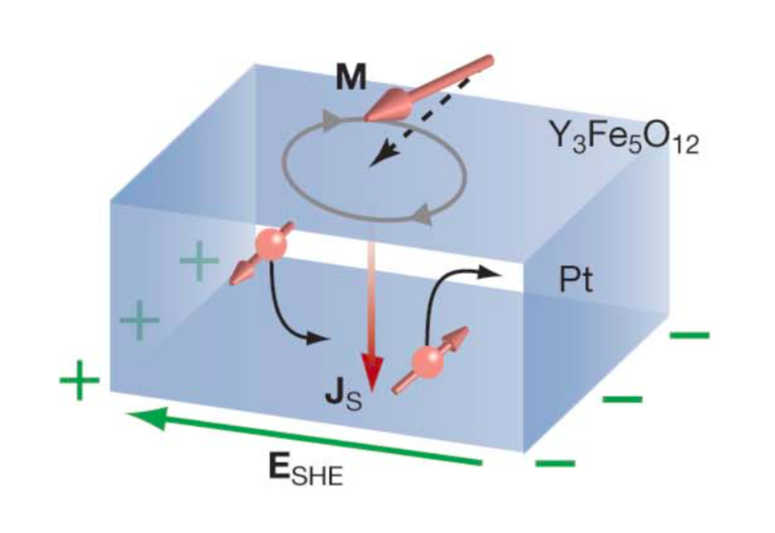}
		\caption{\footnotesize\small Two different spin currents. (\textbf{a}): Conduction electron spin current that is carried by the electron diffusion. (\textbf{b}): Spin wave spin current that is carried by collective magnetic-moment precession. (\textbf{c}): Spin pumping of the spin wave spin current generates electric potential through the inverse spin Hall effect.  Reproduced with permission \cite{Kajiwara2010}.  }
		\label{fig:SpinCurrentsTwoTypes}
	\end{center}
\end{figure}

In solid there are two different types of non-equilibrium spin currents as illustrated in the figure \ref{fig:SpinCurrentsTwoTypes} \cite{Kajiwara2010}. The first is the conduction electron spin current, the flow of net spin angular momentum that is carried by the conduction electrons as in the figure \ref{fig:SpinCurrentsTwoTypes} (\textbf{a}). The electrons with the same polarization move toward the direction of the spin current, while the electrons with opposite polarization move against the current direction.  
The second is the spin wave spin current. The spin angular momentum is carried by the spin wave, which is a low energy excited state that can be described by a collective magnetic moment precession as depicted in the figure \ref{fig:SpinCurrentsTwoTypes} (\textbf{b}). The spin wave is preferred when a system prefers low energy excited states compared to the spin flip that requires higher energy in the ferromagnetic materials. The spin wave spin current (Magnon current) has been shown to persist for much greater distances, especially in magnetic insulators, because its decay via conduction electrons is suppressed.   

There can be mutual conversion of the spin angular momentum between the conduction electron spins in the conducting layer and magnetization in the insulating ferromagnet layer. The spin wave spin current can be converted into electric current (voltage) by the spin pumping that is illustrated in the figure \ref{fig:SpinCurrentsTwoTypes} (\textbf{c}). The spin wave in the $Y_3 Fe_5 O_{12}$ layer transfer spins to the $Pt$ layer that generates the spin current $\vec J_S$, which is the flow of the electrons with the parallel spin with the magnetization of the ferromagnet $Y_3 Fe_5 O_{12}$ layer and opposite flow of the electrons with anti-parallel spins. Due to the spin Hall effect, these electrons in $Pt$ layer move toward right and create the electric potential. The inverse process, spin current of conduction electrons can be converted into spin wave by absorbing the spin current, which is the spin transfer torque in \S \ref{sec:STT0}. It is interesting to recognize that one can transfer electric signals through an insulating layer by devising a suitable multi layers system such as $Pt /  Y_3 Fe_5 O_{12} / Pt$ \cite{Kajiwara2010}. 

\begin{figure}[h]
	\begin{center}
		\includegraphics[width=0.35\textwidth]{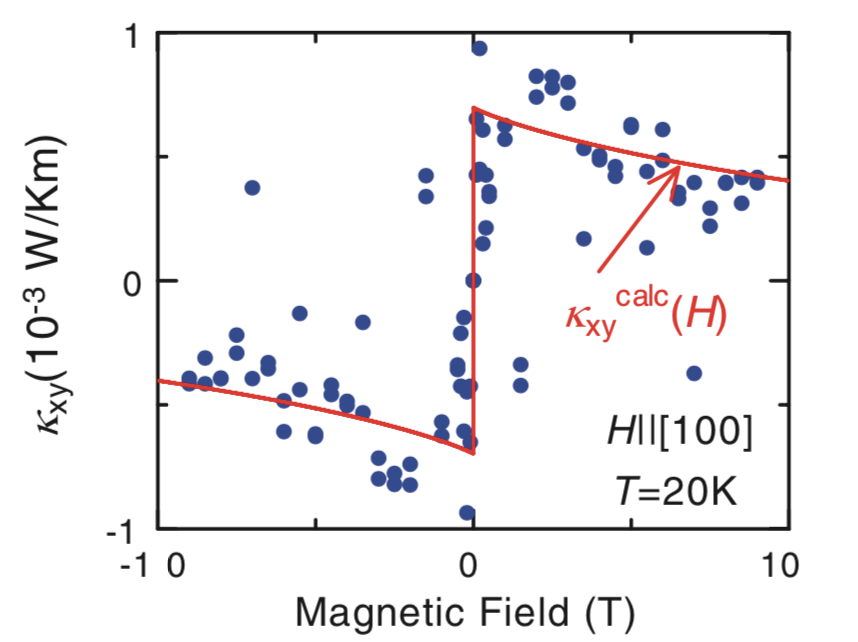}
		\caption{\footnotesize\small  Thermal Hall conductivity $\kappa_{xy}$ measurement of the $Lu_2 V_2 O_7 $ at the low temperature insulating phase. Reproduced with permission \cite{Onose2010}.  }
		\label{fig:MagnonThermalHall}
	\end{center}
\end{figure}

Under a temperature gradient, Magnons, the quanta of the magnetic excitations, diffuse from the hotter side to the colder side. In addition to this longitudinal motion, the transverse motion, Magnon Hall effect, exists and has been reported in \cite{Onose2010} based on the previous theoretical study \cite{Katsura2010}. Novel feature of the observation is its profile as a function of applied magnetic field: the thermal Hall conductivity steeply increases and saturates in the low magnetic field region. See figure \ref{fig:MagnonThermalHall}. This is not related to the normal Hall effect that is proportional to magnetic field, but the anomalous Hall effect that is affected by the spontaneous magnetization.  

\newpage 
\section{Skyrmion Hall Experiments} \label{sec:SkyrmionHallMeasurements}

One needs to be able to control the motion of the Skyrmions if one want to use them in any application. While Skyrmions are electrically neutral object, electric currents turn out to be useful to manipulate the Skyrmion motion due to the interaction between the spins of the Skyrmions and electrons. Majority of the Skyrmion experiments are focused on this aspects that is triggered by the experimental observation: Skyrmions can be moved by ultra low current density which is five or six order of magnitude smaller than that is necessary for domain wall (DW) motion in ferromagnets \cite{Jonietz2010}. Thus, we start with the electric Hall effect measurements.   
\begin{figure}[h]
	\begin{center}
		\includegraphics[width=0.75\textwidth]{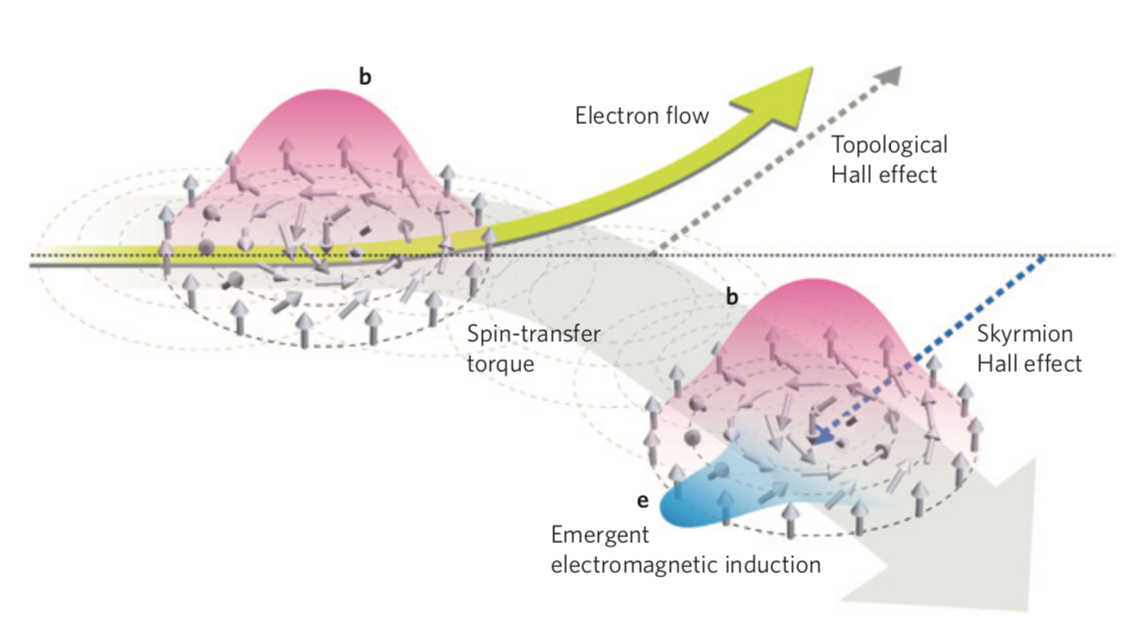}
		\caption{\footnotesize\small  A schematic figure that explains basic elements of the Skyrmions motion under the influence of electric currents. Reproduced with permission \cite{SkyrmionTopologicalReview}.  }
		\label{fig:SkyrmionTHE}
	\end{center}
\end{figure}

\subsection{Topological Hall effect} \label{sec:TopologicalHallEffect}

Mobile spins interact with the localized spins and vice versa. The extended spin configuration of the Skyrmions trades non-trivial physical effects with the conduction electrons through a coupling with the spins of the conduction electrons in conducting materials. For simplicity, we assume Hund’s rule coupling that the conduction electron spin is coupled strongly to, and forced to be parallel to, the localized spin at each atomic site \cite{SkyrmionTopologicalReview}. This produces several interesting physical consequences. For example, this constraint leads to the emergent electromagnetic fields $\vec e$ and $\vec b$ originating from the coupling of the Skyrmion spin textures to the conduction electrons as depicted in the Figure \ref{fig:SkyrmionTHE}. See also \cite{FerroCoupling}\cite{DMinteractionLGModel}\cite{HC4}\cite{FerroCoupling2}.
\begin{align}
e_i = E_i^e = \vec n \cdot (\partial_i \vec n \times \partial_t \vec n) \;, \qquad b_i = B_i^e = \frac{1}{2} \epsilon_{ijk} \vec n \cdot (\partial_j \vec n \times \partial_k \vec n) \;, 
\end{align}
where $i, j$ are spatial indices and  $\vec n = \vec M/M_s $. These emergent electromagnetic fields measure the sold angle for an infinitesimal loop in space time that is captured by the quantum mechanical Barry phase covered by $\vec n$. These effects are captured by the spin transfer torque we studied above in \S \ref{sec:STT0}. Of course, we also have mentioned these explicit higher derivative contributions in the spin torque considered above in \eqref{LLG-EMField}, which produce prominent effects on the magnetic metals. 

The conduction electrons moving to the right (namely along the $\hat x$) pass through the emergent magnetic field $\vec b = b \hat z$ in the figure \ref{fig:SkyrmionTHE}. The conventional Lorenz force makes the electrons bend toward $ \hat y$ direction. This electron flow is called the Topological Hall Effect. Experimentally, this effect is added to the conventional Hall effects due to the physical magnetic field $\vec B$ \cite{HC1}\cite{HC4}\cite{FerroCoupling2}\cite{HC2}\cite{HC3}\cite{HC5}. 

\begin{figure}[h]
	\begin{center}
		\includegraphics[width=0.35\textwidth]{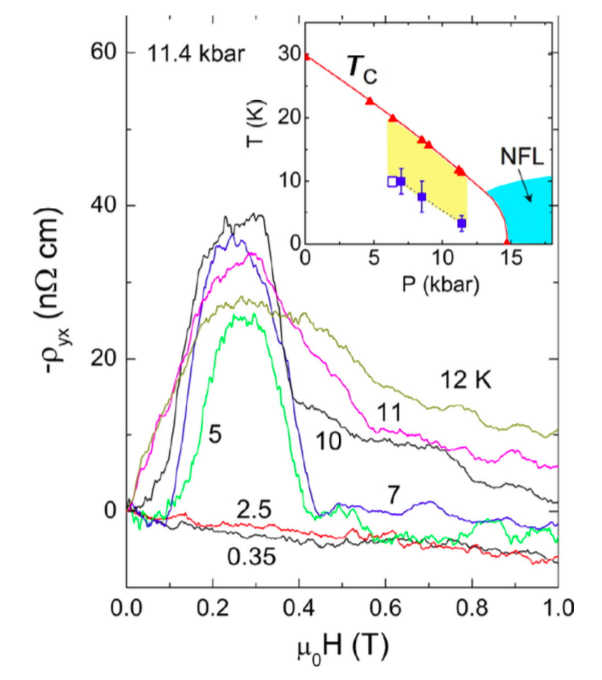} \qquad \qquad 
		\includegraphics[width=0.54\textwidth]{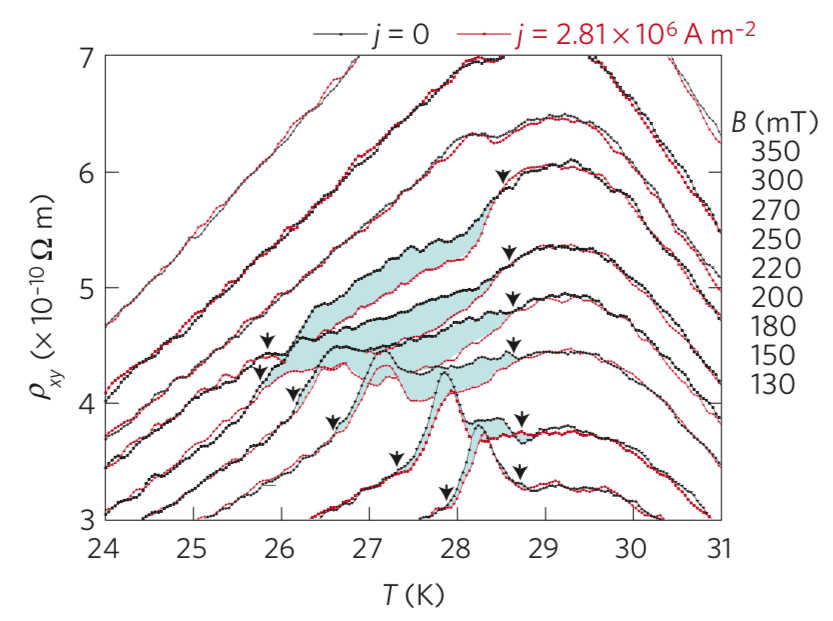}
		\caption{\footnotesize\small  Experimental data for the topological Hall effects. A box function-like contribution in the left figure exists only in the Skyrmion phase (\textbf{a}). This extra contribution to the electric resistivity confirms the existence of the emergent magnetic field $\vec b$ and thus the existence of the Skyrmions. (\textbf{b}) shows more systematic results of the box function-like contribution to the resistivity from the $\vec b$. The ranges of the magnetic field and temperature correspond to the Skyrmion phase. The results of the Hall experiments, of course, take into account of the additional applied magnetic field $\vec B$. Reproduced with permission \cite{HC1}\cite{HC4}.  }
		\label{fig:THE0}
	\end{center}
\end{figure}

Experimental data for the topological Hall effect are depicted in the figure \ref{fig:THE0}. A rectangular-shape function like contribution in the figure \ref{fig:THE0} (\textbf{a}) exists only in the Skyrmion phase. This extra contribution to the electric resistivity confirms the existence of the emergent magnetic field $\vec b$ and thus the existence of the Skyrmions. The ranges of the temperature and magnetic field are slightly different from those reported in \cite{SkyrmionExp2}. This is due to the applied pressure, depicted in the inset in the figure \ref{fig:THE0} (\textbf{a}), that shifts and expands the Skyrmion phase. The figure \ref{fig:THE0} (\textbf{b}) shows more systematic results of the rectangular-shape function like contribution to the resistivity in the presence of the field $\vec b$. The ranges of the magnetic field and temperature correspond to the Skyrmion phase. The results of the Hall experiments, of course, take into account of the additional applied magnetic field $\vec B$. See further details in the review \cite{SkyrmionTopologicalReview} and to find various other physical phenomena that can be described by these emergent gauge fields. 

\subsection{Skyrmion Hall effect} \label{sec:SkyrmionHallEffect}

The well known Hall effect describes the transverse motion of charged particles due to the Lorentz force, which provides a definite way to identify charge carriers in an electric wire. Recently, similar Hall effect for the topological Skyrmion charge $Q$ has been studied theoretically in \cite{FerroCoupling2} and demonstrated experimentally \cite{Jiang2017}\cite{Litzius2017}. These two papers use different spin torques to reveal the Skyrmion Hall effects: \cite{Jiang2017} uses spin Hall torque (SHT) of the current in the heavy metal layer and \cite{Litzius2017} uses spin-orbit torque (SOT) in the multilayer ferromagnetic system with in-layer current. 

As we described above, the Thiele equation \eqref{ThieleEQ} describes the translational motion of the center of a Skyrmion without macroscopic deformation of its structure. 
Following \cite{Jiang2017}, we use the modified version of the 2 dimensional in-plane Thiele equation \eqref{ThieleEQ2CC} that is introduced in the same section.  
\begin{align}\label{ThieleEQ2CC1}
\vec G  \times \vec v_d + \alpha \mathcal {D} \cdot \vec v_d + 4\pi  {\mathcal B} \cdot \vec J_{HM} = 0 \;,
\end{align}
where $\vec v_d= (v_{dx}, v_{dy})^T$ is an in-plane drift velocity of the Skyrmions, $\vec J_{HM}$ is the current in the heavy metal layer without the current in the ferromagnet layer $\vec v_s=0$, $\vec G = C \hat z$ related to the topological charge $C$, given in \eqref{TopologicalChargeDensity}, pointing out of the layer, and 
\begin{align}
\mathcal {D} = \left(\! \begin{array}{cc}
D_{xx} & D_{xy}  \\
D_{yx} & D_{yy}  \end{array}\! \right) \;, \qquad 
{\mathcal B} = \left(\! \begin{array}{cc}
B_{xx} & B_{xy}  \\
B_{yx} & B_{yy}  \end{array}\! \right) \;.
\end{align}
The dissipative force tensor $ \mathcal {D}$ is related to the damping torque. The tensor ${\mathcal B}$ quantifies the efficiency of the spin Hall spin torque over the 2-dimensional spin texture of the skyrmion and is in general non-zero \eqref{LLGEQ-SHE-Thiele}. The first term in \eqref{ThieleEQ2CC1} captures the Magnus force that results in transverse motion of the Skyrmion. This term thus acts equivalently to the Lorentz force for electric charge, and results in a Hall-like response of magnetic skyrmions. 

\begin{figure}[h]
	\begin{center}
		\includegraphics[width=0.55\textwidth]{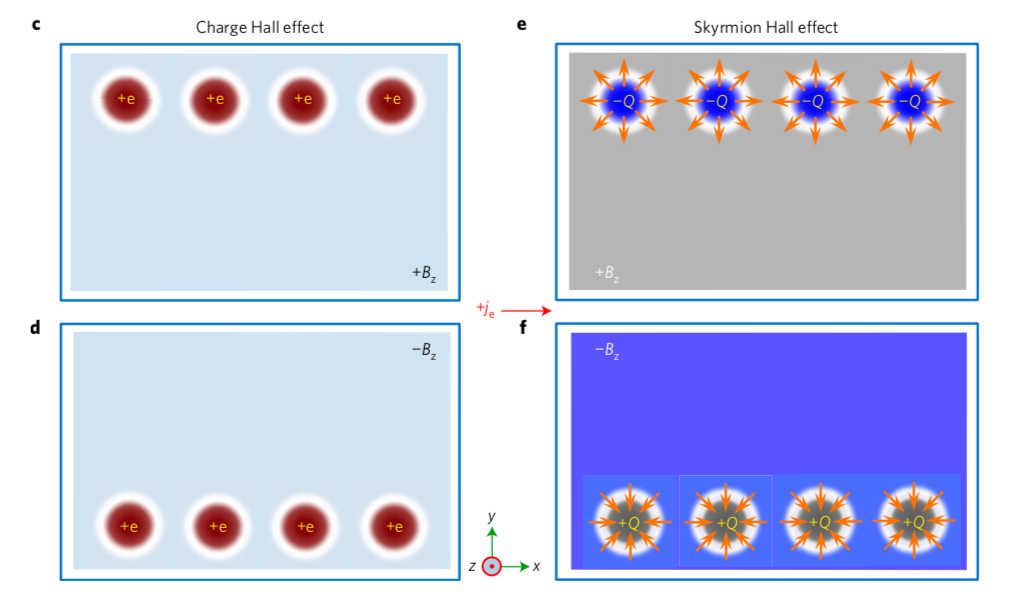} \qquad\qquad
		\includegraphics[width=0.24\textwidth]{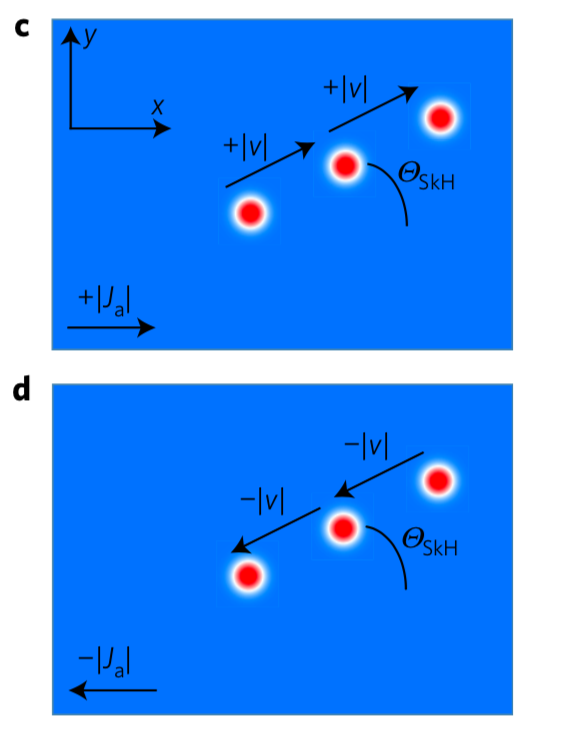}
		\caption{\footnotesize\small  (\textbf{a}): An experimental verification of the Skyrmion Hall effect through the accumulation of the Skyrmions in one side of the film.  c – d illustrate the Hall effect for holes with a unit electronic charge of $+e$, that accumulate at the opposite edges of a device upon reversal of the magnetic field directions. e – f illustrate the skyrmion Hall effect, for which the reversal of the magnetic field direction reverses the sign of the topological charge from $C = -1$ to $C = +1$ leading to the accumulation of skyrmions of opposite topological charges at opposite edges of the device. Reproduced with permission from \cite{Jiang2017}. 
			(\textbf{b}): Real time dynamic imaging of the motion of Skyrmions. It is clear to observe the longitudinal and transverse motion of Skyrmions and thus the Hall effect. The Hall angle $\Theta_{SkH}$ given in \eqref{SkyrmionHallAngle} can be read off directly. Reproduced with permission from \cite{Litzius2017}. }
		\label{fig:SkyrmionHallAngle}
	\end{center}
\end{figure}

We consider an isolated Neel type Skyrmion and the $x$-directional current $\vec J_{HM} = J_0 \hat x$. The Skyrmion is described by $\vec n = \sin \Theta (\rho) \hat \rho + \cos \Theta (\rho) \hat z$ in 2-dimensional space $ (\rho, \phi)$ with a diameter $d_N$ and the size of the domain wall $\gamma_0^{DW}$ \cite{Jiang2017}. Direct computations show that the off-diagonal components of the tensors $ \mathcal {D}$ and ${\mathcal B}$ \eqref{LLGEQ-SHE-Thiele} vanish: $ D_{xy} = D_{yx}=0$ and $B_{xy}=B_{yx} \propto \int_0^{2\pi} d\phi \sin \phi \cos \phi = 0$. Then it is straightforward to solve \eqref{ThieleEQ2CC1} to get $v_{dx} = -4\pi \alpha D B_0 J_0/(C^2 + \alpha^2 D^2) $ and $v_{dy} = 4\pi C B_0 J_0/(C^2 + \alpha^2 D^2) $, where $ D_{xx} = D_{yy}=D=\frac{\pi^2 d}{8 \gamma_0^{DW}}$ and $ B_{xx} = B_{yy}=B_0$. Thus the ratio of in-plane velocities and Skyrmion Hall angle can be written as 
\begin{align} \label{SkyrmionHallAngle} 
\frac{v_{dy}}{v_{dx}} = \frac{-C}{\alpha D} \;, \qquad \Theta_{SkH} = \tan^{-1} \left(\frac{v_{dy}}{v_{dx}} \right)\;.
\end{align}
The top right figure \ref{fig:SkyrmionHallAngle} demonstrates the Skyrmion Hall effect, acquiring the transverse component of velocity, under the electric current in the heavy metal layer. The Skyrmion Hall effect is further confirmed when the magnetic field is reversed. The sign of the topological charge is reversed, which also causes the change in the direction of the transverse velocity. This happens because the topological charge is odd function of the magnetization $\vec n$ which is clear in \eqref{TopologicalChargeDensity}. 

In \cite{Litzius2017}, similar Skyrmion Hall effect is precisely measured at room temperature with the precise angle with a low pinning stack on thin multilayers of $ [Pt (3.2 nm) / CoFe (0.7 nm) / MgO (1.4 nm)]_{15} $. Direct dynamics imaging of the displacement in real time has been taken and has revealed the dynamic velocities and displacement direction. As mentioned in \eqref{SpinTorque-SOTFL} in \S \ref{sec:SpinOrbitTorque}, field like SOT plays an important role for Skyrmion motion. The Skyrmion mass scale is inversely related with the rigidity of the spin structure, which turns out to be very high due to the strong Dzyaloshinskii–Moriya interaction (DMI). Thus the effective mass of the Skyrmion is small and the velocity of the Skyrmions that follows the ac current without any noticeable delay. Thus this material exhibits excellent homogeneity and a very low pinning effect. 

The Skyrmions move very efficiently with the velocities reaching those predicted for perfect pinning-free systems. With the real-time dynamic imaging, it was possible to actually measure the Hall angle $\Theta_{SkH}$ given in \eqref{SkyrmionHallAngle} as in the figure \ref{fig:SkyrmionHallAngle} (\textbf{b}). As we see in the figure, the Skyrmions move along the direction of the current, meaning that move against the flow of the conduction electrons, which is the characteristic behavior for the left-handed Neel domain walls (DW) and Skyrmions as we demonstrated in the figure \ref{fig:SOT-FL} of \S \ref{sec:SpinOrbitTorque}.    

\subsection{Skyrmion Seebeck effect in Insulating magnets}

In addition to the discovery of the Skyrmion crystals in metallic materials, Skyrmion crystals have been also discovered in insulating materials, such as $Cu_2 O Se O_3$ \cite{Insulating1}\cite{Rotation1} and $Ba Fe_{1-x-0.05} Sc_x Mg_{0.05} O_{19}$ \cite{Yu2012}. 

One clear advantage of the Skyrmions in insulating materials is the absence of dissipation due to the conducting current. While the Skyrmion Hall effect has been experimentally measured as discussed in \S \ref{sec:SkyrmionHallEffect}, understanding the motion of the Skyrmions in insulating materials shed insights on their interactions with Magnons, the low energy excitations of magnetization. Here are consider the {\it numerical studies} of the Skyrmion Hall effect in the {\it insulating magnets} \cite{Kong2013}\cite{Lin2014}\cite{Kovalev2015} based on recent development of the spin Seebeck effect \S \ref{sec:SpinSeebeck}. 

\begin{figure}[h]
	\begin{center}
		\includegraphics[width=0.43\textwidth]{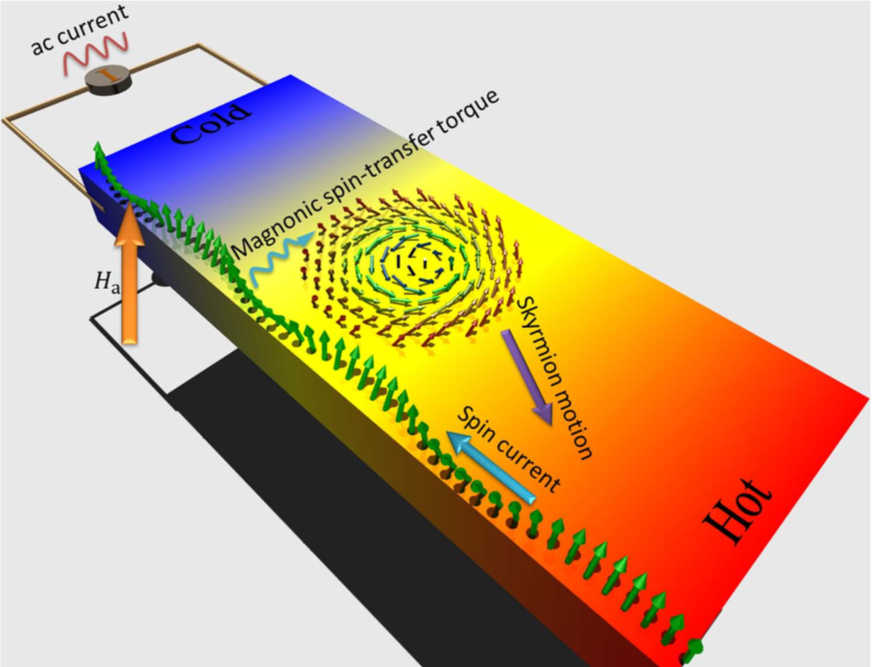} \qquad 
		\includegraphics[width=0.35\textwidth]{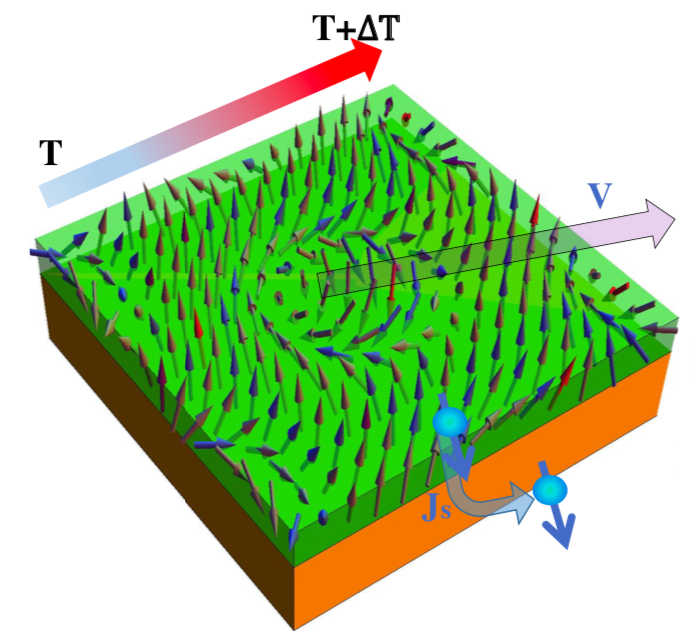} 
		\caption{\footnotesize\small Numerical studies for the Skyrmion motion in the insulating magnets. (\textbf{a}): Skyrmions move from the cold to the hot region, while Magnons diffuse from the hot to the cold region. The Skyrmions exhibit the transverse motion in addition to the longitudinal motion. Reproduced with permission from \cite{Lin2014}. (\textbf{b}): An additional non-magnetic layer, such as $Pt$, can be used to detect the spin pumping resulting from Skyrmion motion. Due to mostly out-of-plane magnetization configuration of the ferromagnet, the ordinary spin Seebeck effect would be suppressed. Reproduced with permission from \cite{Kovalev2015}. }
		\label{fig:InsulatingSkyrmionSeebeck}
	\end{center}
\end{figure}

The numerical studies \cite{Kong2013}\cite{Lin2014}\cite{Kovalev2015} reveal that the Skyrmions move towards the high temperature region as illustrated in the figure \ref{fig:InsulatingSkyrmionSeebeck} (\textbf{a}). This is against our usual intuition based on the Brownian diffusion, particles like electrons move toward the colder regions as discussed as Seebeck effect \S \ref{sec:SeebeckPeltierThompson}. See the figure \ref{fig:Seebeck}. 
To understand this counterintuitive Skyrmion motion, a Magnon assisted theory is employed in \cite{Kong2013}. The result shows that each Magnon carries spin one polarized anti-parallel to the equilibrium direction. Under a temperature gradient, Magnons response more easily and diffuses towards the cold region. Due to the anti-parallel alignments between the Magnon spins and the spins of Skyrmions, the Magnon current provides a negative transfer torque on the Skyrmions, which moves in opposite direction to conserve the total angular momentum. The spin waves in the context of a Skyrmion crystal has been studied in \cite{Petrova2011}\cite{Mochizuki2011}. 

The Magnon quanta has a small wavelength even at low temperature and can be treated as a particle when interacts with Skyrmions. Thus the strong coupling between the Skyrmions and Magnons in insulating magnets has similarities with the Hunt rule coupling between the Skyrmions and conducting electrons in conducting magnets. Thus the formulation due to Zhang-Li \cite{ZhangLi} can be applied to the interaction between the Skyrmions and Magnons \cite{Kovalev2012}. In particular, the above numerical results for the counterintuitive Skyrmion motion can be understood in terms of the analytical Thiele equation \eqref{ThieleEQ2CC} with some changes. We separate the magnetization $\vec m = \vec m_0 + \delta \vec m$ and identify the slow component $\vec m_0$ as the Skyrmion motion and the fast component $\vec \delta m$ as the Magnon motion.  
\begin{align}\label{ThieleEQMagnonSkyrmion}
\vec G \times (\vec J_m^s - \bb{s} \vec v_d) + \eta_{ \bb{s}} (\beta \vec J_m^s - \alpha  \bb{s} \vec v_d) = 0 \;, 
\end{align} 
where $  \bb{s} = M_s/\gamma_0$ is the renormalized spin density with saturation spin density $M_s$, $ \vec J_m^s = - \hbar \vec J_s$ the spin current with polarization along the slow magnetization $\vec m_0$ carried by Magnon, $\eta_{\bb{s}} = \eta_\mu^{\bb{s}} = \int d^2 r (\partial_\mu \vec m_0)^2/4\pi $, for $\mu = 0, 1, 2$, the form factor of Skyrmions, and $\vec v_d$ is the drift velocity of the Skyrmions. 
Solving this equation \eqref{ThieleEQMagnonSkyrmion} gives 
\begin{align}
v_x^d = \frac{G^2+\alpha\beta \eta_{\bb{s}}^2 }{\bb{s} (G^2+ \alpha^2\eta_{\bb{s}}^2)}  J_m^s \;, \qquad\quad 
v_y^d = \eta_{\bb{s}} G \frac{\alpha - \beta  }{\bb{s} (G^2+ \alpha^2\eta_{\bb{s}}^2)}  J_m^s \;.
\end{align}	 
Now the Skyrmion will move towards the hot region. \eqref{ThieleEQMagnonSkyrmion} demonstrate that $\vec v_d$ should have the same sign as $\vec J_m^s$, which is opposite the direction of the current $\vec J_s$ given the parameters $ G, \eta_\mu^{\bb{s}}$ of order $1$ and $\beta \sim 1.5 \alpha > 0$ \cite{Kovalev2015}. While the direction of the longitudinal motion of the Skyrmion is independent of the sign of $G$ and the relative magnitude of the parameters $\alpha$ and $\beta$, the direction of the Hall effect, the transverse motion, reverses when the sign og $G$ and $\alpha - \beta$ changes. 
It is proposed to detect temperature induced Skyrmion dynamics by employing spin pumping into the neighboring non-magnetic metallic layer, such as $Pt$ \cite{Kovalev2015}. The ordinary spin Seebeck effect should be suppressed since the polarization of the moving electron in the $Pt$ layer is parallel to the magnetization of the ferromagnet layer which is pointing out of plane. See the figure \ref{fig:InsulatingSkyrmionSeebeck} (\textbf{b}). Tunable Magnon thermal Hall effect in Skyrmion crystal of Ferrimagnets in the vicinity of the angular momentum compensation point has been proposed recently \cite{KimSK2018}.

\subsection{Skyrmion rotations and Magnon Hall effect} \label{sec:SkyRotationMagnonHall}

\begin{figure}[h]
	\begin{center}
		\includegraphics[width=0.3\textwidth]{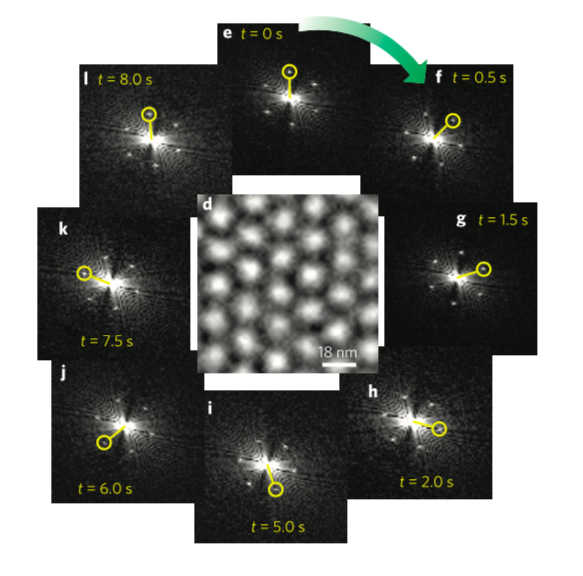} \qquad 
		\includegraphics[width=0.6\textwidth]{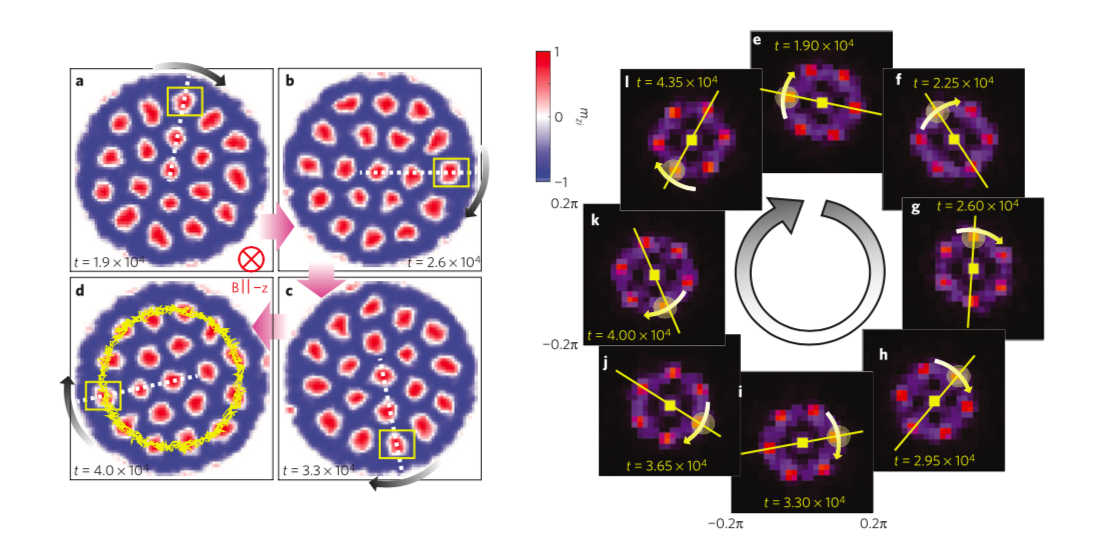} 
		\caption{\footnotesize\small (\textbf{a}): Observed Lorentz Transmission electron microscopy image of the Skyrmion rotation motion in the chiral-lattice magnets $MNSi$.  Skyrmions show clockwise rotation under the temperature gradient along with the magnetic field pointing into the page. (\textbf{b}): Simulation of the thermally driven rotational motion of the Skyrmion microcrystal. Reproduced with permission from \cite{ThermalHC1}. }
		\label{fig:SkyrmionRotation}
	\end{center}
\end{figure}

Our final topic for the review is the experimental realization of the rotational motion of the Skyrmion and the topological Magnon Hall effect in the presence of temperature gradient. Rotation motion of Skyrmions happens universally under the temperature gradient in the conducing conducting materials such as $MnSi$ \cite{Jonietz2010}\cite{ThermalHC1}  and insulating materials $ Cu_2 O Se O_3$ \cite{Rotation1}\cite{ThermalHC1}. See the figure \ref{fig:SkyrmionRotation}. In numerical studies \cite{ThermalHC1}, the rotation motions of the Skyrmions are systematically studied, for both conducting and insulating materials, to find that the rotational motion is driven purely by the thermal gradient. 

The universal rotation turns out to be explained by the interaction between the Skyrmions and Magnons. Magnons can be treated as particles, and thus the rotational motion of the Skyrmion interacting with the Magnons can be understood similarly. To include the rotational motion and the interaction with the Magnons, one can consider \cite{ThermalHC1}
\begin{align}
\vec n (t, \vec x) = \vec n_0 (\vec x - \vec R(t)) + \vec n_1 + \delta \vec n \;,
\end{align}	
where the magnetization $\vec n$ is separated into the slow part $ \vec n_0 $ for the Skyrmion from the fast part $ \delta \vec n $ for Magnon. $\vec R(t)= (X(t), Y(t))$ is the coordinate for the center of the Skyrmion. The remaining middle part $  \vec n_1$ captures the deformation of a moving Skyrmion that is proportional to $\dot {\vec R}(t)$. Once we include the second term, the Thiele equation has another contribution that is proportional to the mass $\mathcal M = \frac{1}{2} \int d^2 x \vec n_0 \cdot \left[ ({\partial \vec n_1}/{\partial \dot R_i}) \times ({\partial \vec n_0}/{\partial \dot r_i}) \right]$ multiplied by $\dot {\vec R}(t)$. Then, focusing on the rotational motion, the analogue of the Thiele equation \eqref{ThieleEQ2CC01} becomes \cite{ThermalHC1}
\begin{align}\label{ThieleEQ2CCRot}
&\mathcal{M} \ddot{X} + G (\dot Y +J_y^{M} ) + \alpha D \dot X =  - \frac{\partial U}{\partial X} \;, \\
&\mathcal{M} \ddot{Y} - G (\dot X +J_x^{M} ) + \alpha D \dot Y =  - \frac{\partial U}{\partial Y} \;,
\end{align} 
where the force has the form $\vec F = (F_x, F_y) = (- G J_x^{M} - \frac{\partial U}{\partial X}, G J_y^{M} - \frac{\partial U}{\partial X}) $ with the reaction force $\vec F_M$ due to the Magnon currents $ \vec J^M$ and the other force $\vec \nabla U$ due to the repulsion of Skyrmions from disk edge, magnetic fields and impurities \cite{ThermalHC1}\cite{FerroCoupling2}. Under the temperature gradient along the radial coordinate $r$, the skew scattering of the Magnons off Skyrmions gives rise to the topological Magnon Hall effect. 
\begin{align}
J^M_r = \kappa_{xx}^{Magnon} \left(-\frac{dT}{dr}\right) \;, \qquad  J^M_\theta = \kappa_{xy}^{Magnon} \left(-\frac{dT}{dr}\right) \;.
\end{align}	 
The latter is transverse to the temperature gradient, corresponding to counter-clockwise rotation of Magnon gas. Conservation of momentum makes the Skyrmions rotates along the clockwise direction. 

Theoretical computation of the thermal Hall conductivity of the Magnon under the influence of the emergent magnetic fields due to the Skyrmion crystals in insulating magnets \cite{ThermalHC2}. Skyrmion rotation under the influence of conduction electrons through the spin transfer torque was studied previously \cite{Everschor2011}.

\section{Outlook} \label{sec:Outlook}

Hall effects of magnetic Skyrmions are essential part of their motion due to the broken parity invariance. Physical situations are much more complicated due to the existence of other ubiquitous players such as conduction electrons in metallic magnets and low energy spin waves (Magnons) in insulating magnets. On the other hands, these electrons and Magnons provide essential tools for us to control the motion of the magnetic Skyrmions. Extensive theoretical and experimental studies of chiral magnetic Skyrmions for almost a decade since the first observation of the Skyrmions have provided large amount of information and corresponding understanding of the nature of Skyrmions as reviewed in Part \ref{sec:PartII}. In particular, direct observation of Skyrmion Hall and topological (Magnon) Hall effects has been a central piece of our understanding of the Skyrmions motion in conducting (insulating) materials, that has required so many different technological innovations in addition to the newly developed physical knowledge.

Now after the extensive and exciting studies, do we feel that we are close at getting a full picture of the Skyrmion physics? The complex and extended nature of Skyrmions that involve many spins collectively demands guides from top down first principle methods not to mention extensive phenomenological approaches in both theoretical and experimental sides. This brings the subject of Part \ref{sec:PartI}. Quantum field theory Ward identities are first principle methods that relates various physical quantities in a highly non-trivial way and has been extensively studied in the context of Quantum Hall systems. When the Ward identity methods is generalized to integrate the topological charge of Skyrmions, the identities inform us that there is a new physical quantity that can play an important role in Skyrmion dynamics, especially the Hall viscosity. This Hall viscosity has attracted much attention in Quantum Hall systems, yet never measured. Our Ward identities have provided us simple ways to confirm the existence of the Hall viscosity in terms of Hall conductivities as discussed in this review. After spending time trying to put the Part \ref{sec:PartII} together in a coherent fashion, there are a few concrete things that come into our mind. 

In an elementary level, there is a parallel story between the Skyrmions in conducting magnet and those in insulating magnet. The Skyrmions in conducting magnet mainly interact with the conduction electron, while those in insulating magnets with Magnons. The low energy spin wave excitations, particle-like Magnons, exist in both materials. Thus one can not completely ignore the Magnon contributions even in conducting materials. Thus the Skyrmions in the insulating materials provide a simpler and clearer physical picture for Skyrmion motion. Recent progress in controlling Magnon dynamics add more leverages toward this avenue, in addition to the known advantages that there is no dissipation due to the conduction electrons. With suitable conditions, one can consider the conservation of Magnon current similar to the conservation of electric current. Then the Ward identity approach outlined in Part \ref{sec:PartI} can be applied with suitable modifications. This seems to be a promising future avenue.   

While we try to organize various spin torques that we have considered in Part \ref{sec:PartII}, the effective field description proves to be useful. With that, we are able to organize the spin torques into two categories: one with derivative of magnetization and the other that is proportional to various currents. Spin transfer torque belongs to the former, and its effects only operate on the portion of the Domain Walls or the internal structures of the Skyrmions without affecting the ferromagnet backgrounds. The other tow torques, the spin-orbit and spin Hall torques, belong to the latter, and they operate uniformly on the ferromagnet layer whether the driving currents are in the same ferromagnetic layer or in a different layer. It is an interesting question whether there are other types of spin torque that can drive Skyrmions. It would be also interesting to understand the Magnon spin torque acting on Skyrmions in a systematic way. See a recent study on Magnon and Skyrmion scattering \cite{MagnonScattering}.  

It is interesting to find that there are systems with Skyrmions that have very low pinning or impurity effects \cite{Litzius2017}\cite{Impurity}. As discussed in \S \ref{sec:MagnonHallEffect}, the multilayer $Pt (3.2 nm) / CoFeB (0.7 nm) / MgO (1.4 nm)$ materials is homogeneous and show very low pinning so that Skyrmion motions follow the velocity profile of perfectly pinning free systems. There has been also numerical studies that support this observations. Unlike to the case of domain wall, the Skyrmion motion driven by the spin transfer torque is unaffected by impurities due to the Magnus force and flexible shape-deformation of individual Skyrmions, which enable them to avoid pinning centers \cite{Impurity}. These results allow us to apply the Ward identities without considering the impurity or pinning effects that has been discussed in \S \ref{sec:WIInsulating} \cite{Kim:2015qsa}. This low pinning system $Pt (3.2 nm) / CoFeB (0.7 nm) / MgO (1.4 nm)$ seems to be an ideal candidate for future study of Hall viscosity. A simple way to verify the Hall viscosity has been discussed for the insulating Skyrmions in \S \ref{sec:WIInsulating} and for the conducting materials in \S \ref{sec:HallViscosityConductor} based on \cite{Kim:2015qsa}. It will be interesting to make the experimental setup more concretely and to estimate the magnitude of the Hall viscosity.

\begin{figure}[h]
	\begin{center}
		\includegraphics[width=0.45\textwidth]{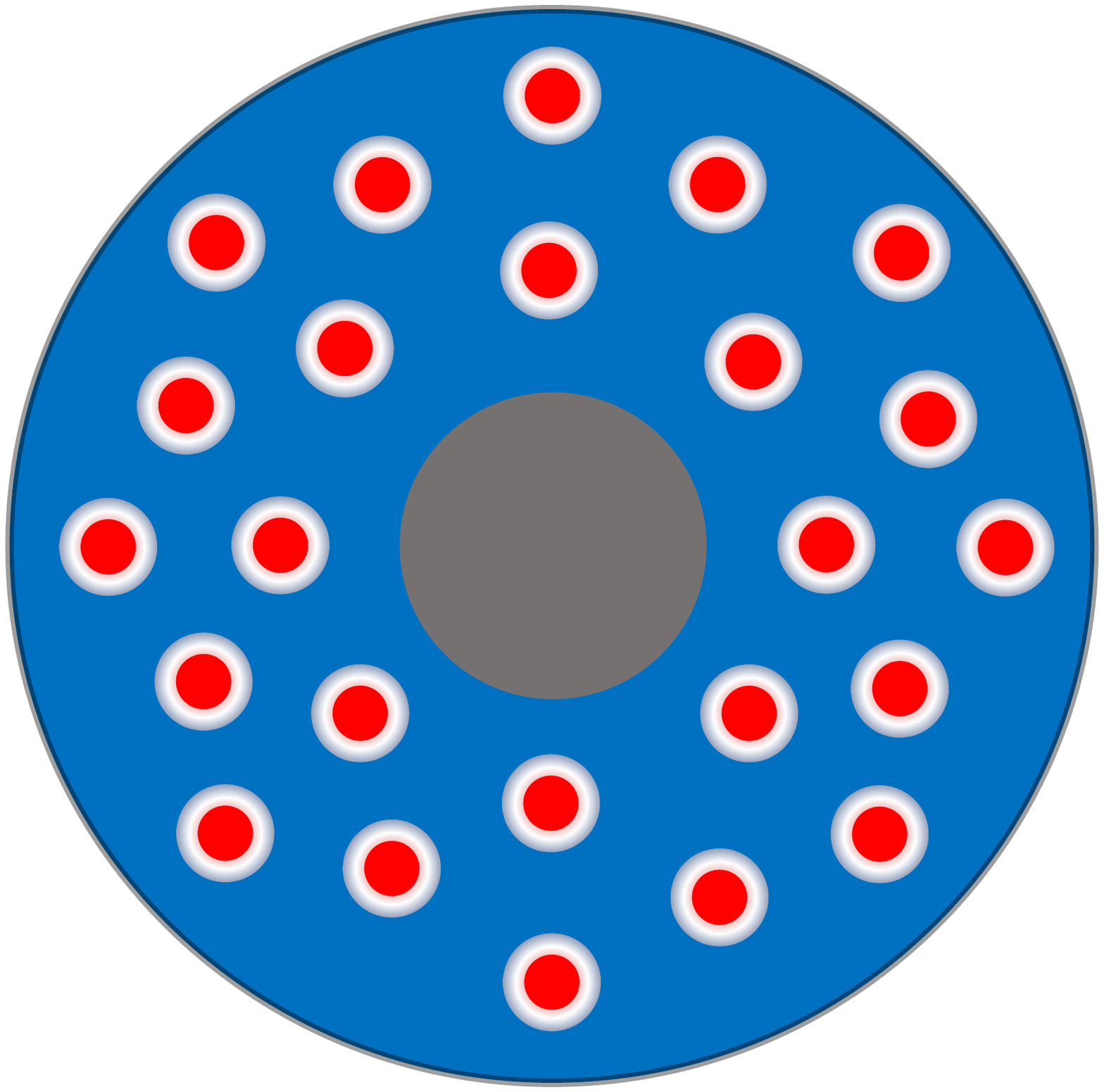} \qquad 
		\includegraphics[width=0.45\textwidth]{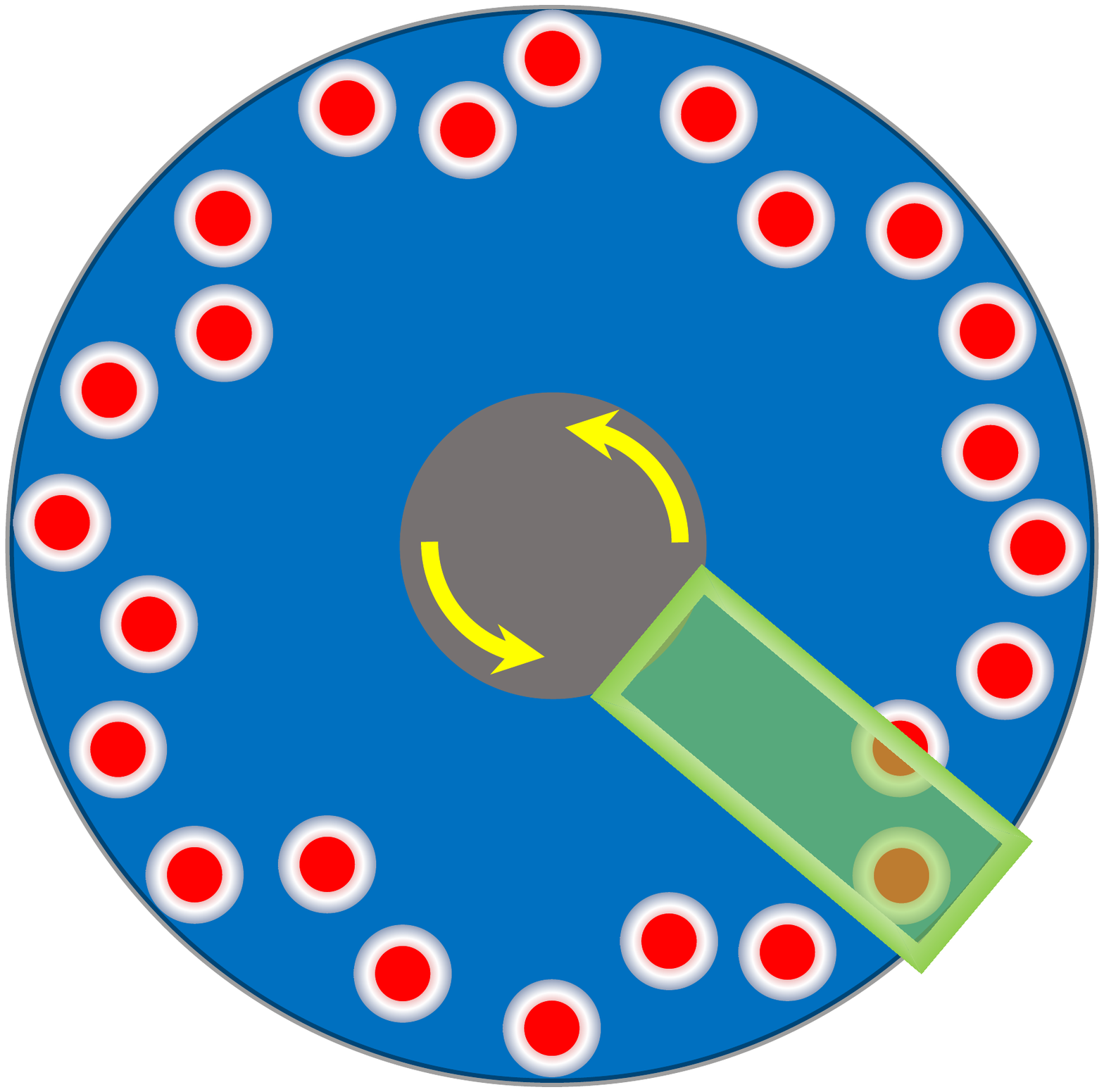} 
		\caption{\footnotesize\small Illustration of possible measurement of the Hall viscosity of Skyrmion system in contact with a rotating cylinder, with neither electric current nor temperature gradient. (\textbf{a}): Initial setup. (\textbf{b}): the rotation of the cylinder contacting with the ferromagnet will produce the radially outward motion of Skyrmions. This produce the spin imbalance that can be measured by an overlaid non-magnet metal through inverse spin Hall effect.}
		\label{fig:SkyrmionHallViscosity}
	\end{center}
\end{figure}

Finally, we would like to speculate a possible way to measure Hall viscosity in Skyrmion systems neither with an electric current nor with a temperature gradient. As discussed in \S \ref{sec:HallViscosity}, the Hall viscosity is transverse and dissipationless component of the shear tensor given in \eqref{HallViscosityTensor2}. Pictorially the direction of the Hall viscosity is given in the figure \ref{fig:HallViscosityDroplet} (\textbf{b}). Thus we can think about a similar experimental setup with an insulating magnet at a very low temperature that suppresses thermal excitations. Possible Magnon contributions can be isolated as we see below. Of course there is no conduction electrons in the system. The initial setup is depicted in the figure \ref{fig:SkyrmionHallViscosity} (\textbf{a}). The Skyrmions are populated with an appropriate magnetic field and temperature. Blue color represents spin up background of the ferromagnetic material. Red blobs are the Skyrmions. In the middle, there is a solid object that can be rotated.  

When the object at the center rotates, the Skyrmions move radially inward or outward depending on the rotation direction due to the Hall viscosity. See the figure \ref{fig:SkyrmionHallViscosity} (\textbf{b}). This will produce the spin imbalance between the inner boundary and the outer boundary. These spin imbalance can be measure by the inverse spin Hall effect by attaching a non magnetic layer, for example a $Pt$ layer that is represented as a green rectangle. Because majority of the spin background in the ferromagnetic layer is up spin, the Magnon contribution for the inverse spin Hall effect of the transport will be suppressed. Thus what we observe for the inverse spin Hall effect on the $Pt$ layer is from the Skyrmions motion due to Hall viscosity!  \\

\acknowledgments{We are grateful to Sumit Das, Ori Ganor, Jung Hoon Han, Petr Ho\v{r}ava, Carlos Hoyos, Elias Kiritsis, Hitoshi Murayama, Yaron Oz, and Alfred Shapere for the collaborations, helpful discussions and valuable comments. Various parts of this work have been presented over a period of time in numerous places, including the invited talk at the APS March Meeting 2018, Los Angeles, CA; the invited talk at the 62nd Annual Conference on Magnetism and Magnetic Materials (the MMM conference) 2017, Pittsburgh, PA; the Berkeley Center for Theoretical Physics, Berkeley, CA; Korea Institute for Advanced Study, Seoul, Korea; Great Lakes Strings Conference, Ann Arbor, MI; SPOCK regional string meeting, Cincinnati, OH. We thank to the members of the institutes and organizers of the meetings for the invitations and opportunities to present the work. 
}

\reftitle{References}

\newpage
\appendixtitles{no} 
\appendixsections{multiple} 
\appendix
\section{Details of thermo-electromagnetic effects} \label{appsec:TEME}

\noindent In this appendix, we identify various transport coefficients that have been known in the literature. We quote the definition and work out the relations between the coefficients $L_{ij}$ and $L_{ij}'$ discussed in the main body. For generality, we present the results in the presence of magnetic field. \\

\noindent {\bf (a) Absolute Thermoelectric Power} is defined as $\epsilon = - \frac{E_x}{\nabla_x T} $ with $J_{e,x}=J_{e,y}=\nabla_y T =0$. Thus $\epsilon = L_{13}$, which can be trivially solved from the first equation in \eqref{TransportFormula2DAlternative} which is written in terms of $L_{ij}$. In the \eqref{TransportFormula2D} with the coefficients $L_{ij}'$, we need to solve the first two equations simultaneously. 
\begin{align} 
\begin{split}
0&= L_{11}' \frac{e^2}{T} E_x + L_{12}' \frac{e^2}{T} E_y + L_{13}' \frac{e}{T^2} (- \nabla_x T)   \;, \\
0&= -L_{12}' \frac{e^2}{T} E_x + L_{11}' \frac{e^2}{T} E_y - L_{14}' \frac{e}{T^2} (-\nabla_x T)    \;.
\end{split}
\end{align}
By subtracting these two equations to remove the second terms, one can get the direct relation between $E_x$ and $\nabla_x T$.Thus the definition $\epsilon = - \frac{E_x}{\nabla_x T} $ gives 
\begin{align}
\epsilon = L_{13} = \frac{L_{11}' L_{13}' + L_{12}' L_{14}'}{e T(L_{11}'^2 + L_{12}'^2)}  \;. 
\end{align}
This tells that \eqref{TransportFormula2DAlternative} is more convenient to evaluate the known transport coefficients than \eqref{TransportFormula2D}. As we demonstrate the procedure with an explicit example, we present the other results without explicit computations. \\ 
\noindent {\bf (b) Isothermal Electric Conductivity} is defined as $\sigma_i = -\frac{J_{e,x}}{E_x} $ with $J_{e,y}=\nabla_x T=  \nabla_y T =0$.  
\begin{align}
\sigma_i = \frac{1}{L_{11}} = - \frac{e^2 (L_{11}'^2 + L_{12}'^2)}{T L_{11}' }  \;. 
\end{align}
\noindent {\bf (c) Adiabatic Electric Conductivity} is defined as $\sigma_a = -\frac{J_{e,x}}{E_x} $ with $J_{e,y}=\nabla_x T=  Q_y =0$.  
\begin{align}
\sigma_a = \frac{L_{33}}{L_{11}L_{33} - T L_{14}^2} = - \frac{e^2 (-L_{11}' L_{13}'^2 - 2 L_{12}' L_{13}' L_{14}' + 
	L_{11} L_{14}'^2 + (L_{11}'^2 + L_{12}'^2) L_{33}')}{(-L_{13}'^2 + L_{11}' L_{33}') T}  \;. 
\end{align}
There is a typo in the expression in \cite{CallenThermodynamics}. 

\noindent Let us move on to the thermal conductivity. \\
{\bf (d) Isothermal Heat Conductivity} is defined as $\kappa_i = -\frac{Q_x}{\nabla_x T} $ with $J_{e,x}= J_{e,y}=  \nabla_y T =0$.  
\begin{align}
\kappa_i = L_{33} = \frac{-L_{11}' L_{13}'^2 - 2 L_{12}' L_{13}' L_{14}' + 
	L_{11}' L_{14}'^2 + (L_{11}'^2 + L_{12}'^2) L_{33}'}{(L_{11}'^2 + L_{12}'^2) T^2}  \;. 
\end{align}
{\bf (e) Adiabatic Heat Conductivity} is defined as $\kappa_a = -\frac{Q_x}{\nabla_x T} $ with $J_{e,x} = J_{e,y}=  Q_y =0$.  
\begin{align}
\sigma_a = \frac{L_{33}^2 + L_{34}^2}{L_{33}} = - \frac{Numerator}{(-L_{11}' L_{13}'^2 - 2 L_{12}' L_{13}' L_{14}' + L_{11}' L_{14}'^2 + (L_{11}'^2 + L_{12}'^2) L_{33}') T^2}  \;,
\end{align}
where $numerator = 
L_{13}'^4 + L_{14}'^4 - 4 L_{13}' L_{14}' (L_{12}' L_{33}' + L_{11}' L_{34}') + 
2 L_{14}'^2 (L_{11}' L_{33}' - L_{12}' L_{34}') + 
2 L_{13}'^2 (L_{14}'^2 - L_{11}' L_{33}' + L_{12}' L_{34}') + (L_{11}'^2 + 
L_{12}'^2) (L_{33}'^2 + L_{34}'^2)
$. This expression is fairly complicated. 

\noindent Due to the presence of magnetic field, there are Hall effects. \\
{\bf (f) Isothermal Hall Effect} is defined as $R_i = \frac{E_y}{H J_{e,x}} $ with $\nabla_x T= \nabla_y T=  J_{e,y} =0$.  
\begin{align}
R_i = \frac{L_{12}}{H} = \frac{T L_{12}' }{H e^2 (L_{11}'^2 + L_{12}'^2)}  \;. 
\end{align}
{\bf (g) Adiabatic Hall Effect} is defined as $R_a = \frac{E_y}{H J_{e,x}} $ with $\nabla_x T= Q_y =  J_{e,y} =0$.   
\begin{align}
R_a = \frac{1}{H} \Big(L_{12} + T\frac{L_{13} L_{14}}{L_{33}} \Big) =  \frac{(-L_{13}' L_{14}' + L_{12}' L_{33}') T}{e^2 (-L_{11}' L_{13}'^2 - 2 L_{12}' L_{13}' L_{14}' + L_{11}' L_{14}'^2 + (L_{11}'^2 + L_{12}'^2) L_{33}') }  \;.
\end{align}

\noindent Nernst effect measures that generation of transverse electric field upon applying the temperature gradient. \\
{\bf (h) Isothermal Nernst Effect} is defined as $\eta_i = \frac{E_y}{H \nabla_x T} $ with $\nabla_y T= J_{e,x} =  J_{e,y} =0$.  
\begin{align}
\eta_i = \frac{L_{14}}{H} = \frac{L_{12}' L_{13}' - L_{11}' L_{14}' }{H e T (L_{11}'^2 + L_{12}'^2)}  \;. 
\end{align}
{\bf (i) Adiabatic Nernst Effect} is defined as $\eta_a = \frac{E_y}{H \nabla_x T} $ with $Q_y = J_{e,x} =  J_{e,y} =0$.   
\begin{align}
\eta_a = \frac{1}{H} \Big(L_{14} - T\frac{L_{13} L_{34}}{L_{33}} \Big) =  \frac{L_{11}'L_{13}'L_{34}' - L_{11}'L_{14}' L_{33}' + L_{12}' L_{13}' L_{33}' +L_{12}'L_{14}'L_{34}' - L_{13}'^2 L_{14}' - L_{14}'^3}{e H T (-L_{11}' L_{13}'^2 - 2 L_{12}' L_{13}' L_{14}' + L_{11}' L_{14}'^2 + (L_{11}'^2 + L_{12}'^2) L_{33}') }  \;.
\end{align}
{\bf (j) Ettingshausen Effect} is given by $E_t = - \frac{\nabla_y T}{H Jx} $ with $\nabla_x T= Q_y =  J_{e,y} =0$.  
\begin{align}
E_t= \frac{T L_{14}}{H L_{33}} = -\frac{T^2 (L_{12}' L_{13}' - L_{11}' L_{14}') }{e H (-L_{11}' L_{13}'^2 - 2 L_{12}' L_{13}' L_{14}' + L_{11}' L_{14}'^2 + (L_{11}'^2 + L_{12}'^2) L_{33}')}  \;. 
\end{align}
{\bf (k) Leduk-Righi Effect} is defined as $\mathcal {L} = \frac{\nabla_y T}{H \nabla_x T} $ with $Q_y = J_{e,x} =  J_{e,y} =0$.   
\begin{align}
\mathcal {L}= \frac{L_{34}}{H L_{33}} =  \frac{L_{11}'(L_{11}'L_{34}' -2 L_{13}'L_{14}')+ L_{12}'^2 L_{34}' + L_{12}' (L_{13}' - L_{14}') (L_{13}' + L_{14}') }{H (-L_{11}' L_{13}'^2 - 2 L_{12}' L_{13}' L_{14}' + L_{11}' L_{14}'^2 + (L_{11}'^2 + L_{12}'^2) L_{33}') }  \;,
\end{align}

\noindent We summarize the result by combining together in a matrix form. 
\begin{align} \label{TransportFormula2DAlternativeMatrix}
\left( \begin{array}{c}
E_x \\ E_y \\Q_x\\ Q_y
\end{array} \right) =
\left( \begin{array}{cccc}
1/\sigma_i & H R_i & -\epsilon & - H \eta_i \\ 
-H R_i & 1/\sigma_i & H \eta_i&  -\epsilon\\ 
-T\epsilon & - T H \eta_i & -\kappa_i & - H \kappa_i \mathcal{L}_R \\ 
T H \eta_i & - T\epsilon& H \kappa_i \mathcal{L}_R  & - \kappa_i 
\end{array} \right)
\left( \begin{array}{c}
J_{e,x} \\ J_{e,y} \\ \nabla_x T \\ \nabla_y T
\end{array} \right) 
\end{align}
As discussed in the main body, there are various relations among the transport coefficients. Here are some of the relations. 
\begin{align}
\kappa_i E_{tth} &= T \eta_i \;,  \label{EttRelation}\\ 
\kappa_a- \kappa_i &= H \kappa_i^2 \mathcal{L}_R^2 \;, \\
\rho_a - \rho_i &= H^2 \eta_i E_{tth} \;, \\
R_a - R_i &= \epsilon E_{tth} \;, \\
\eta_ i - \eta_a &= \epsilon \mathcal{L}_R \;.
\end{align}	
The first relation \eqref{EttRelation} is presented in \eqref{EtthRelation0}. There we use $\kappa$ and $\eta_N$ instead of $\kappa_i$ and $\eta_i$. 

\end{document}